\documentclass[12pt]{article}

\usepackage[latin1]{inputenc}
\usepackage[authoryear,nonamebreak,elide]{natbib}
\usepackage[pagebackref,colorlinks=true,urlcolor=red,citecolor=blue,linkcolor=red,breaklinks=true]{hyperref}

\usepackage{graphicx, epsfig}
\usepackage{amsmath,amssymb}
\usepackage{color}
\usepackage{subfigure} % uso de várias figuras numa só
\usepackage{xytree} %pacote para diagrama de árvore
\usepackage[para]{threeparttable} %pacote para fazer rodapé de tabela
\usepackage{lscape} %pacote para landscape de páginas específicas usar \begin{landscape} e \end
\usepackage{rotating}

\usepackage{paralist} % pacote para  change the amount of vertical space between items in list environments - use \begin{compactitem}, \begin{compactenum}, \begin{compactdesc}

\usepackage{chngcntr}
\counterwithin*{section}{part}

\usepackage{titlesec}
\makeatletter
\@addtoreset{section}{part}
\makeatother

\newcommand*\Hide{
\titleformat{\part}{\LARGE\bfseries}{}{0pt}{}
}

\begin{document}

\title{A comparative review of generalizations of \\ the Gumbel extreme value distribution \\ with an application to wind speed data}
\author{Eliane C. Pinheiro\thanks{Corresponding author. Email: elianecpinheiro@gmail.com}
\ \ and \ \ Silvia L.P. Ferrari \\
{\small {\em Department of Statistics, University of S\~ao Paulo,
%\\Rua do Mat\~ao, 1010, S\~ao Paulo/SP, 05508-090, 
Brazil}}
}

\maketitle
\begin{abstract}
The generalized extreme value distribution and its particular case, the Gumbel extreme value distribution, are widely applied for extreme value analysis. 
The Gumbel distribution has certain drawbacks because it is a non-heavy-tailed distribution and is characterized by constant skewness and kurtosis.
The generalized extreme value distribution is frequently used in this context because it encompasses the three possible limiting distributions for a normalized maximum of infinite samples of independent and identically distributed observations.
However, the generalized extreme value distribution might not be a suitable model when each observed maximum does not come from a large number of observations. Hence, other forms of generalizations of the Gumbel distribution might be preferable.
Our goal is to collect in the present literature the distributions that contain the Gumbel distribution embedded in them and to identify those that have flexible skewness and kurtosis, are heavy-tailed and could be competitive with the generalized extreme value distribution. 
The generalizations of the Gumbel distribution are described and compared using an application to a wind speed data set and Monte Carlo simulations. 
We show that some distributions suffer from overparameterization and coincide with other generalized Gumbel distributions with a smaller number of parameters, i.e., are non-identifiable. 
Our study suggests that the generalized extreme value distribution and a mixture of two extreme value distributions should be considered in practical  applications.

%\noindent {\it Key words:} Extreme value distribution; Generalized extreme value distribution; Gumbel distribution; Heavy-tailed distribution; Non-identifiable model; Kurtosis; Mixture of extreme value distributions; Skewness; Wind speed.

\end{abstract}

\section{Introduction}
\label{introduction}

%Jeong, Murshed, Seo & Park 2014 - Stoch Environ Res Risk Asses - 
%A three-parameter kappa distribution with hydrologic application - a generalized gumbel distribution
%. In practice, however, results
%are sometimes unsatisfactory when the GEVD is fitted to
%finite samples, and thus a more flexible model or other
%forms of GEVD might be helpful. In this perspective, a
%four-parameter kappa distribution (K4D) was introduced
%by Hosking (1994) to modeling the maximum precipitation
%data, and then it has been used in many fields including
%environmental sciences (Hosking and Wallis 1997; Paridia
%1999; Park and Jung 2002; Singh and Deng 2003, for
%example). The

%Cox, Isham & Northrop 2002 - Philosophical Transactions of the Royal Society of London A - 
%Floods some probabilistic and statistical approaches.
%In spite of such well-established theory, extreme-value distributions are not always
%preferred in studies of empirical data on extremes. For example, Jones (1983) argues
%for a totally empirical approach to ­ tting, while the FEH (Institute of Hydrology
%1999, vol. 3, x 17.3.2) recommends a particular (three-parameter) generalized logistic
%distribution as preferable to a GEV distribution for UK annual ®ood maxima. A

Extreme value data usually exhibit excess kurtosis and/or heavy right tails. 
This is particularly common in environmental data, e.g., maximum water level \citep{BRUXER},
maximum wind speed \citep[Examples 6.1 and 9.14]{CASTILLO}, spatial and temporal variability of turbulence \citep{SANFORD}, 
daily maximum ozone measurement \citep{GILLELAND}, and 
largest lichen measurements \citep{COOLEY}. 
The generalized extreme value distribution (GEV) is fairly well-accepted as a standard working model. 
Despite such well-established theory, extreme-value distributions are not always
preferred in studies of empirical data that do not contemplate the conditions to use extreme value theory results.
Sometimes, the fit for finite samples is poor. 
To surpass these issues, other generalizations of the Gumbel distribution were proposed. 
For instance, \cite{HOSKING} proposed a four-parameter distribution to model the maximum precipitation
data that has been used in many fields including
environmental sciences, see \cite{HOSKING1997}, \cite{PARIDA}, \cite{PARK}, and \cite{SINGH};
%Cox et al.(2002) pág 1393 ...we consider data for annual maxima for the Thames
%at Kingston (site 39 001) taken from vol. 3 of the Flood estimation handbook (FEH)
%(Institute of Hydrology 1999), for which a record of 112 years is available.
% pág 1394 :
\citet[\S 17.3.2]{REED} recommends a particular three-parameter generalized logistic
distribution as preferable to a GEV distribution for UK annual flood maximum. 

The Gumbel distribution, also known as the extreme value distribution or the Gumbel extreme value distribution, is also used to model extreme values \citep{COLES,CASTILLO,PINHEIRO2012,PINHEIRO2015}. However, its skewness and kurtosis coefficients 
are constant, and its right tail is light. Generalizations of the Gumbel distribution with flexible skewness and kurtosis coefficients 
could provide better fits for extreme value data. 

We present a comprehensive comparative review of distributions that contain the Gumbel distribution as a special or limiting case.
We note that certain generalizations of the Gumbel distribution proposed in the literature are not identifiable.~\footnote{
A family of distributions with probability density function $f(x;\theta), \theta \in \Theta$, 
is said to be identifiable if, for any $\theta$ and $\theta^*$ in the parameter 
space $\Theta$, $f(x;\theta)=f(x;\theta^*) \Leftrightarrow \theta=\theta^*$.}
Some distributions suffer from overparameterization and
coincide with other generalized Gumbel distributions with a smaller number of parameters.
As noted by \cite{HUANG} {\it ``when applying a nonidentifiable model, different people may draw different conclusions 
from the same model of the observed data. Before one can meaningfully discuss the estimation of a model, model 
identifiability must be verified.''} Therefore, we distinguish between the identifiable and nonidentifiable models and limit our 
study to the identifiable family of distributions only.

We investigate and compare the relevant properties of the selected distributions. In particular, 
we derive their coefficients of skewness and kurtosis, which are invariant under location-scale transformations and are 
primarily controlled by the extra parameters. We graphically illustrate their flexibility relative to the Gumbel distribution
and highlight those that can achieve high values of skewness and kurtosis with a heavy right tail.

\cite{DANIELSSON} stated {\it ``heavy-tailed distributions are often defined in terms of higher than normal kurtosis. 
However, the kurtosis of a distribution may be high if either the tails of the cumulative distribution function are heavier 
than the normal or the center is more peaked or both.''} Moment-based measures suffer from effects from an extreme tail of the 
distribution, which may have negligible probability. These characteristics motivated us to study the tail behavior of the 
distributions specifically. To mathematically classify the tail behavior of distributions, we employ regular variation theory 
\citep{HAAN} and a criterion proposed by \cite{RIGBY} based on an approximation of the logarithm of 
the probability density function.  

Additionally, we conduct a comprehensive simulation study to evaluate the flexibility of each selected distribution in fitting 
data sets generated from the Gumbel distribution and its different generalizations.
The simulated data sets cover a reasonable range of skewness, kurtosis and tail heaviness behaviors.
We compare the different distributions through the analysis of a data set on the maximum monthly wind speed in 
West Palm Beach, Florida, for the years 1984-2014. 

The paper is organized as follows. In Section~\ref{gumbel_generalizations}, we present the Gumbel distribution and its generalizations.
In Section~\ref{tail}, we study the right tail heaviness of the identifiable distributions. Monte Carlo simulations 
are presented in Section~\ref{simulation}, and an application to a real data set is provided in Section~\ref{application}. 
The paper ends with conclusions in Section~\ref{conclusion}. Technical details are given in the Supplement.

\section{The Gumbel distribution and its generalizations}
\label{gumbel_generalizations}

We present selected characteristics of the Gumbel distribution and distributions that contain the Gumbel distribution as a special or limiting case.
For the identifiable distributions, the moments, $p$-quantile ($x_p$), skewness ($\gamma_1$) and kurtosis ($\gamma_2$) coefficients are summarized in the Supplement.
Random draws from distributions with closed-form $p$-quantiles can be generated by replacing $p$ with a standard uniform distributed observation.
For the others, generating methods are given.

\paragraph{Gumbel distribution (EV).}  
Let $X \sim {\rm EV}_{max}(\mu,\sigma)$ be a continuous random variable with a maximum extreme value distribution. 
The probability density function (pdf) and cumulative distribution function (cdf) are, respectively,
\begin{equation}
\label{extreme value-density}
f_{{\rm EV}_{max}}(x; \mu, \sigma) = \frac{1}{\sigma}\exp\left(-\frac{x-\mu}{\sigma}\right)\exp\left\{-\exp\left(-\frac{x-\mu}{\sigma}\right)\right\},\quad x \in I\!\! R,
\end{equation}
\begin{equation}\label{extreme value-cumulative}
F_{{\rm EV}_{max}}(x; \mu, \sigma) = \exp\left(-\exp\left(-\frac{x-\mu}{\sigma}\right)\right) ,\quad x \in I\!\! R,
\end{equation}
where $\mu \in I\!\! R$ is the location parameter and $\sigma > 0$ is the scale parameter.
This distribution is also known as the Gumbel or type I extreme value distribution. 
The distribution in (\ref{extreme value-density}) is one of the three possible limiting laws 
of the standardized maximum of independent and identically distributed random variables \citep{GNEDENKO}.
It is frequently invoked to model extreme events; see, e.g., \citet[Table 9.16]{CASTILLO} and \citet[Section 3.4.1]{COLES}.
We refer to this distribution as the maximum extreme value distribution to distinguish it from the minimum extreme value distribution, 
which is also often known as the Gumbel or type I extreme value distribution in the statistical literature. 
%The pdf of the minimum extreme value distribution with location parameter $\mu$ and dispersion parameter $\sigma$ is given by
%\begin{equation}
%\label{extreme value-density-min}
%f_{{\rm EV}_{min}}(x; \mu, \sigma) = \frac{1}{\sigma}\exp\left(\frac{x-\mu}{\sigma}\right)\exp\left\{-\exp\left(\frac{x-\mu}{\sigma}\right)\right\} ,\quad x \in I\!\! R,
%\end{equation}
%and we write $X \sim {\rm EV}_{min}(\mu,\sigma)$. 
%A useful property is that
%$$
%X \sim {\rm EV}_{min}(\mu,\sigma) \Longleftrightarrow -X \sim {\rm EV}_{max}(-\mu,\sigma).
%$$
%The distribution in (\ref{extreme value-density-min}) is the distribution of the logarithm of a Weibull distributed random variable and is often used in reliability and survival analysis to model log-lifetimes \citep{LAWLESS}.

The coefficients of skewness and kurtosis of the Gumbel distribution are constant $\gamma_{1,{\rm EV}}=1.14$ and $\gamma_{2,{\rm EV}}=5.4$, respectively, i.e., parameter independent. 
This restriction motivates more flexible and useful generalizations of the Gumbel distribution to fit real data.

Hereafter, the maximum extreme value or Gumbel variable will be referred to as Gumbel and denoted by EV.

\paragraph{Generalized extreme value distribution (GEV).} 
The generalized extreme value distribution (GEV) was defined for the first time by \citet{JENKINSON}, and the three possible limiting distributions of the maximum/minimum of random variables are embedded within it.
This distribution is also known as the von Mises extreme value, von Mises-Jenkinson, and Fisher-Tippet distribution.
A historical review of extreme value theory, the main results, and a list of several areas of application are provided in \cite{KOTZ2000}.

Let $X \sim {\rm GEV}(\mu,\sigma,\alpha)$ be a generalized extreme value distributed random variable. Its pdf and cdf are, respectively,
\begin{eqnarray*}
\label{fdpGEV}
&&f_{{\rm GEV}}(x; \mu, \sigma,\alpha)=\frac{1}{\sigma}
{\rm exp} \left( -\left[ 1+\alpha \left( \frac{x-\mu}{\sigma}  \right) \right]^{-1/\alpha} \right)
\left[ 1+\alpha \left( \frac{x-\mu}{\sigma}  \right) \right]^{-1/\alpha-1},
%\nonumber\\ &&\{x:1+\alpha ( x-\mu) / \sigma >0 \},
\end{eqnarray*}
and
$$
F_{{\rm GEV}}(x; \mu, \sigma,\alpha)={\rm exp} \left( -\left[ 1+\alpha \left( \frac{x-\mu}{\sigma}  \right) \right]^{-1/\alpha} \right),\quad \{x:1+\alpha ( x-\mu) / \sigma >0 \},
$$
where  $\alpha \in I\!\! R$.
The Gumbel distribution is a particular case of the GEV distribution when $\alpha \rightarrow 0$.

\begin{figure}[!ht]
\centering
\subfigure{\includegraphics[height=38mm,width=38mm]{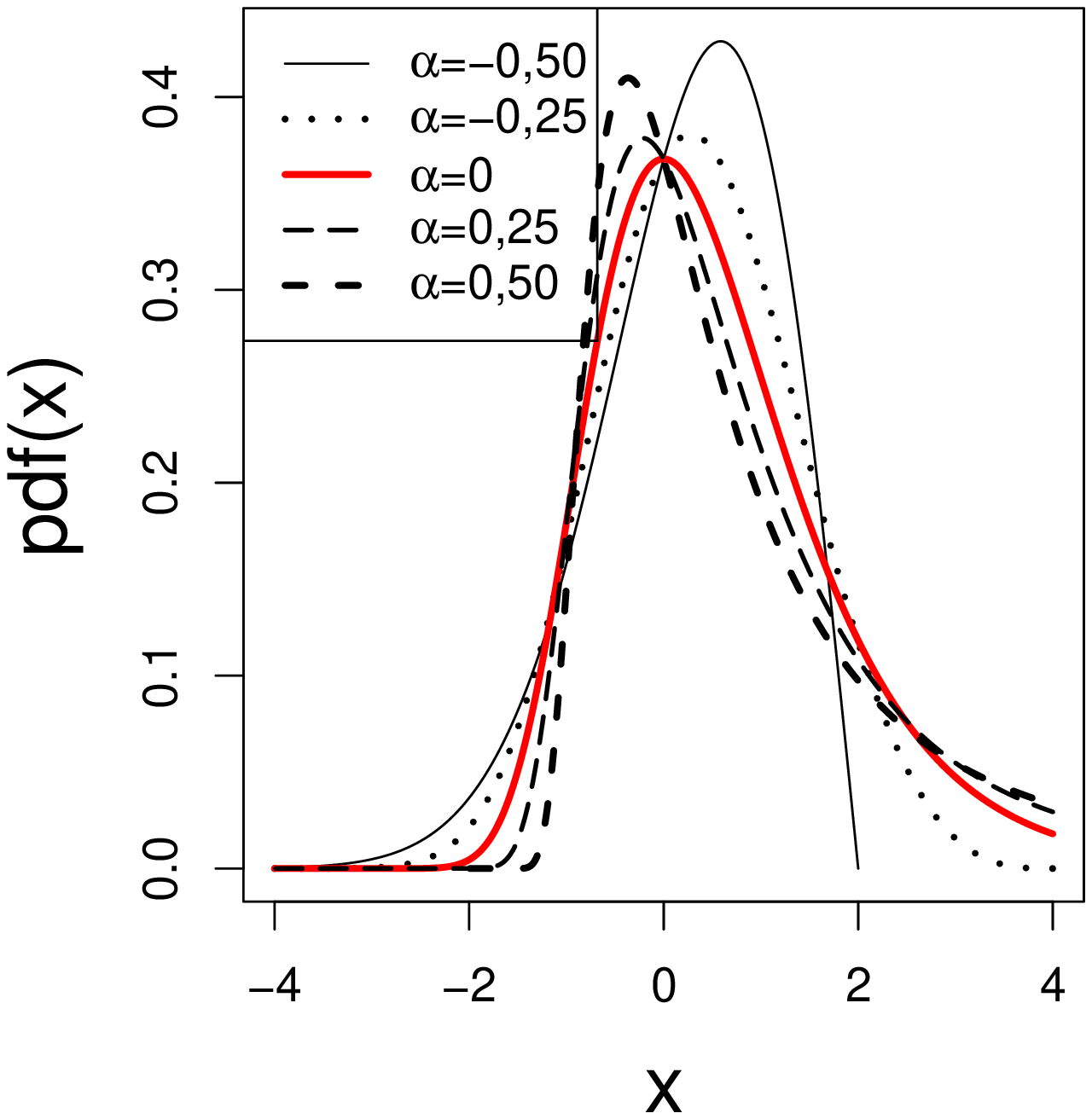}} \quad
\subfigure{\includegraphics[height=38mm,width=38mm]{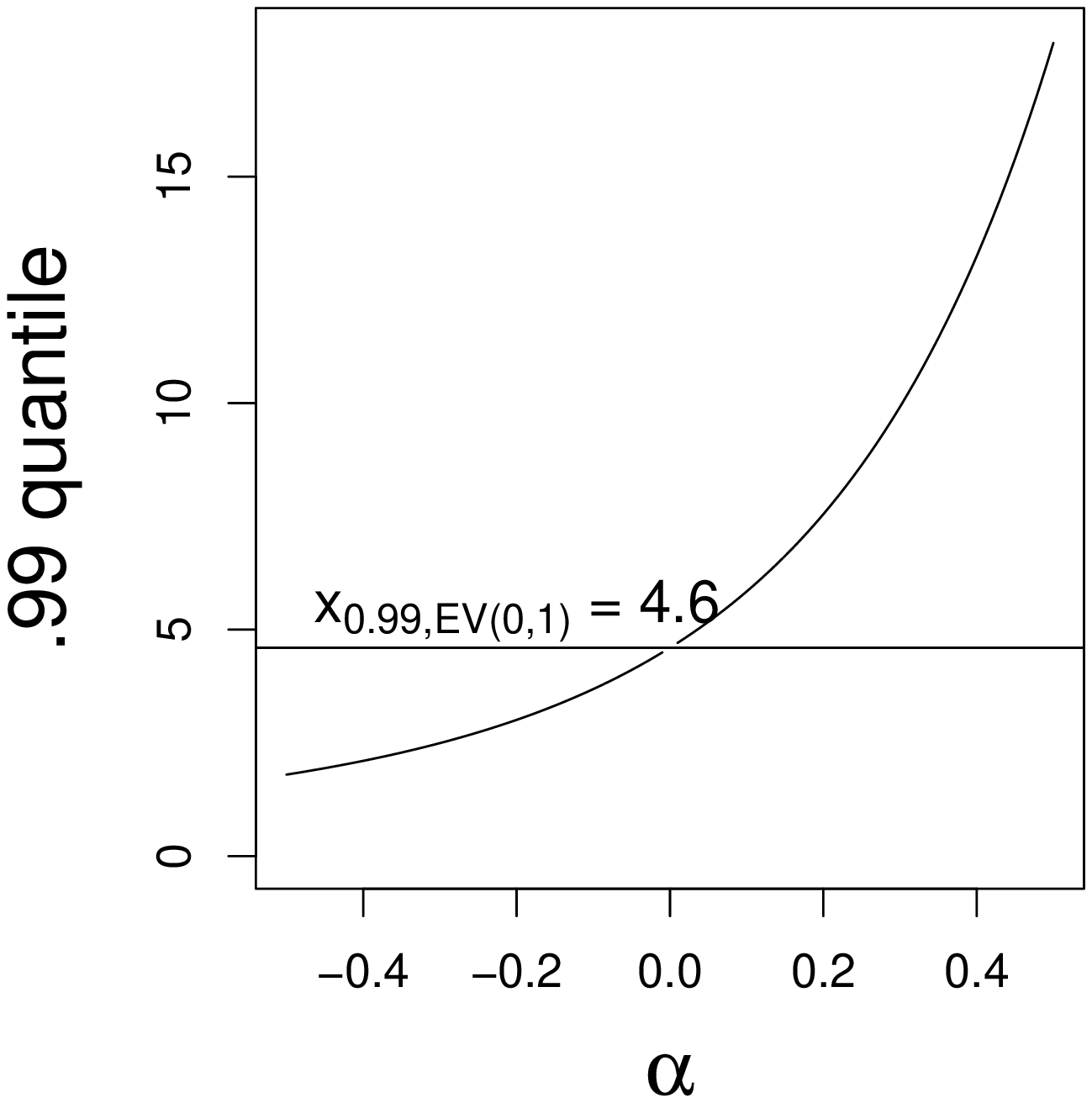}} \quad
\subfigure{\includegraphics[height=38mm,width=38mm]{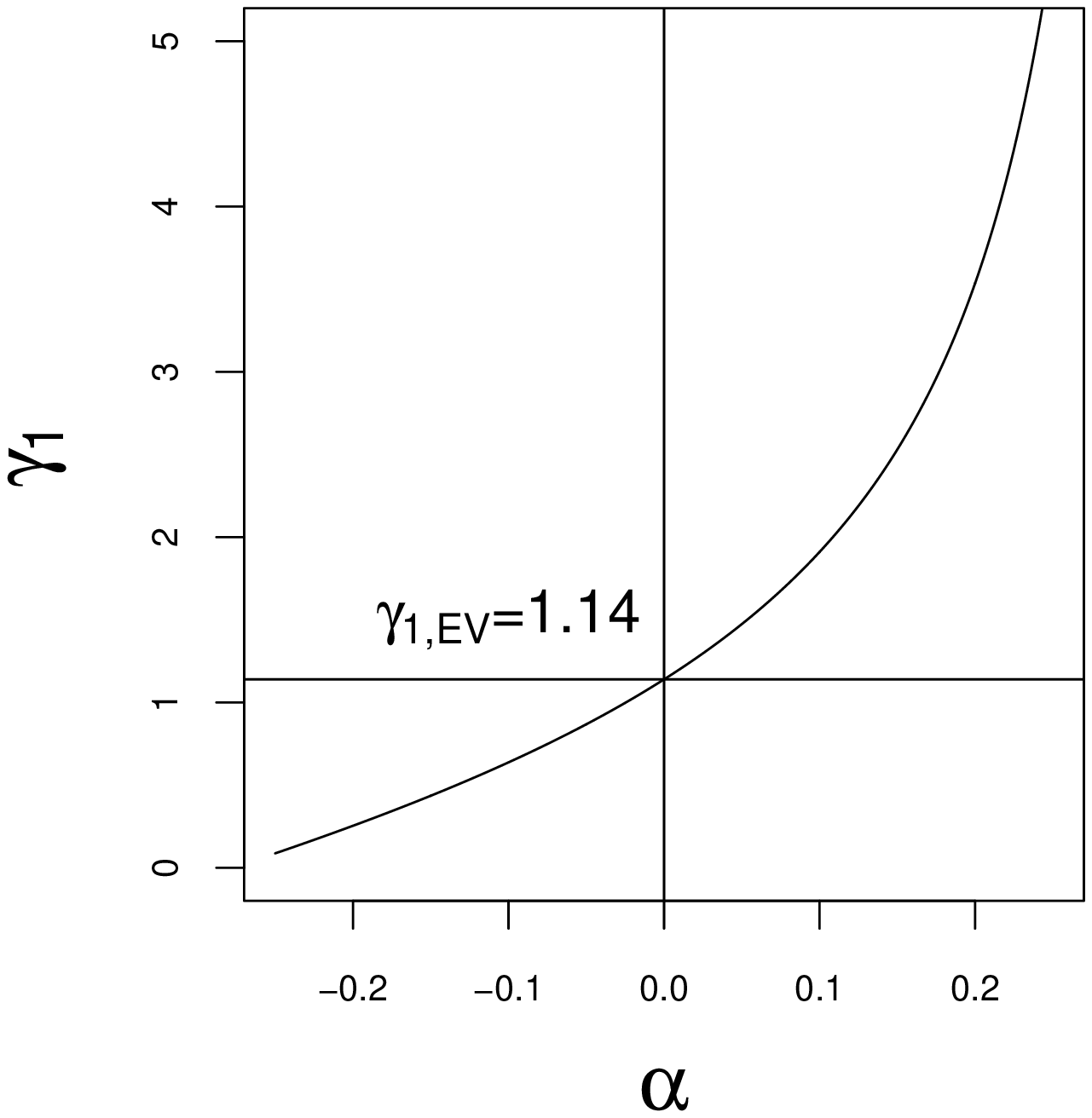}} \quad
\subfigure{\includegraphics[height=38mm,width=38mm]{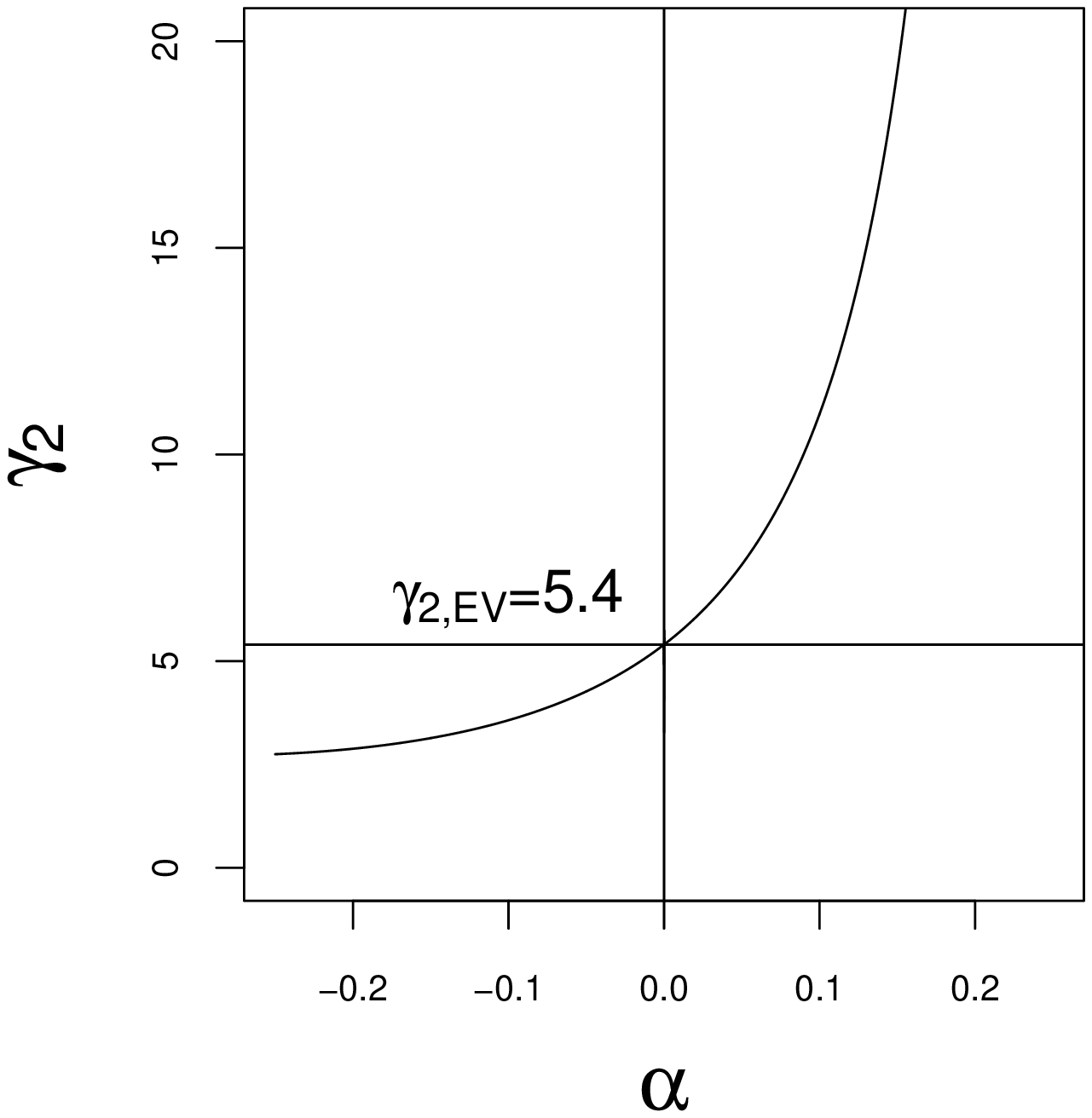}}
%\vspace{-1.2cm}
\caption{Density function and .999 quantile - GEV$(0,1,\alpha)$; skewness and kurtosis - GEV$(\mu,\sigma,\alpha)$}
\label{fig:GEV}
\end{figure}

Plots of the pdf and the .99 quantile of GEV$(0,1,\alpha)$, and the skewness and kurtosis of GEV$(\mu,\sigma,\alpha)$ for 
selected values of $\alpha$ are shown in Figure~\ref{fig:GEV}.
The GEV distribution is quite versatile, and $\alpha$ has a substantial effect on its skewness and kurtosis.
The parameter $\alpha$ affects location, dispersion, skewness and kurtosis. 
Increasing values of $\alpha$ increases the quantiles, skewness and kurtosis coefficients, and right-tail heaviness.
Skewness is defined for $\alpha<1/3$ and kurtosis for $\alpha<1/4$.
Skewness and kurtosis can assume different values from those of the Gumbel distribution.

\paragraph{Exponentiated Gumbel distribution (EGu).}  
Let $X \sim {\rm EGu}(\mu,\sigma,\alpha)$ be an exponentiated Gumbel distributed random variable. Its pdf and cdf are, respectively,
$$
f_{{\rm EGu}}(x; \mu, \sigma,\alpha)=\frac{\alpha}{\sigma}
\exp\left(-\frac{x-\mu}{\sigma}  \right)
\exp  \left(-\exp\left(-\frac{x-\mu}{\sigma}  \right)\right)
\left[  1-\exp \left(-\exp\left(-\frac{x-\mu}{\sigma}  \right)\right)\right]^{\alpha-1},
$$
$x \in I\!\! R,$ and
$$
F_{{\rm EGu}}(x; \mu, \sigma,\alpha)=1-\left[  1-\exp \left(-\exp\left(-\frac{x-\mu}{\sigma}  \right)\right)\right]^\alpha,\quad x \in I\!\! R,
$$
where 
%$\mu \in I\!\! R$ is the location parameter, $\sigma > 0$ is the dispersion parameter and  
$\alpha>0$ \citep{NADARAJAH2006}.
The Gumbel distribution is a special case of the EGu distribution when
$\alpha=1$.

The pdf can be written as 
$$
f_{{\rm EGu}}(x; \mu, \sigma,\alpha)=
$$
$$
\frac{\alpha}{\sigma}\exp\left(-\frac{x-\mu}{\sigma}  \right)
\exp  \left(-\exp\left(-\frac{x-\mu}{\sigma}  \right)\right)
\; _2 F_1\left(1-\alpha,1;1;\exp \left(-\exp\left(-\frac{x-\mu}{\sigma}  \right)\right)\right),
$$
where $_2 F_1(a,b;c;z)=\sum_{k=0}^\infty[(a)_k(b)_k/(c)_k] [z^k/k!]$ for $|z|<1$ is the hypergeometric function, $(a)_k=\Gamma(a+k)/\Gamma(a)$ is the Pochhammer symbol
and 
$\Gamma(\cdot)$ is the gamma function. 
Thus
$_2 F_1(a,1;1;z)=\sum_{k=0}^\infty(a)_k z^k/k!=(1-z)^{-a}$ for $|z|<1$. 
Note that 
$\left|\exp \left(-\exp\left(-(x-\mu)/(\sigma) \right)\right)\right|<1$.
This form of the pdf is computationally highly efficient for evaluating moments of the EGu distribution if using software that contains an optimized implementation of the hypergeometric function.

\begin{figure}[!ht]
\centering
\subfigure{\includegraphics[height=38mm,width=38mm]{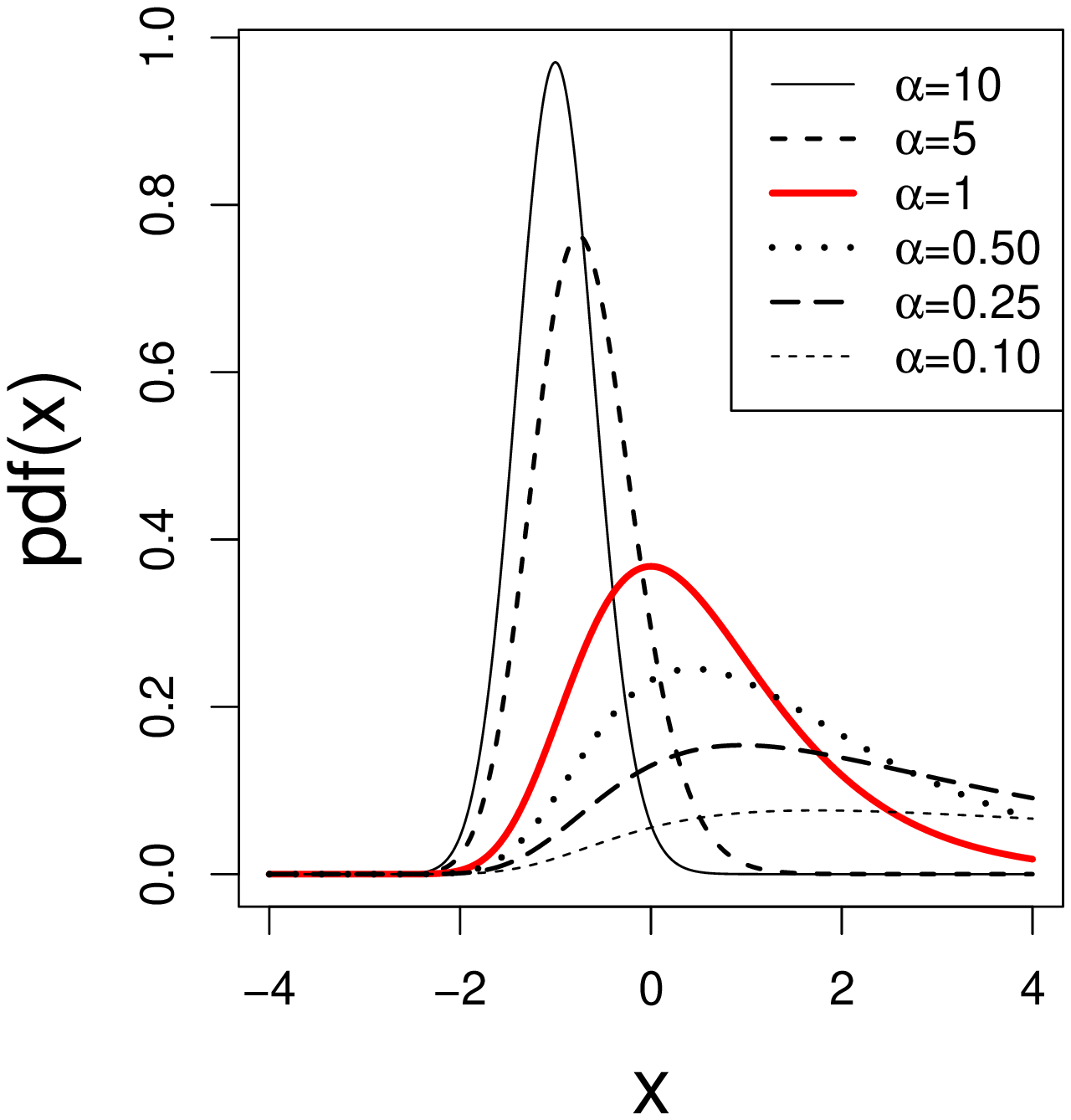}} \quad
\subfigure{\includegraphics[height=38mm,width=38mm]{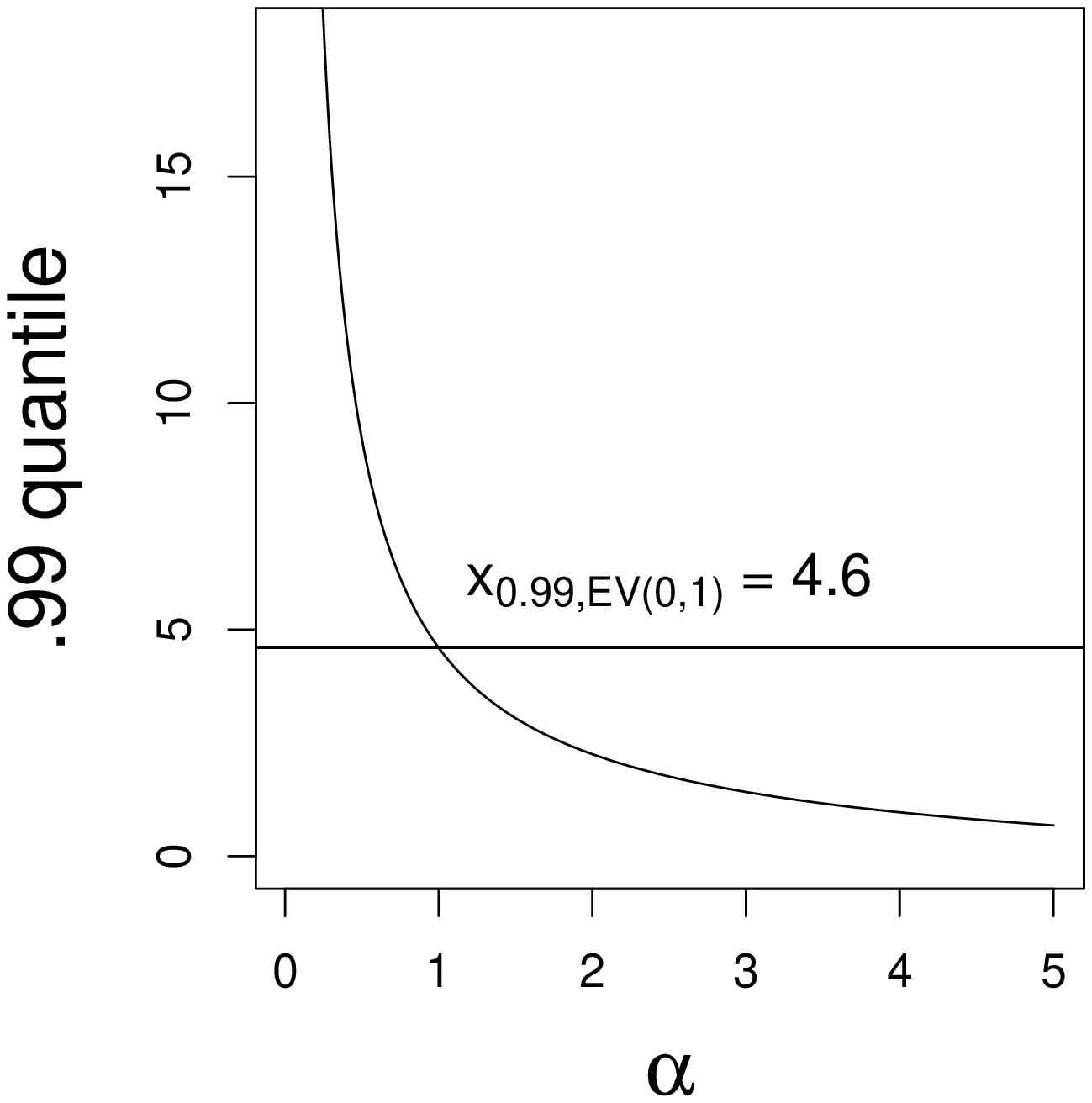}} \quad
\subfigure{\includegraphics[height=38mm,width=38mm]{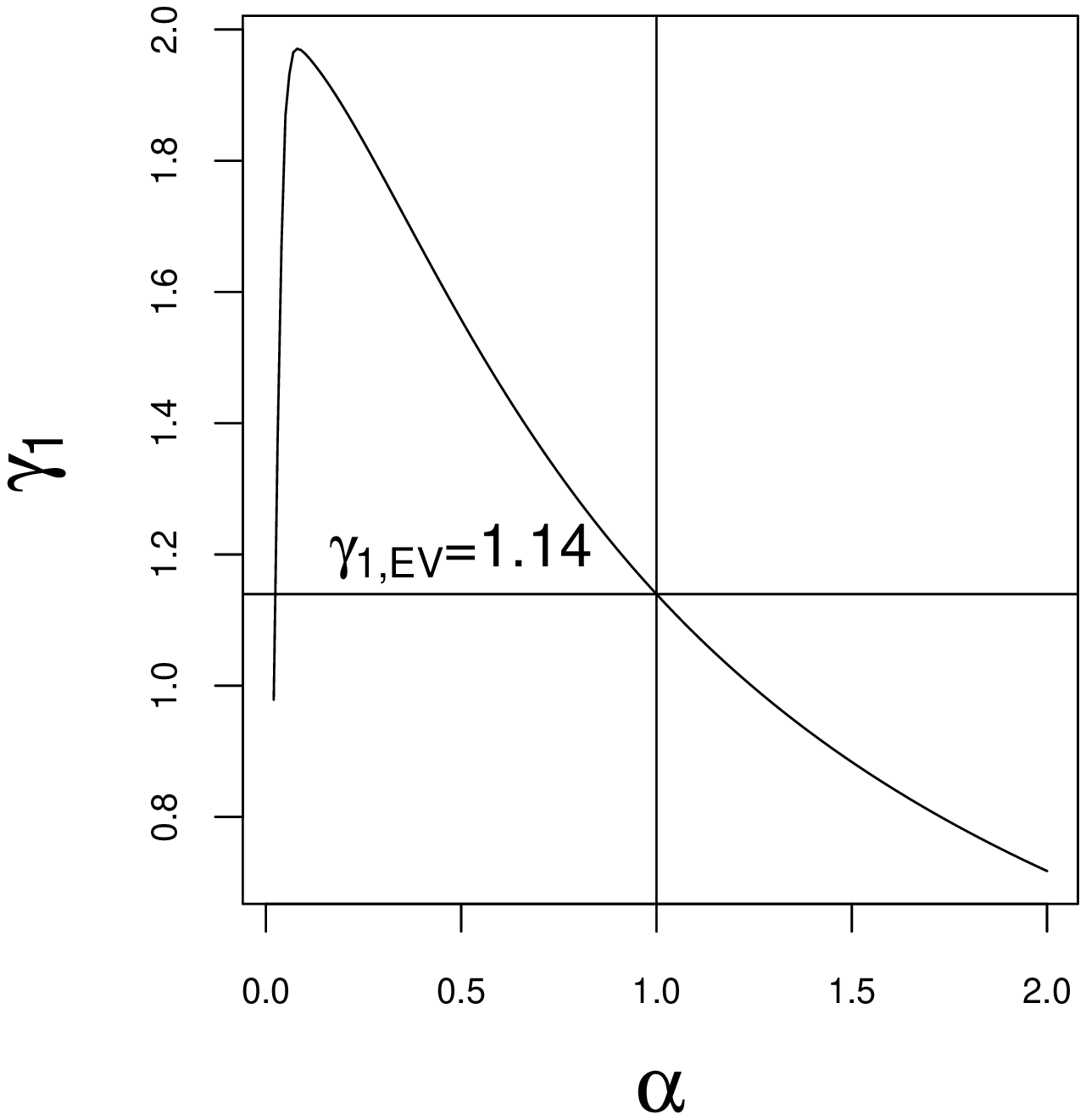}} \quad
\subfigure{\includegraphics[height=38mm,width=38mm]{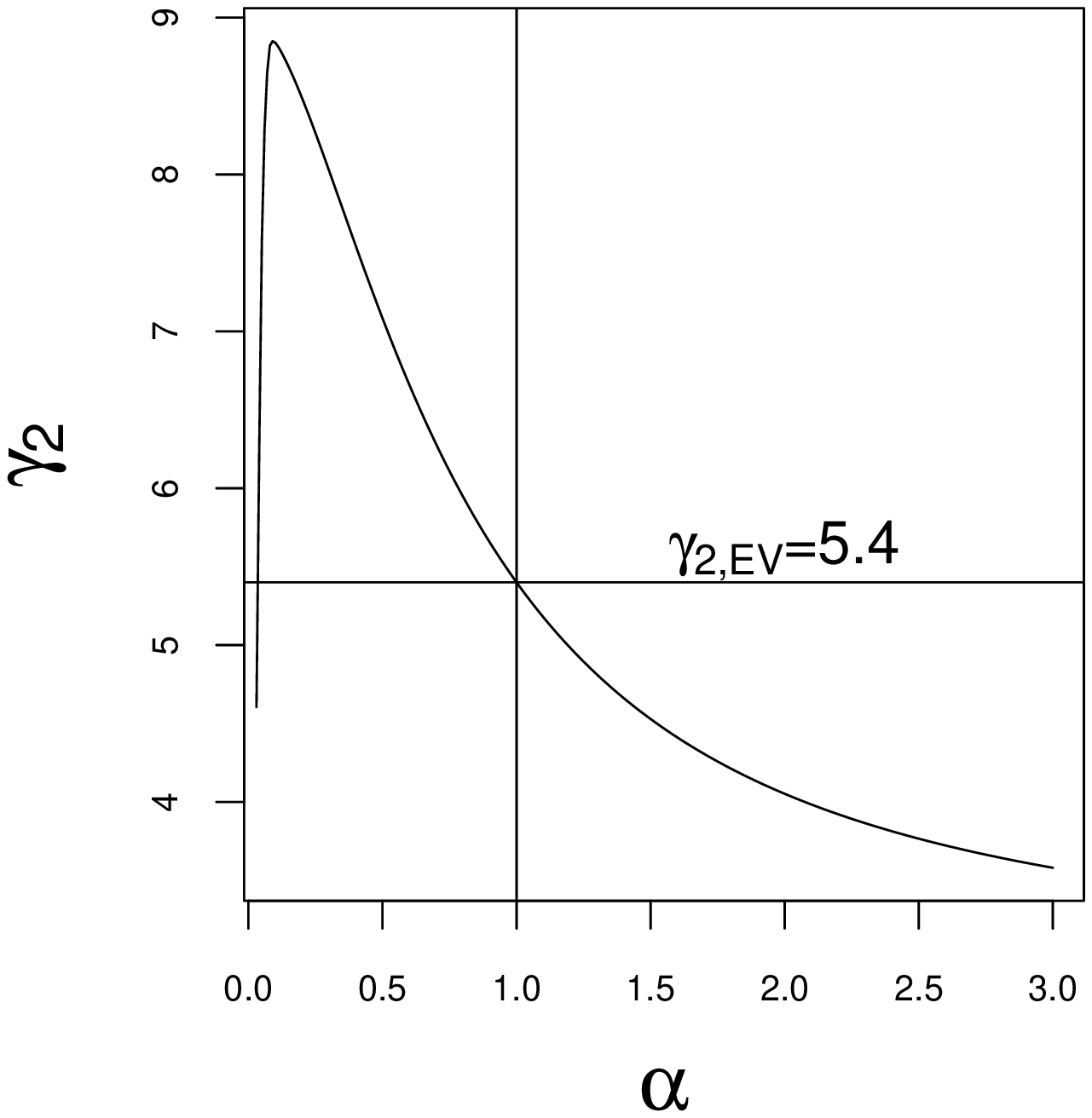}}
%\vspace{-1cm}
\caption{Density function and .99 quantile - EGu$(0,1,\alpha)$; skewness and kurtosis - EGu$(\mu,\sigma,\alpha)$}
\label{fig:EGu}
\end{figure}

The right tail is heavier for smaller values of $\alpha>0$ (Figure~\ref{fig:EGu}). When $\alpha$ is 
close to zero, minor changes in $\alpha$ lead  to significant changes in the quantile values. 
The skewness and kurtosis can reach values close to 2 and 9, respectively, indicating that the EGu distribution is more 
flexible than the Gumbel distribution.

\paragraph{Transmuted extreme value distribution (TEV).} 
\citet{SHAW} defined a transformation known as the rank transmutation map, with the aim of obtaining distributions with skewness and kurtosis distinct from those of the normal distribution. 

Transmutation is a composite map of a cumulative distribution function with a quantile function of another distribution defined on the same sample space. 
A particular case of rank transmutation map is derived by considering
$$
T_R(u)=F_2(F_1^{-1}(u))=u+\alpha u(1-u),
$$
which leads to
$$
F_2(x)=(1+\alpha)F_1(x)-\alpha F_1^2(x),
$$
for $|\alpha|\leq 1$, known as the quadratic rank transmutation.
There are two important boundary cases.
When $\alpha=-1$, $F_2(x)=F_1(x)^2$, i.e., $F_2$ 
is the distribution of the maximum of two independent variables with distribution $F_1$.
Analogously, when $\alpha=1$, $F_2$ is the distribution of the minimum.
Motivated by the various applications of the extreme value theory, particularly the Gumbel distribution, \citet{ARYAL} defined a new distribution known as the transmuted extreme value distribution (TEV) by replacing $F_1$ with a Gumbel cdf.

Let $X \sim {\rm TEV}(\mu,\sigma,\alpha)$ be a transmuted extreme value distributed random variable. Its pdf and cdf are, respectively, 
\begin{eqnarray*}
&&f_{{\rm TEV}}(x; \mu, \sigma,\alpha)=\frac{1}{\sigma}\exp\left[ -\frac{x-\mu}{\sigma} -\exp\left(-\frac{x-\mu}{\sigma} \right)\right]\left[1+\alpha -2\alpha \exp\left(  -\exp\left(-\frac{x-\mu}{\sigma} \right)\right)  \right] ,\\
&& x \in I\!\! R,
\end{eqnarray*}
and 
$$
F_{{\rm TEV}}(x; \mu, \sigma,\alpha)=(1+\alpha)\exp\left[  -\exp\left(-\frac{x-\mu}{\sigma} \right)\right]-\alpha\exp\left[  -2\exp\left(-\frac{x-\mu}{\sigma} \right)\right],\quad x \in I\!\! R,
$$ 
where $|\alpha|\leq 1$. 
The Gumbel distribution is a particular case of the TEV distribution when $\alpha=-1$ or $\alpha=0$.
Note that to make the TEV$(\mu,\sigma,\alpha)$ family of distributions identifiable, it is sufficient to restrict $\alpha$ to the set $(-1,1]$.
 
\begin{figure}[!ht]
\centering
\subfigure{\includegraphics[height=38mm,width=38mm]{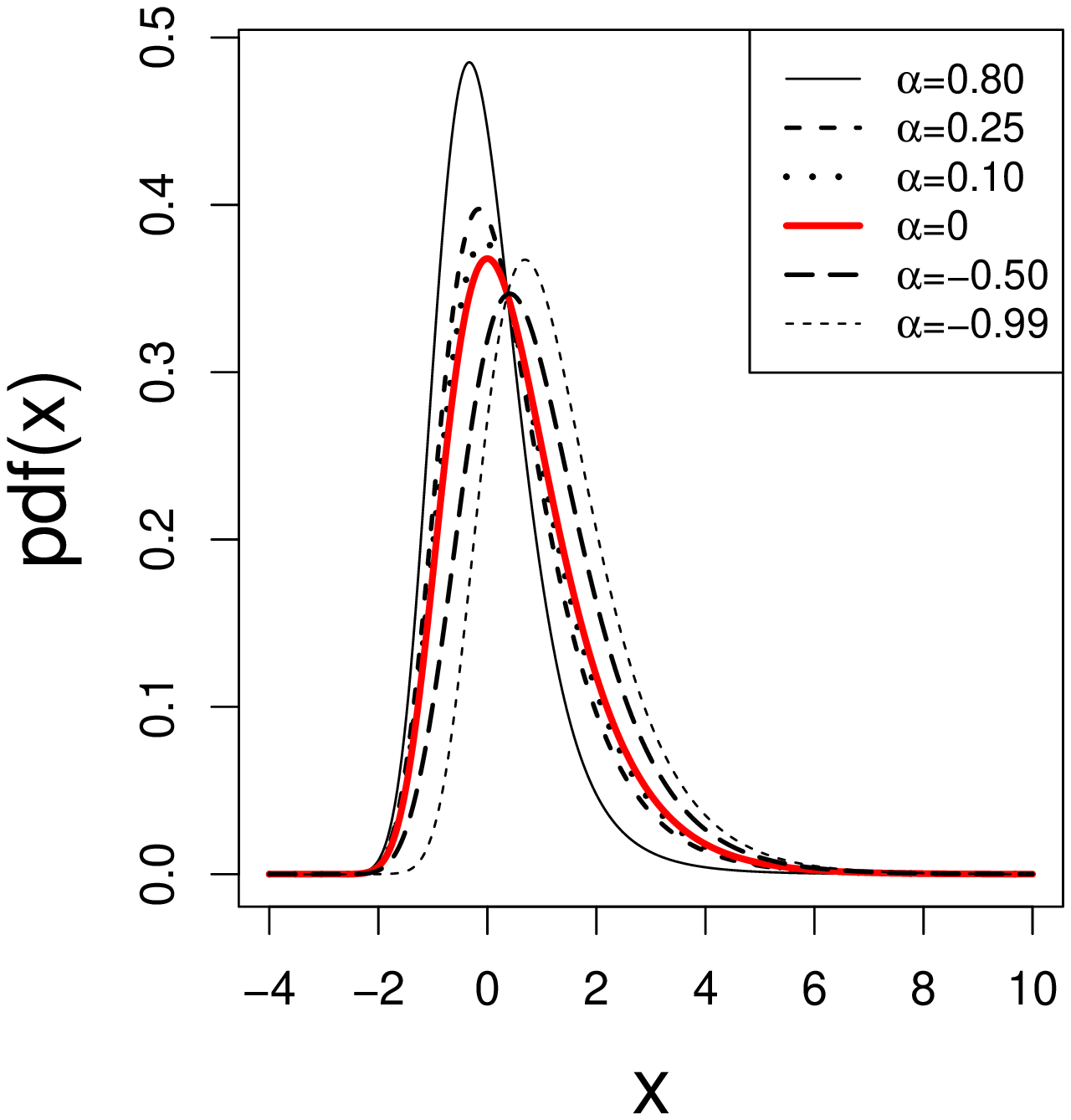}} \quad
\subfigure{\includegraphics[height=38mm,width=38mm]{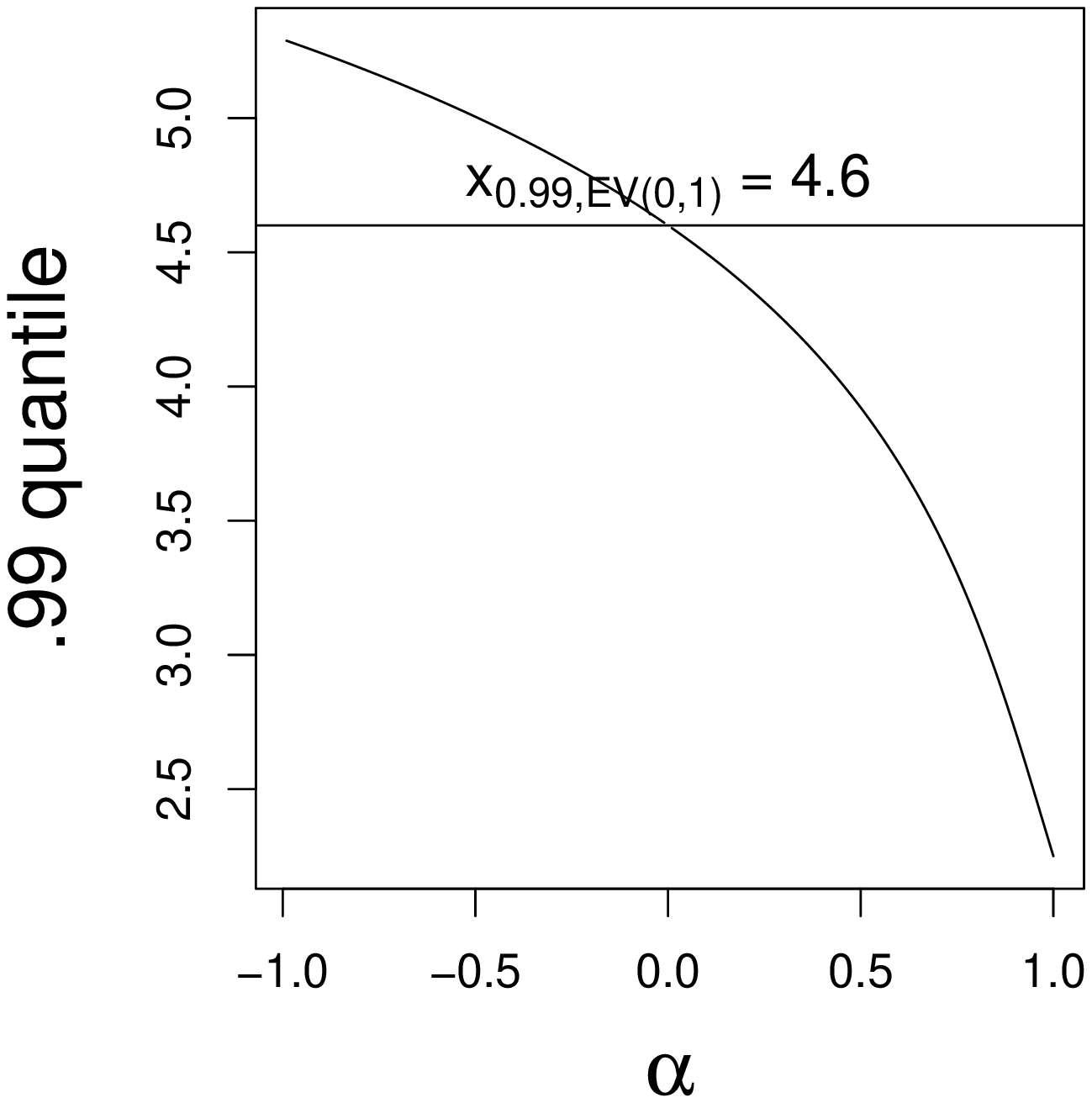}} \quad
\subfigure{\includegraphics[height=38mm,width=38mm]{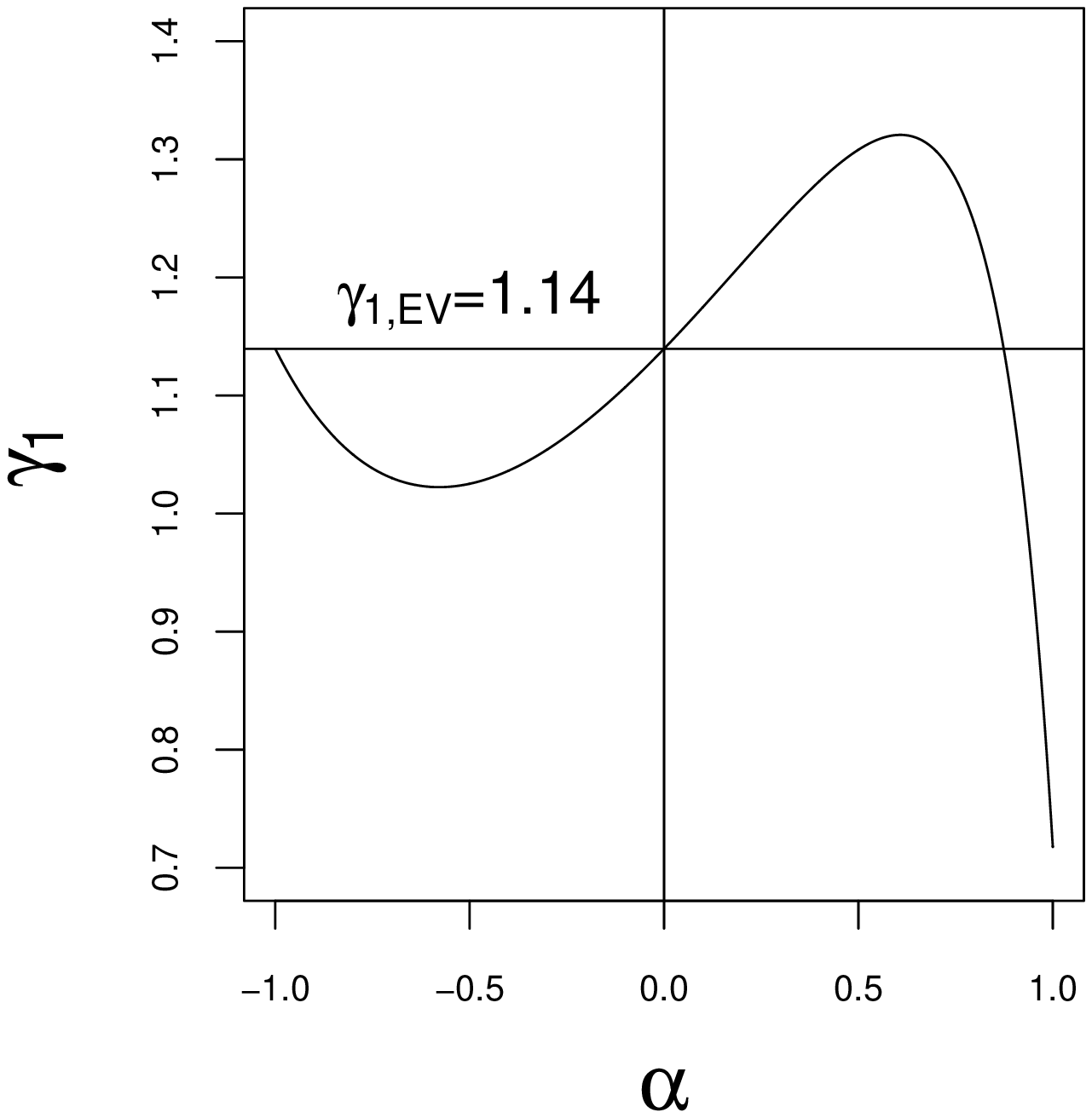}} \quad
\subfigure{\includegraphics[height=38mm,width=38mm]{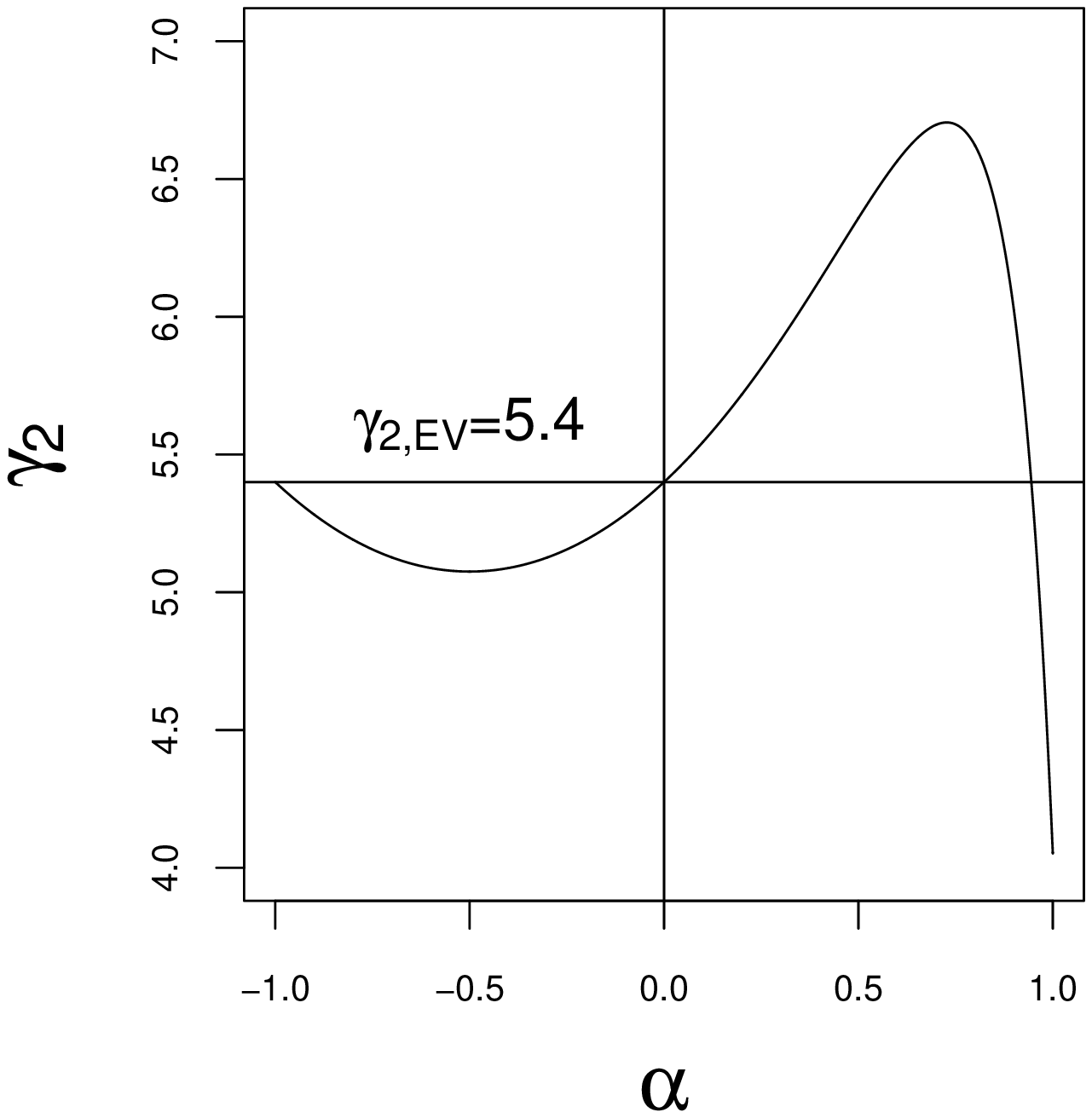}}
%\vspace{-0.6cm}
\caption{Density function and .99 quantile - TEV$(0,1,\alpha)$; skewness and kurtosis - TEV$(\mu,\sigma,\alpha)$}
\label{fig:TEV}
\end{figure}

The TEV distribution is more flexible relative to the Gumbel distribution but less flexible than the GEV and EGu distributions, 
with maximum .99 quantile (for $\mu=0$ and $\sigma=1$) and coefficients of skewness and kurtosis lower than 6, 2 and 7, respectively (Figure~\ref{fig:TEV}).
Note that from the pdf and .99 quantile plots, the right tail gets heavier for smaller values of $-1<\alpha\leq 1$.

\paragraph{Kumaraswamy Gumbel distribution.} 
\citet{CORDEIRO2012} defined a generalization of a cdf $G(x)$ from the Kumaraswamy distribution, which they referred to as Kum-G. The cdfs of the Kumaraswamy and Kum-G distributions are given, respectively, by
$$
F_{{\rm Kum}}(x;\alpha,\beta)=1-\{1-x^\alpha \}^\beta,\quad x \in (0,1),
$$
and
$$
F_{{\rm Kum}{\texttt -}{\rm G}}(x;\alpha,\beta)=1-\{1-G(x)^\alpha \}^\beta,\quad x \in I\!\! R,
$$
where $\alpha>0$ and $\beta>0$.
% are additional parameters closely related to skewness and kurtosis.
 If the distribution $G(x)$ is EV$(\mu,\sigma)$, the cdf is defined by
$$
F_{{\rm KumGum}}(x; \mu, \sigma,\alpha,\beta)=1-\left[1-\left(  \exp\left(  -\exp\left(-\frac{x-\mu}{\sigma}  \right) \right)  \right)^\alpha \right]^\beta,\quad x \in I\!\! R, 
$$
where $\alpha>0$ and $\beta>0$.

Note that if $X \sim {\rm KumGum}(\mu,\sigma,\alpha,\beta)$, then
\begin{eqnarray*}
F_{{\rm KumGum}}(x; \mu, \sigma,\alpha,\beta)
&=&1-[1-\exp(-\exp(- (x-\mu^*)/\sigma))]^\beta \\
&=&F_{{\rm KumGum}}(x; \mu^*, \sigma,1,\beta)=F_{{\rm EGu}}(x; \mu^*, \sigma,\beta),
\end{eqnarray*}
where $\mu^*=\mu+\sigma \ln \alpha$. 
Therefore, the Kumaraswamy Gumbel family of distributions KumGum $(\mu,\sigma,\alpha,\beta)$, where $\alpha>0$ and $\beta>0$, is nonidentifiable. 
It coincides with the exponentiated Gumbel family of distributions EGu$(\mu^*,\sigma,\beta)$, 
where $\mu^* \in I\!\! R$, $\sigma>0$ and $\beta>0$. 
In other words, the Kumaraswamy Gumbel family of distributions has four parameters but corresponds to a family with only three parameters.
This is a typical case of parameter redundancy, i.e., overparameterization \citep{CATCHPOLE}. Therefore, this distribution will not be contemplated hereafter.

\paragraph{Generalized three-parameter Gumbel distribution (GTIEV3).}  
\citet{DUBEY} built a generalization of the Gumbel distribution which is known as the generalized type I extreme value 
or type I generalized logistic distribution, and we denote it by GTIEV$(\mu,\sigma,\alpha,\beta)$. Its cdf is given by
$$
F_{{\rm GTIEV}}(x; \mu, \sigma,\alpha,\beta)=\left(1+ \frac{\sigma}{\beta} \exp\left(-\frac{x-\mu}{\sigma}\right)\right)^{-\alpha} ,\quad x \in I\!\! R,
$$
where 
$\alpha>0$ and $\beta>0$.
This distribution was first defined by \citet{HALD}. Note that
$$
F_{{\rm GTIEV}}(x; \mu,\sigma,\alpha,\beta)=\left(1+\frac{1}{\alpha}\exp\left(-\frac{x-\mu^*}{\sigma}\right)\right)^{-\alpha},\quad x \in I\!\! R,
$$
where $\mu^*=\mu+\sigma\ln( \sigma \alpha \beta^{-1}) \in I\!\! R$.
Therefore, the generalized Gumbel family GTIEV$(\mu,\sigma,\alpha,\beta)$, 
where  $\mu \in I\!\! R$, $\sigma > 0$, $\alpha>0$, and $\beta>0$, is nonidentifiable. It coincides with a family of distributions with 
only three parameters, say GTIEV3$(\mu,\sigma,\alpha)$, where $\mu \in I\!\! R$, $\sigma > 0$ and $\alpha>0$.

Let $X \sim  {\rm GTIEV}3(\mu,\sigma,\alpha)$ be a generalized three-parameter Gumbel random variable. Its pdf and cdf are defined, respectively, as  
$$
f_{{\rm GTIEV}3}(x; \mu, \sigma,\alpha)=
\frac{1}{\sigma}\left(1+ \frac{1}{\alpha} \exp\left(-\frac{x-\mu}{\sigma}\right)\right)^{-\alpha-1}\exp\left(-\frac{x-\mu}{\sigma}\right),\quad x \in I\!\! R,
$$
and
$$
F_{{\rm GTIEV}3}(x; \mu, \sigma,\alpha,\beta)=
\left(1+ \frac{1}{\alpha} \exp\left(-\frac{x-\mu}{\sigma}\right)\right)^{-\alpha} 
,\quad x \in I\!\! R,
$$
where $\alpha>0$. The Gumbel distribution is a limiting case of GTIEV3 when $\alpha \rightarrow \infty$.
The three-parameter kappa distribution defined in \citet[eq. 2]{JEONG} with positive shape parameter coincides with the GTIEV3 distribution in a different parameterization.

\begin{figure}[!ht]
\centering
\subfigure{\includegraphics[height=38mm,width=38mm]{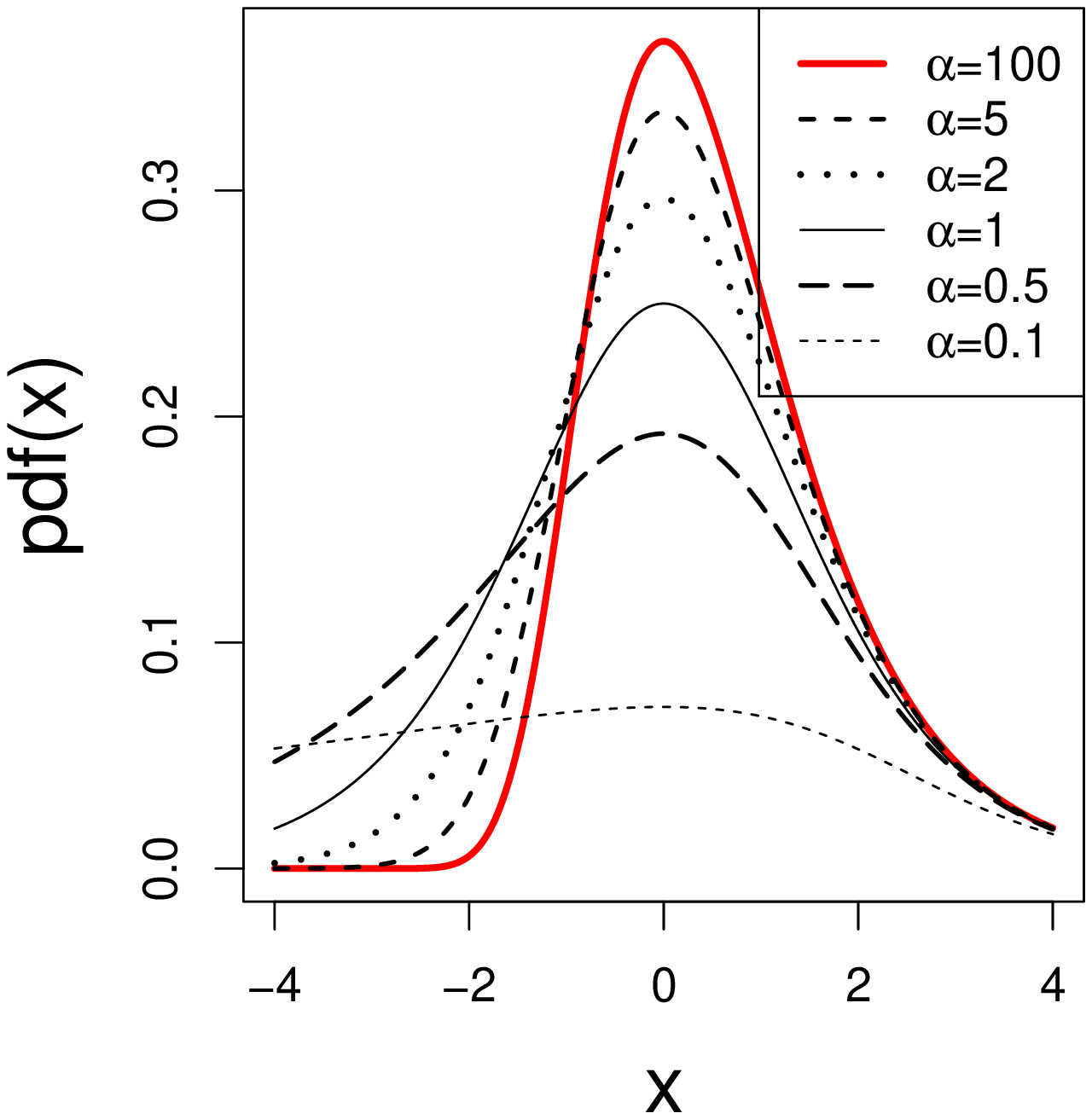}} \quad
\subfigure{\includegraphics[height=38mm,width=38mm]{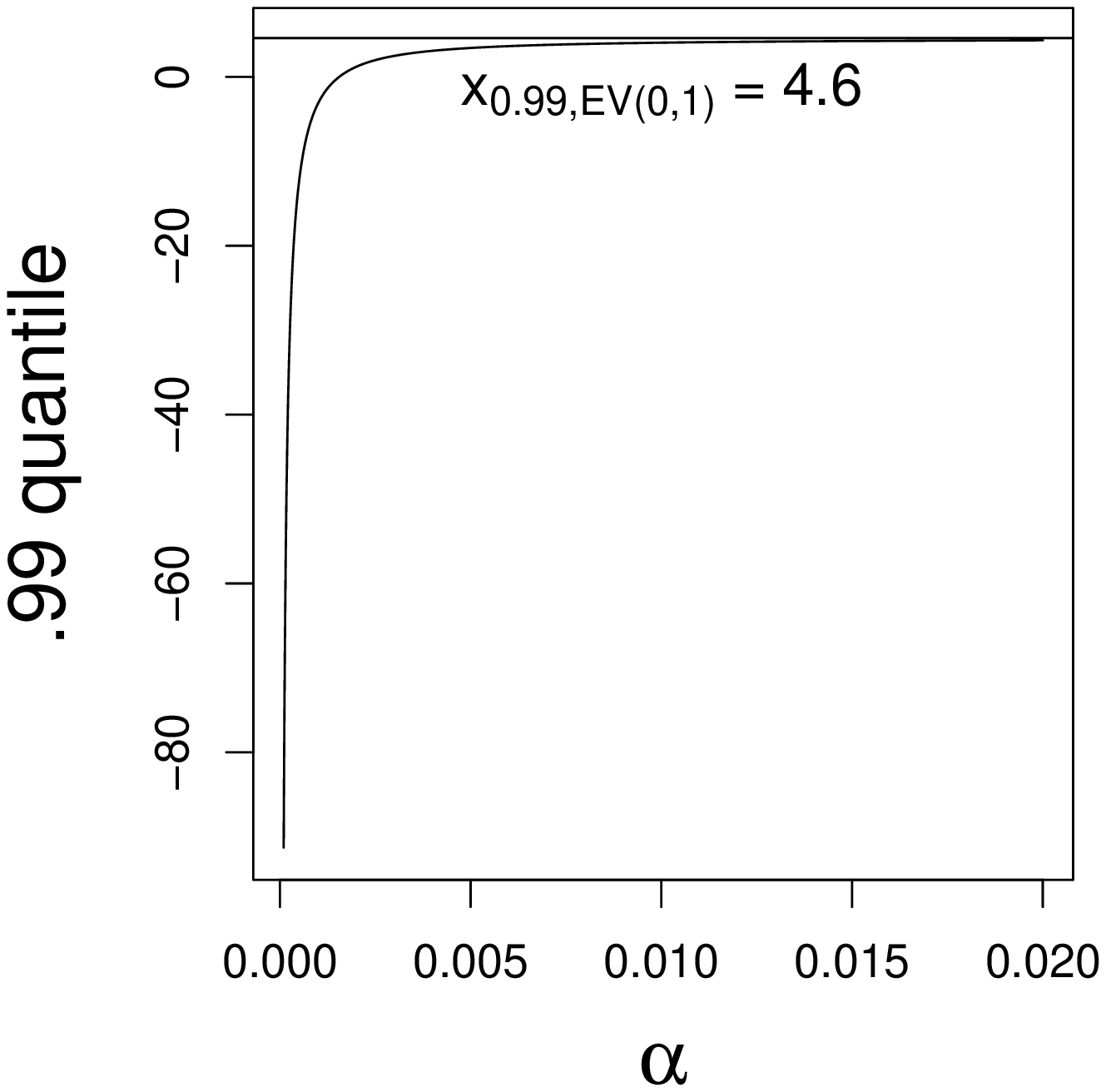}} \quad
\subfigure{\includegraphics[height=38mm,width=38mm]{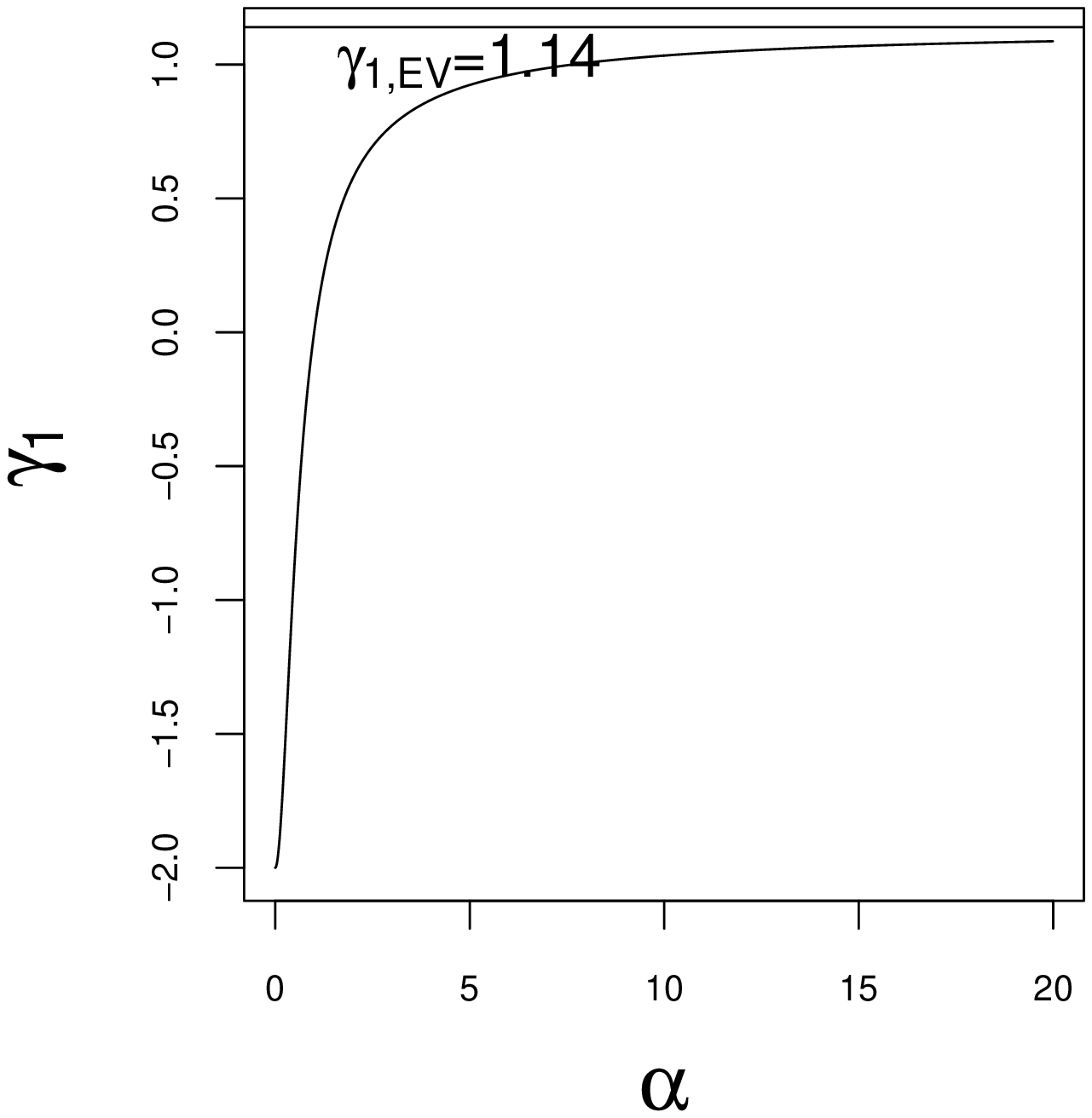}} \quad
\subfigure{\includegraphics[height=38mm,width=38mm]{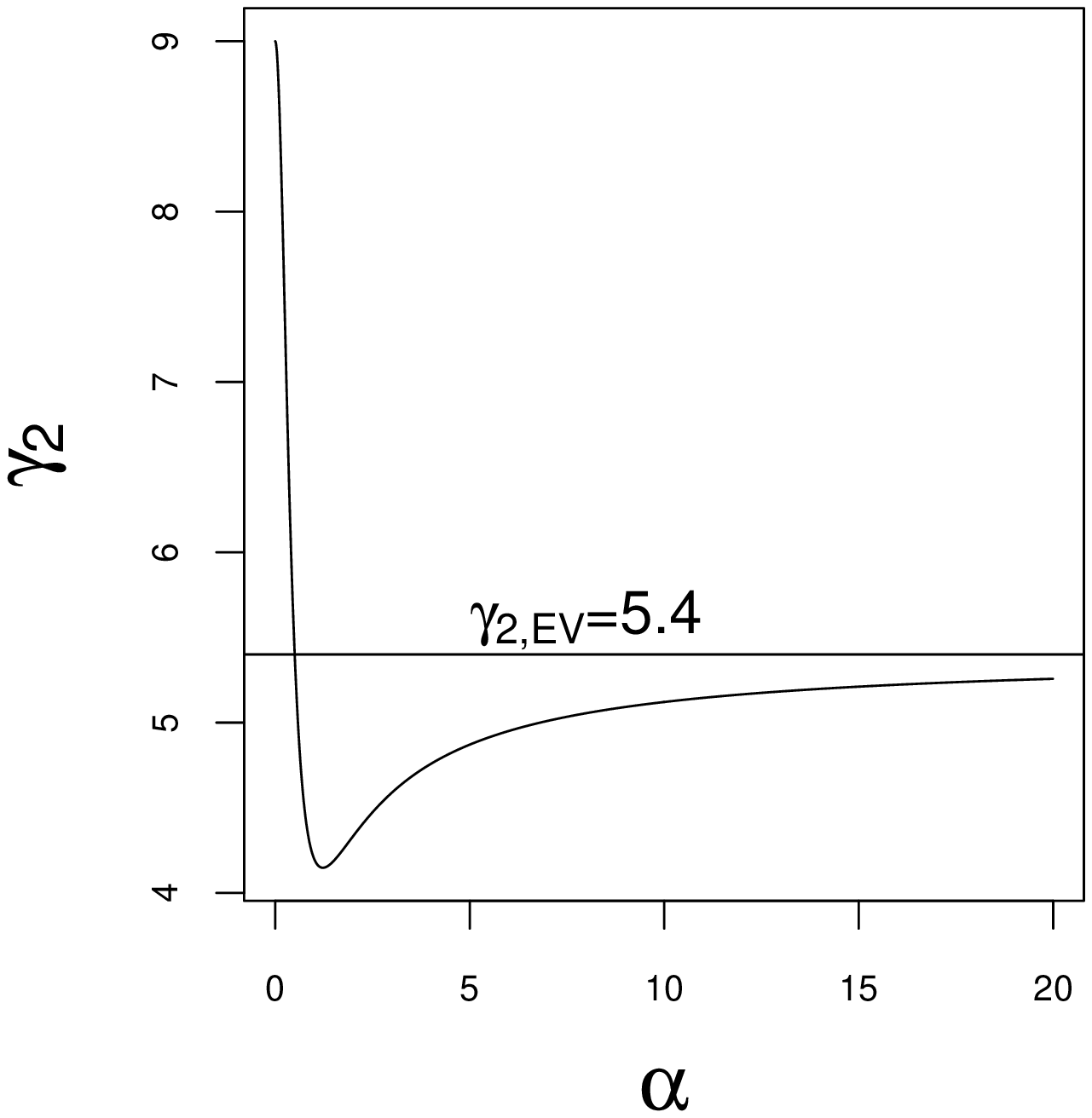}}
%\vspace{-0.6cm}
\caption{Density function and .99 quantile - GTIEV3$(0,1,\alpha)$; skewness and kurtosis - GTIEV3$(\mu,\sigma,\alpha)$}
\label{fig:GTIEV3}
\end{figure}

The GTIEV3 distribution is more flexible than the Gumbel distribution but less flexible relative to the GEV and EGu distributions
(Figure~\ref{fig:GTIEV3}).
The .99 quantile and skewness coefficient are always lower than the corresponding Gumbel values whereas the kurtosis can be greater.
The right tail of the GTIEV3 distribution can not be heavier than that of the Gumbel distribution, in contrast to its left tail. 
This observation suggests that the GTIEV3 distribution is not useful for modeling right-skewed data.

\vspace{1cm}

\paragraph{Three-parameter exponential-gamma distribution (EGa).}  \citet{OJO} presents a generalization of the Gumbel distribution, with three parameters $\mu$, $\sigma$, and $\alpha$. We refer to it as the three-parameter exponential-gamma distribution and denote it by EGa$(\mu,\sigma,\alpha)$.

Let $X \sim \rm{EGa}(\mu,\sigma,\alpha)$ be an exponential-gamma distributed random variable. Its pdf and cdf are, respectively, 
$$
f_{{\rm EGa}}(x; \mu,\sigma,\alpha)=
\frac{1}{\Gamma(\alpha)}\frac{1}{\sigma}\exp\left(-\exp\left(-\frac{x -\mu}{\sigma}\right)\right)\exp\left(-\alpha \frac{x -\mu}{\sigma}\right) 
,\quad x \in I\!\! R,
$$
and
$$
F_{{\rm EGa}}(\mu,\sigma,\alpha)=\frac{1}{\Gamma(\alpha)}\Gamma\left(\alpha,\exp\left(-\frac{x-\mu}{\sigma} \right) \right), \quad x \in I \!\! R,
$$
where $\alpha >0$, and $\Gamma(s,x)=\int_x^\infty t^{s-1} \exp(-t) dt$ 
is the incomplete gamma function. The Gumbel distribution is a particular case of EGa when $\alpha=1$.
To generate $X \sim {\rm EGa}(\mu,\sigma,\alpha)$, we write $X=\mu-\sigma\ln(Y)$, where $Y \sim {\rm gamma}(\alpha,1)$.\footnote{The 
parameterization for the gamma distribution is such that, if $W \sim {\rm gamma}(\alpha,\beta)$, its pdf is $f(w)=(\beta^\alpha / \Gamma(\alpha))w^{\alpha-1}\exp(-\beta w)$, $w>0$.}

\begin{figure}[!ht]
\centering
\subfigure{\includegraphics[height=38mm,width=38mm]{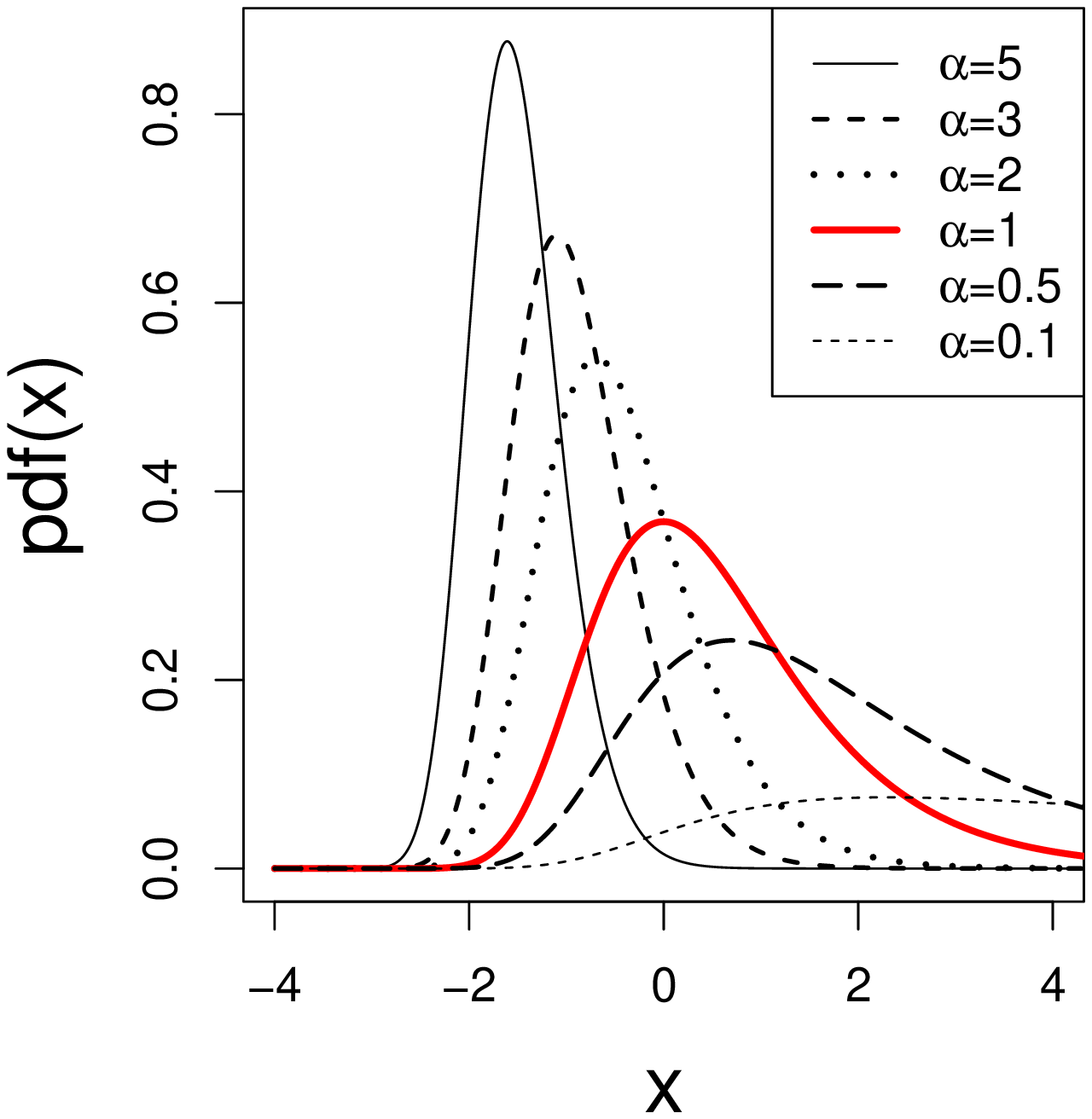}} \quad
\subfigure{\includegraphics[height=38mm,width=38mm]{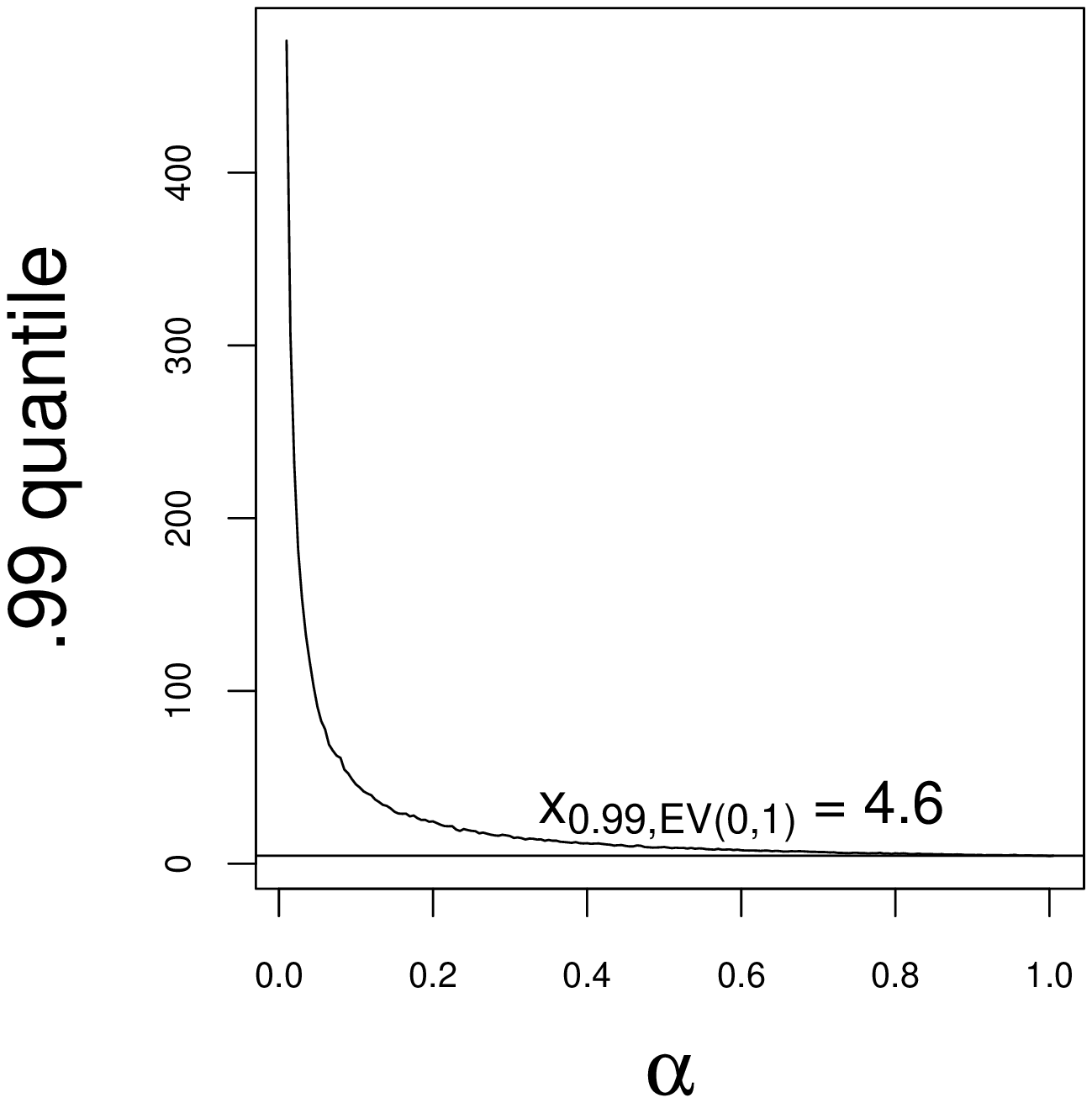}} \quad
\subfigure{\includegraphics[height=38mm,width=38mm]{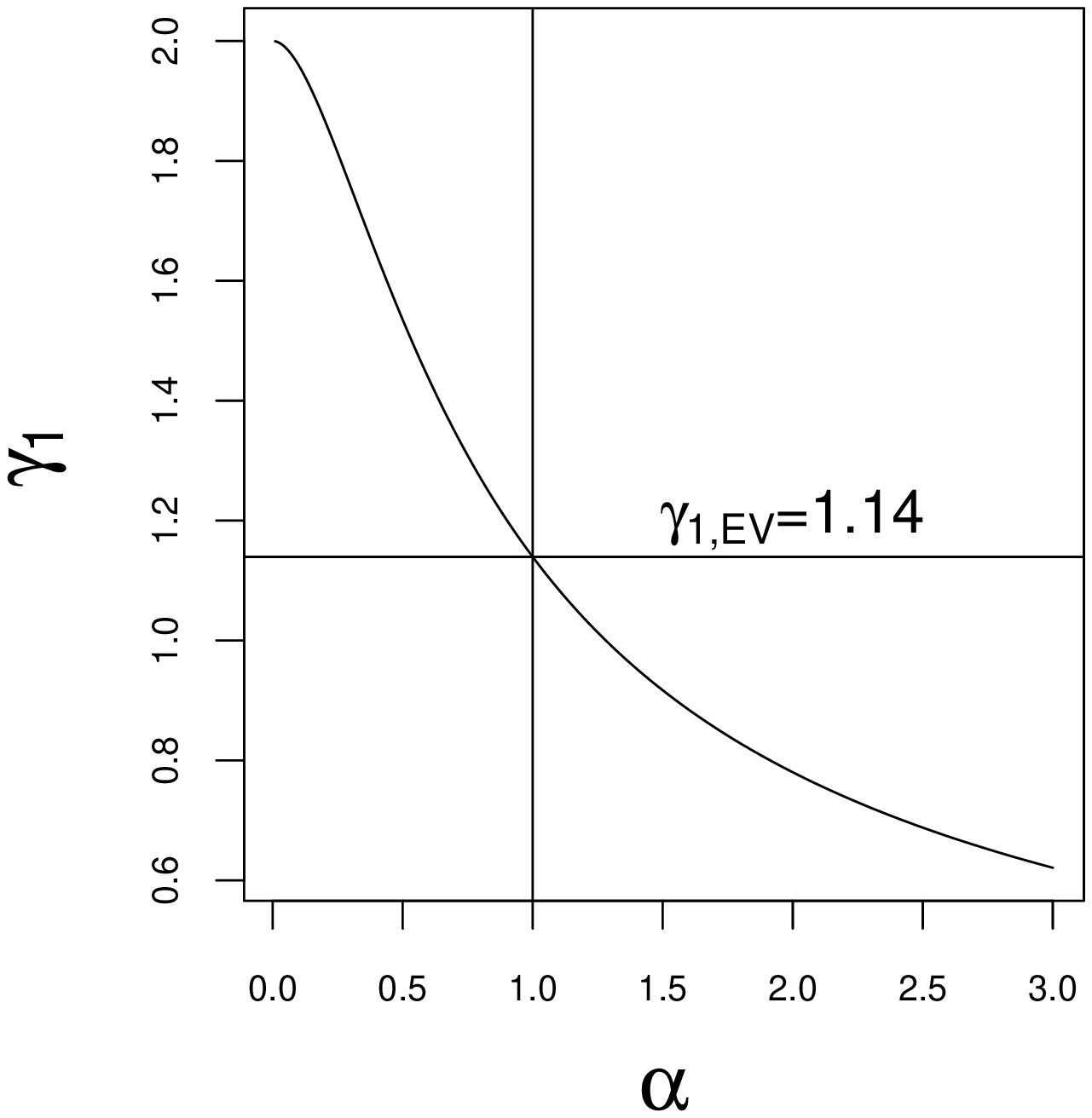}} \quad
\subfigure{\includegraphics[height=38mm,width=38mm]{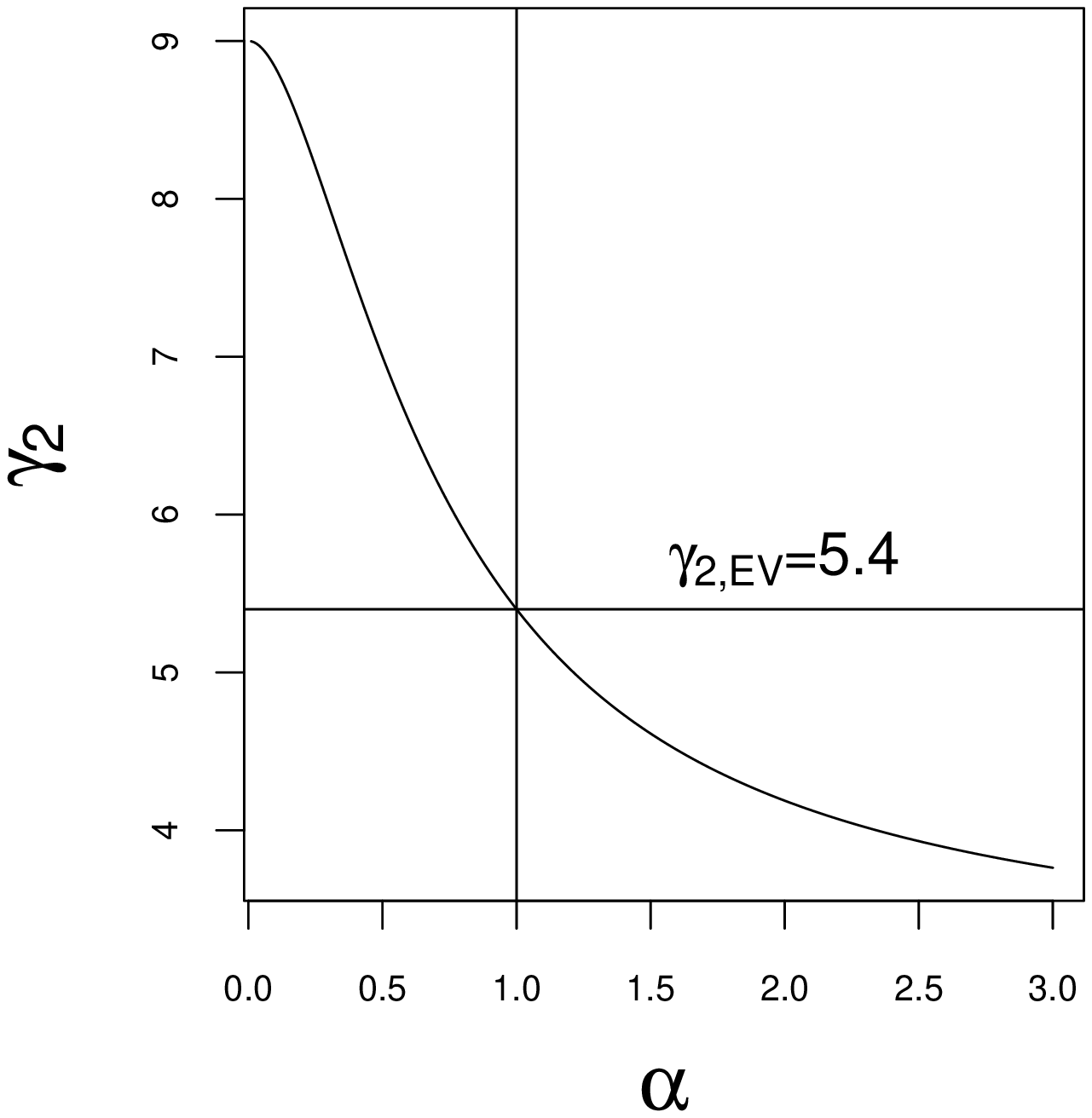}}
%\vspace{-0.6cm}
\caption{Density function and .99 quantile - EGa$(0,1,\alpha)$; skewness and kurtosis - EGa$(\mu,\sigma,\alpha)$}
\label{fig:EGa}
\end{figure}

Similarly to the EGu distribution, the right tail gets heavier for smaller values of $\alpha>0$ (Figure~\ref{fig:EGa}). 
For $\alpha$ close to zero, the .99 quantile can be greater than that of the Gumbel, TEV and GTIEV3  distributions. 
The .99 quantile plots indicate that when $\alpha$ is close to zero, small changes in $\alpha$ lead to significant changes in the quantile values;
similarly to the EGu distribution, the skewness and kurtosis can reach values close to 2 and 9, respectively, indicating more flexibility than the Gumbel distribution. 

\paragraph{Generalized Gumbel distribution (GGu)}
\cite{COORAY} derived a distribution which is referred to as the generalized Gumbel distribution (GGu).

\begin{figure}[!ht]
\centering
\subfigure{\includegraphics[height=38mm,width=38mm]{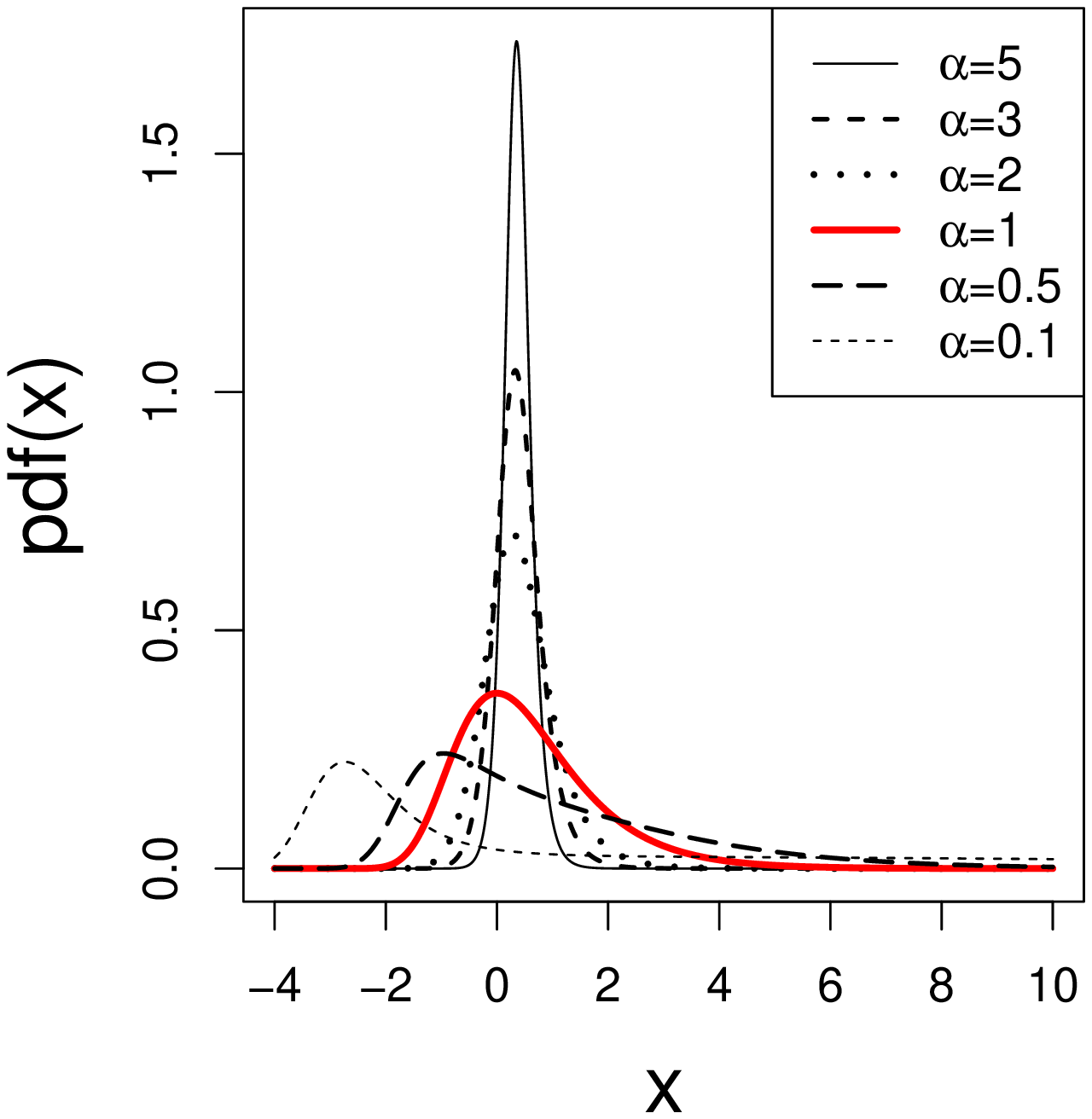}} \quad
\subfigure{\includegraphics[height=38mm,width=38mm]{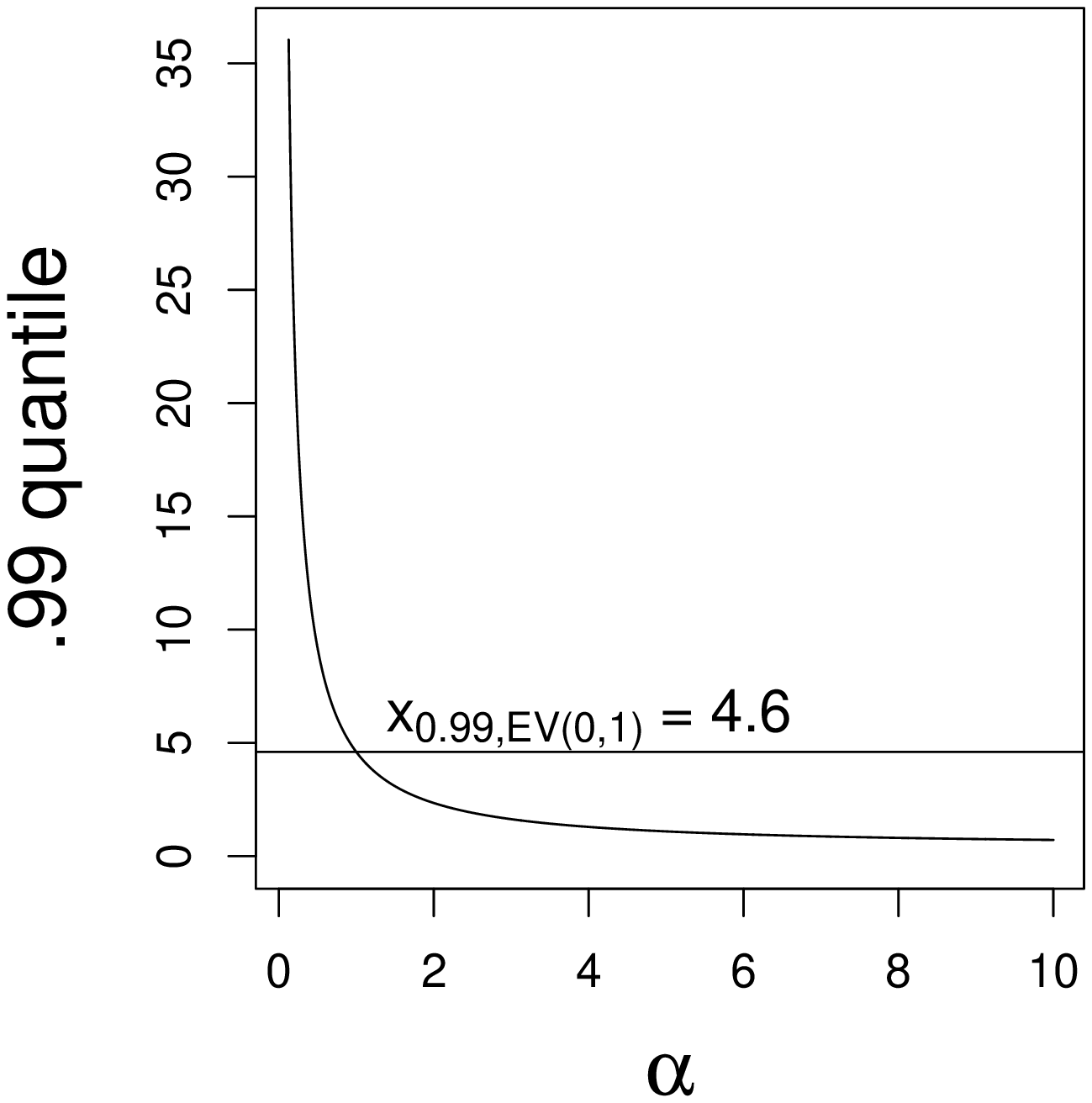}} \quad
\subfigure{\includegraphics[height=38mm,width=38mm]{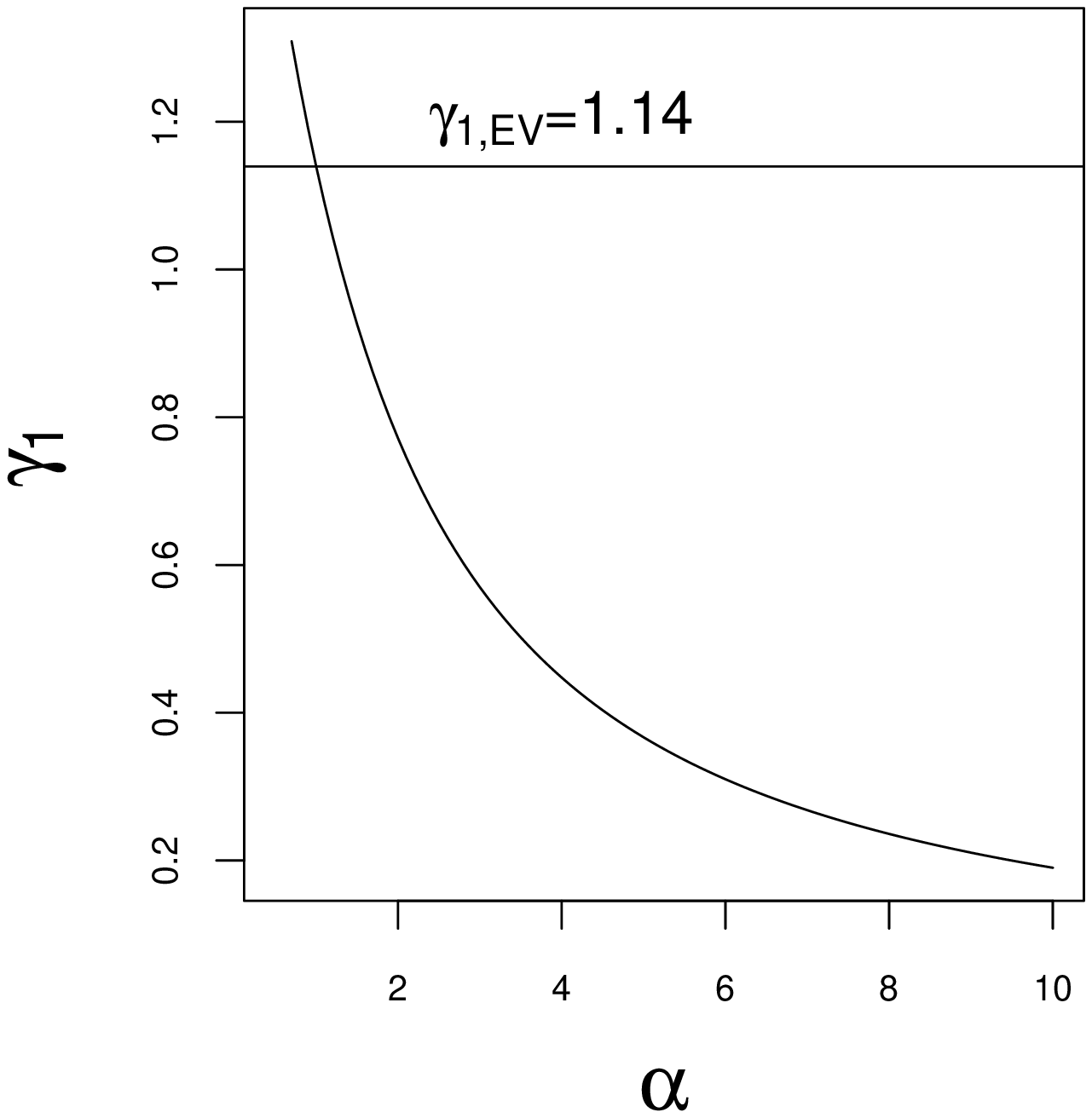}} \quad
\subfigure{\includegraphics[height=38mm,width=38mm]{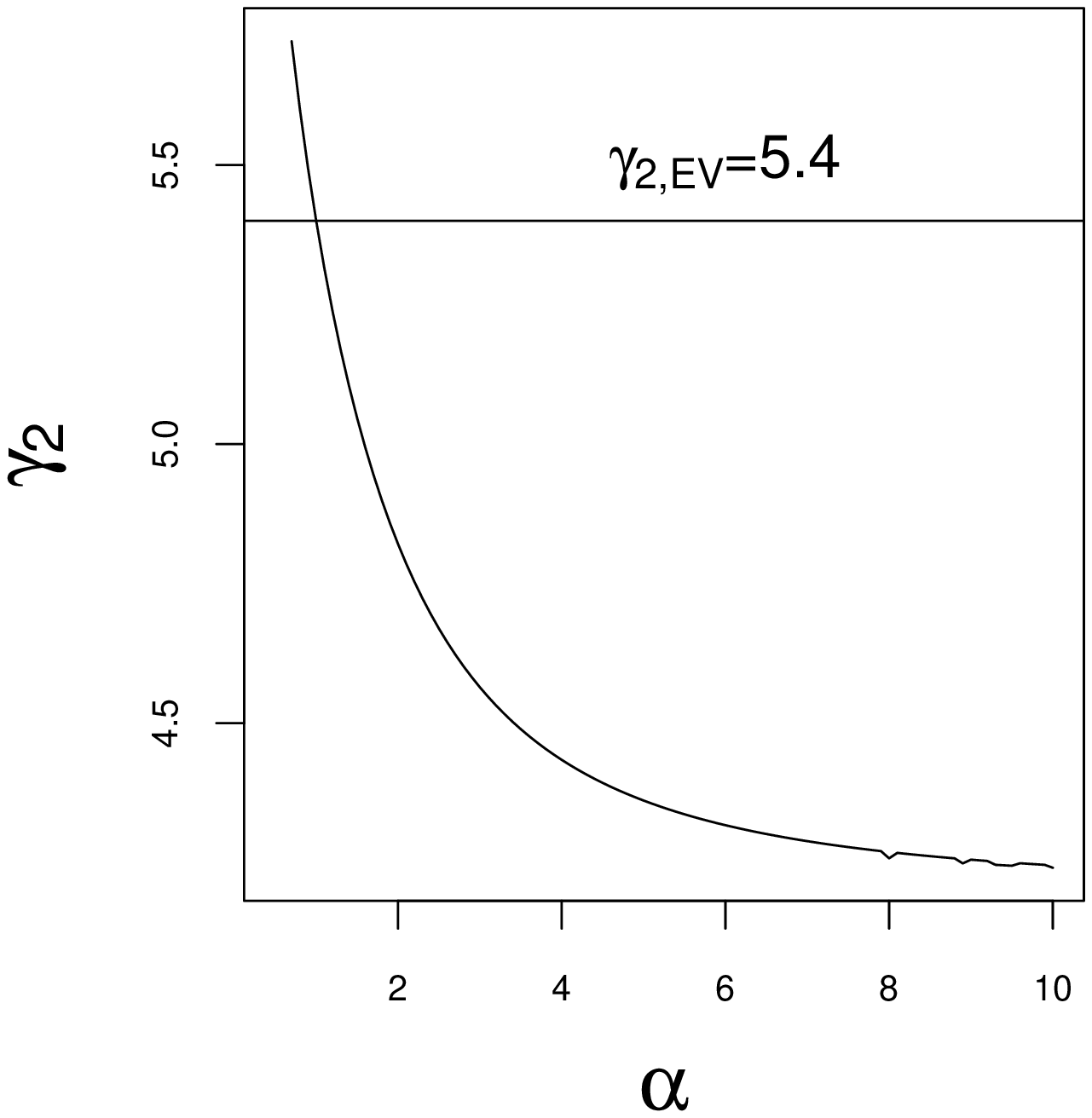}}
%\vspace{-0.6cm}
\caption{Density function, .99 quantile, skewness and kurtosis - GGu$(0,1,\alpha)$}
\label{fig:GGu}
\end{figure}

Let $X \sim {\rm GGu}(\mu,\sigma,\alpha)$ be a generalized Gumbel distributed random variable. Its pdf and cdf are, respectively,
\begin{eqnarray*}
f_{GGu}(x;\mu,\sigma,\alpha)&=&\frac{\alpha}{\sigma}\exp\left( - \frac{x-\mu}{\sigma} \right)\exp\left( \exp\left( - \frac{x-\mu}{\sigma} \right) \right)\left(\exp\left( \exp\left( - \frac{x-\mu}{\sigma} \right) \right) - 1 \right)^{-\alpha -1} \\
&& \left(1 + \left(\exp\left( \exp\left( - \frac{x-\mu}{\sigma} \right) \right) - 1 \right)^{-\alpha} \right)^{-2}, \;\; x \in I\!\! R,
\end{eqnarray*}
and
$$
F_{GGu}(x;\mu,\sigma,\alpha)=1-\left(1 + \left(\exp\left( \exp\left( - \frac{x-\mu}{\sigma} \right) \right) - 1 \right)^{-\alpha}\right)^{-1}, \;\; x \in I\!\! R, 
$$
where $\mu \in I\!\! R$, $\sigma>0$ and $\alpha>$\footnote{\cite{COORAY} considers the parameter space $\mu \in I\!\! R$ and $0< \alpha \sigma <\infty$.}.
When $\alpha=1$, the GGu distribution reduces to a Gumbel distribution.
Figure~\ref{fig:GGu} shows the plots of the pdf for selected parameters, and the .99 quantile, skewness and kurtosis of GGu$(0,1,\alpha)$.
Similarly to the EGu and EGa distributions, for $\alpha$ close to zero, the .99 quantile can be greater than that of the Gumbel, TEV and GTIEV3 distributions. 
The .99 quantile plot indicate that when $\alpha$ is close to zero, small changes in $\alpha$ lead to significant changes in the quantile values; the skewness and kurtosis can reach values greater than those of the Gumbel distribution.

\paragraph{Exponential-gamma distribution.}  
\begin{sloppypar} As a generalization of the Gumbel distribution, \citet{ADEYEMI} proposed  the asymptotic distribution of the $r$-th maximum extremes obtained by \citet{GUMBEL1935}, whose pdf is \end{sloppypar}
$$
f(x; \mu, \sigma,\alpha)=\frac{r^r}{\Gamma(r)}\exp(-r\exp(-x))\exp(-r x), \quad x \in I\!\! R,
$$
for $r>0$, the shape parameter.
When $r=1$, this distribution reduces to a Gumbel distribution.
Its generalized form is known as the exponential-gamma distribution ExpGama$(\mu,\sigma,\alpha,\beta)$ \citep[p. 34]{BALAKRISHNAN1988} and is defined by the pdf and cdf given by, respectively,
$$
f_{{\rm ExpGama}}(x; \mu, \sigma,\alpha,\beta)=\frac{\alpha ^\beta}{\sigma \Gamma(\beta)}\exp\left(-\alpha \exp\left(-\frac{x-\mu}{\sigma}\right) \right) \exp\left(-\beta \frac{x-\mu}{\sigma}\right), \quad x \in I\!\! R ,
$$
and
\begin{eqnarray*}
F_{{\rm ExpGama}}(x; \mu, \sigma,\alpha,\beta)
=\frac{1}{\Gamma(\beta)}\Gamma\left(\beta,\alpha\exp\left(-\frac{x-\mu}{\sigma} \right) \right), \quad x \in I \!\! R,
\end{eqnarray*}
where $\alpha \in I\!\! R$ and $\beta > 0$.
When $\alpha = \beta=1$, the exponential-gamma distribution reduces to a Gumbel distribution, and it reduces to the EGa distribution when $\alpha=1$.
Note that, if $X \sim {\rm ExpGama}(\mu,\sigma,\alpha,\beta)$ then
$$
F_{{\rm ExpGama}}(x; \mu, \sigma,\alpha,\beta)=\frac{\Gamma(\beta, \exp\left(-\frac{x-\mu^*}{\sigma}\right))}{  \Gamma(\beta)} 
= F_{{\rm ExpGama}}(x; \mu^*, \sigma,1,\beta) 
= F_{{\rm EGa}}(x; \mu^*, \sigma,\beta),
$$
where $ $ $\mu^*=\mu+\sigma \ln \alpha \in I\!\! R$. $ $ Hence, the $ $ exponential-gamma $ $ family  $ $ of $ $ distributions ExpGama$(\mu,\sigma,\alpha,\beta)$, where $\mu \in I\!\! R$, $\sigma > 0$, $\alpha >0$ and $\beta > 0$, is nonidentifiable. It coincides with the three-parameter exponential-gamma family of distributions EGa$(\mu,\sigma,\beta)$, where $\mu \in I\!\! R$, $\sigma > 0$ and $\beta >0$. Therefore, this distribution will not be contemplated hereafter.

\paragraph{Type IV generalized logistic distribution (GLIV).}  
\citet{PRENTICE1975} proposed the type IV generalized distribution (GLIV).
Let $X \sim {\rm GLIV}(\mu,\sigma,\alpha,\beta)$ be a type IV generalized distributed random variable. Its pdf and cdf are, respectively,  
$$
f_{{\rm GLIV}}(x; \mu, \sigma,\alpha,\beta)= \left(\frac{\alpha}{\beta}\right)^\alpha\frac{1}{\sigma B(\alpha,\beta)}\frac{[\exp(-(x-\mu) / \sigma)]^\alpha}{[1+ (\alpha / \beta) \exp(-(x-\mu) / \sigma)]^{\alpha+\beta}},
\;\; 
x \in I\!\! R,
$$
and
\begin{eqnarray}
&& F_{{\rm GLIV}}(x; \mu, \sigma,\alpha)
= \nonumber\\
&&\frac{1}{\beta B(\alpha,\beta)}
\left(\frac{\beta}{\alpha} \exp\left(\frac{x-\mu}{\sigma}\right)\right)^{\beta}
\! _2F_1\left(\beta, \alpha+\beta; 1 + \beta; -\frac{\beta}{\alpha} \exp\left(\frac{x-\mu}{\sigma}\right)\right), 
\;\; 
x \in I\!\! R, \nonumber
\end{eqnarray}
where 
$\alpha>0$, $\beta>0$ and $ _2F_1(a,b;c;z)$ is the hypergeometric function mentioned previously.
When $\alpha =1$ and $\beta \rightarrow \infty$, the type IV generalized logistic distribution reduces to a Gumbel distribution, and it reduces to a generalized three-parameter Gumbel distribution (GTIEV3) when $\alpha=1$.
To generate $X \sim {\rm GLIV}(\mu,\sigma,\alpha,\beta)$, write $X=\mu-\sigma\ln Y$, where $Y \sim {\rm F}(2\alpha,2\beta)$.\footnote{
If $W \sim {\rm F}(a,b)$ its pdf is $f(w)=(1/B(a,b))(a/b)^a w^{a-1} / (1+(a/b)w)^{(a+b)}$, for $w>0$ .}

\begin{figure}[!ht]
\centering
\subfigure{\includegraphics[height=38mm,width=38mm]{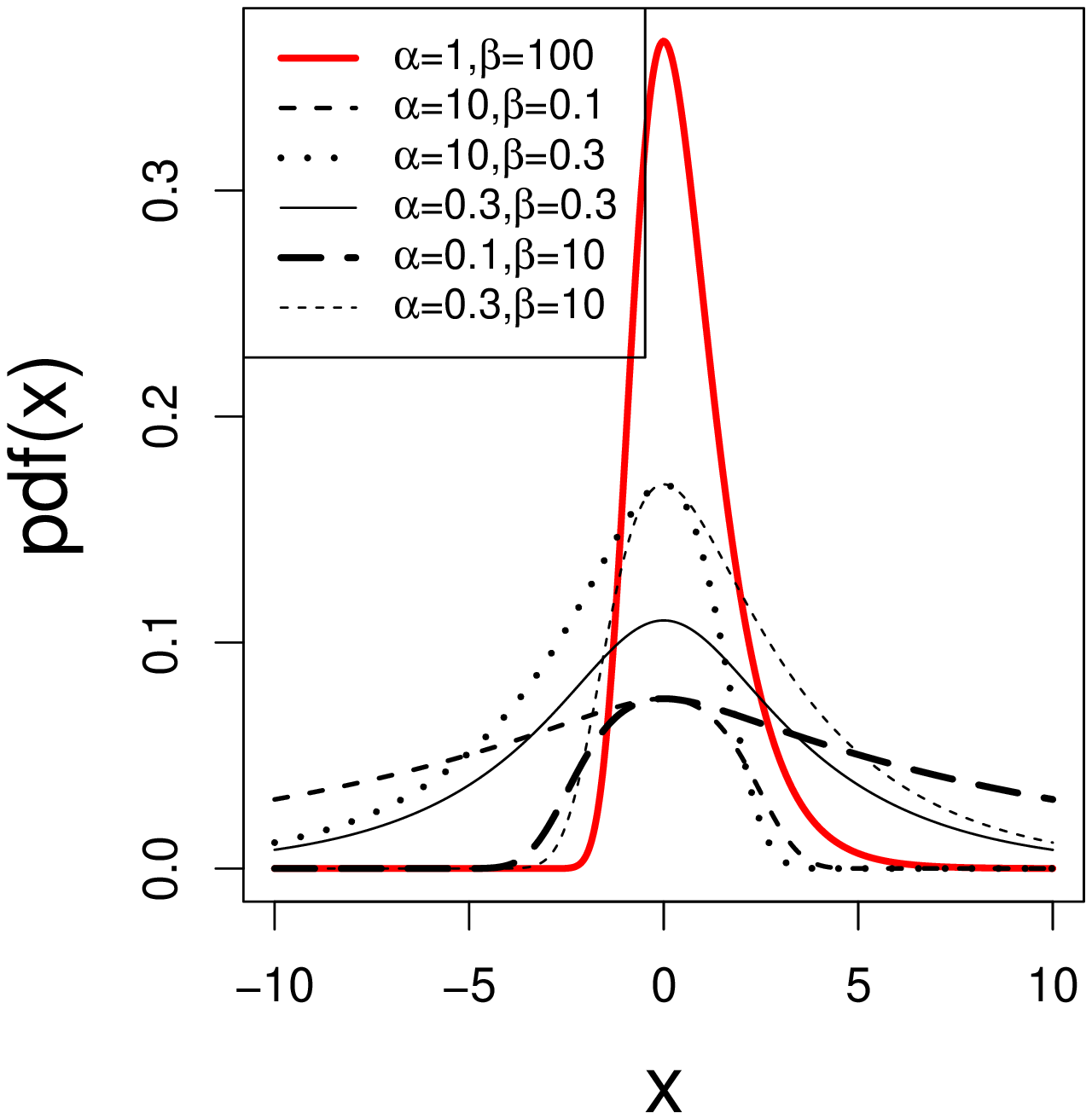}} \quad
\subfigure{\includegraphics[height=38mm,width=38mm]{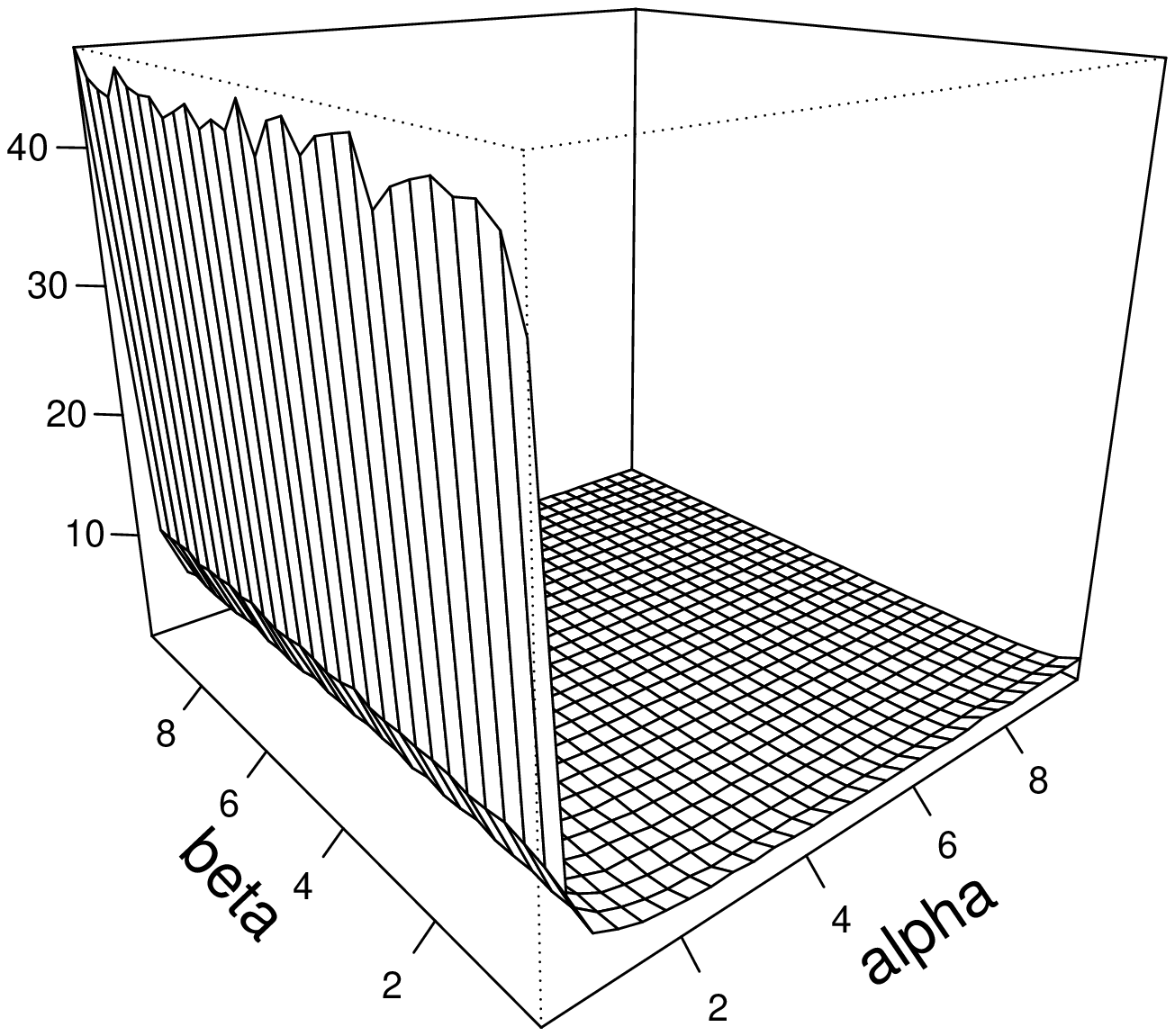}} \quad
\subfigure{\includegraphics[height=38mm,width=38mm]{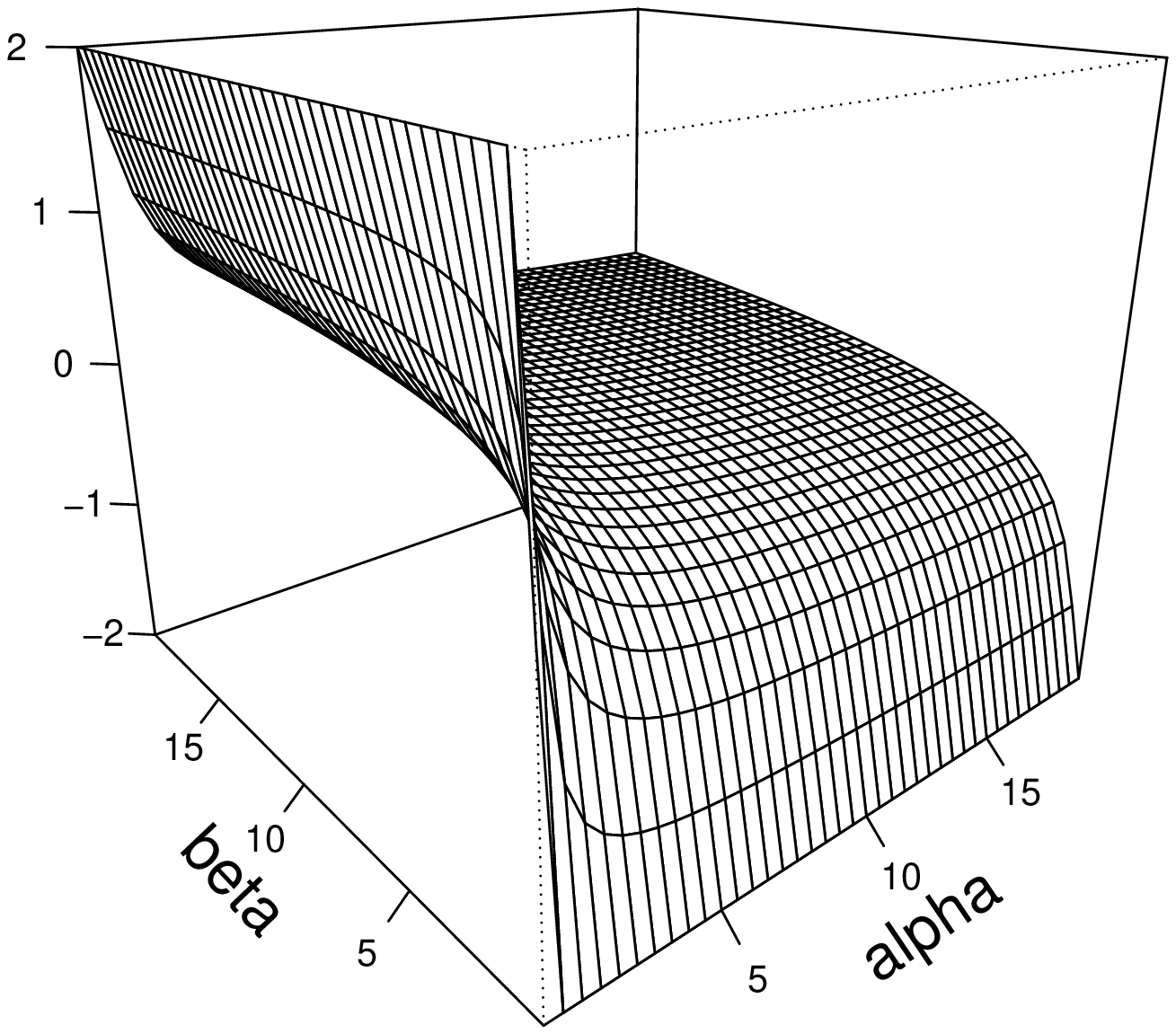}} \quad
\subfigure{\includegraphics[height=38mm,width=38mm]{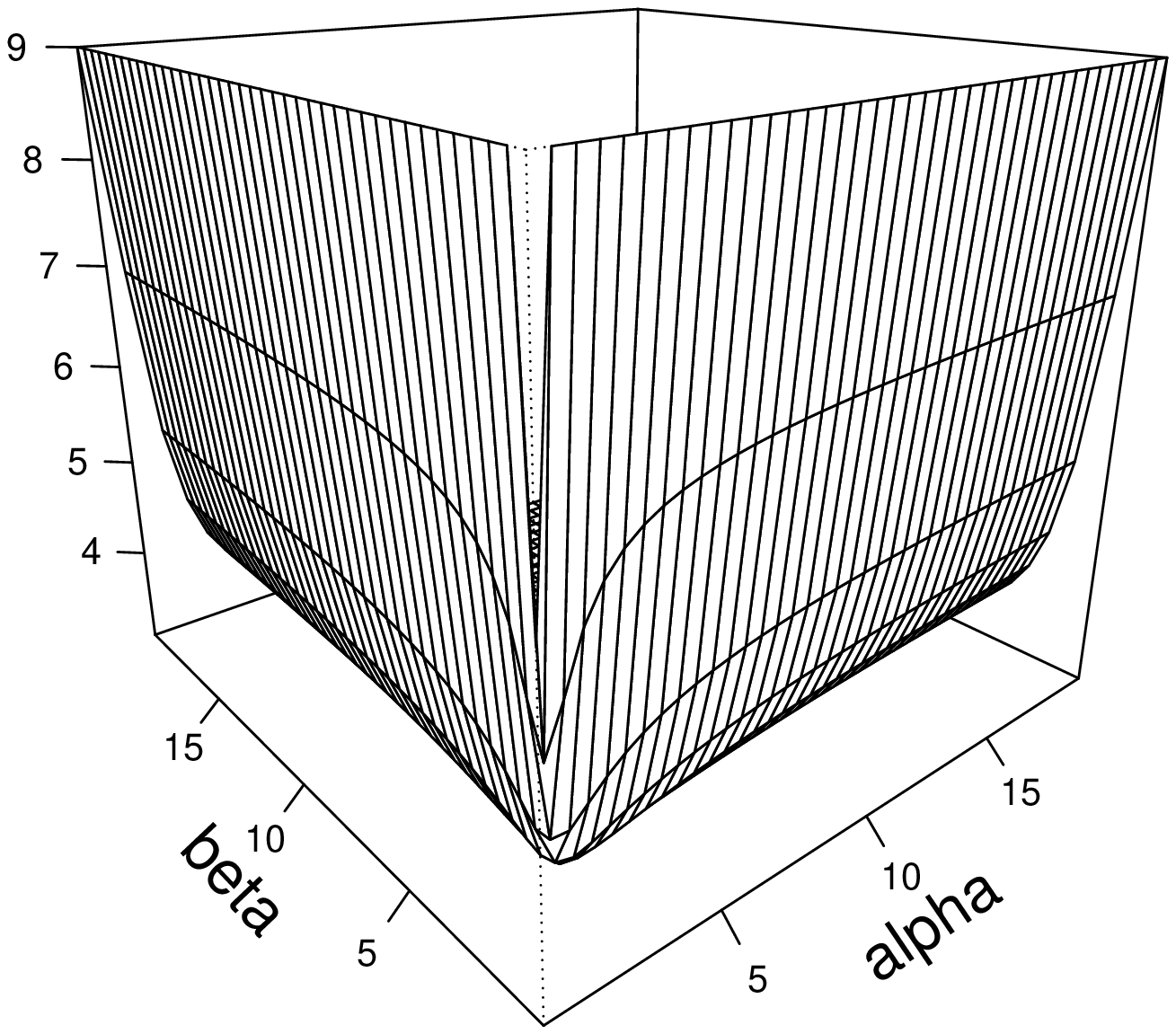}}
%\vspace{-0.6cm}
\caption{Density function and .99 quantile - GLIV$(0,1,\alpha,\beta)$; skewness and kurtosis - GLIV$(\mu,\sigma,\alpha,\beta)$}
\label{fig:GLIV}
\end{figure}

Similarly to EGu and EGa, for fixed $\beta$, the right tail gets heavier for smaller $\alpha>0$ (Figure~\ref{fig:GLIV}).
For $\alpha$ values close to zero, the .99 quantile can be greater than the Gumbel, TEV, and GTIEV3 values. 
The quantile plots indicate that, when $\alpha$ is close to zero, small  changes 
in $\alpha$ lead to significant changes in the quantile values. The skewness and kurtosis can reach values close to 2 and 9, 
respectively, indicating that the GLIV distribution is more flexible than the Gumbel distribution.
We can verify that $f_{{\rm GLIV}}(x,\alpha,\beta)=f_{{\rm GLIV}}(-x,\beta,\alpha)$,
and thus, for fixed $\alpha$, the left tail is heavier for small values of $\beta$.

\citet{PRENTICE1976} presents a simplified form of this distribution.
When $\alpha=\beta$, the type IV generalized logistic distribution is symmetric about $x=\mu$, and the distribution is known as the type III generalized logistic distribution.
 
\paragraph{Exponentiated generalized Gumbel distribution.} 
\citet{CORDEIRO2013} defined a class of distributions known as the exponentiated generalized distribution (EG), by
$$
F(x)=[1-\{1-G(x)  \}^\alpha]^\beta,
$$
where $\alpha>0$ and $\beta>0$ are two additional shape parameters and $G(x)$ is a continuous cdf.
When $G(x)$ is the Gumbel cdf, EG becomes the exponentiated generalized Gumbel distribution (EGGu).
Let $X \sim {\rm EGGu}(\mu,\sigma,\alpha,\beta)$ be an exponentiated generalized Gumbel distributed random variable with 
cdf 
$$
F_{{\rm EGGu}}(x; \mu,\sigma,\alpha,\beta)=\left[1-\left(1-\exp\left(-\exp\left(-\frac{x-\mu}{\sigma}\right)\right)\right)^\alpha \right]^\beta , \quad x \in I\!\! R,
$$
where $\alpha>0$ and $\beta>0$.
The Gumbel distribution is a particular case of EGGu when $\alpha=\beta=1$ and the aforementioned exponentiated Gumbel  distribution EGu is a special case when $\beta=1$.
Note that, if $X \sim {\rm EGGu}(\mu,\sigma,1,\beta)$, then
$$
F_{{\rm EGGu}}(x; \mu, \sigma,1,\beta)
=\exp(-\exp(- (x-(\mu+\sigma\ln \beta))/\sigma)) 
=F_{{\rm EGGu}}(x; \mu^*, \sigma,1,1)
=F_{{\rm EV}}(x; \mu^*, \sigma),
$$
where $\mu^*=\mu+\sigma \ln \beta$. 
Hence, the exponentiated Gumbel family of distributions EGGu$(\mu,\sigma,\alpha,\beta)$ where $\mu \in I\!\! R$, $\sigma>0$, $\alpha>0$ and $\beta>0$, is nonidentifiable. It coincides with the Gumbel family of distributions EV$(\mu^*,\sigma)$, where $\mu^* \in I\!\! R$ and $\sigma>0$, when $\alpha=1$.
Therefore, this distribution will not be considered further.

\paragraph{Beta Gumbel distribution.}  
\citet{NADARAJAH2004} proposed a generalization of the Gumbel distribution, which they referred to as the beta Gumbel distribution (BG), from a generalized class of distributions defined by  
$$
F(x)=\frac{B_{G(x)}(\alpha,\beta)}{B(\alpha,\beta)},
$$
for $\alpha>0$ and $\beta>0$, where $G(x)$ is a cdf, $B(\alpha,\beta)$ is the beta function and
$$
B_w(a,b)=\int_0^w t^{a-1}(1-t)^{b-1} dt
$$
is the incomplete beta function, by taking $G(x)$ as the Gumbel cdf.

Let $X \sim {\rm BG}(\mu,\sigma,\alpha,\beta)$ be a beta Gumbel distributed random variable with cdf
$$
F_{{\rm BG}}(x;\mu,\sigma,\alpha,\beta)=\frac{1}{B(\alpha,\beta)}\int_0^{\exp(-\exp(-(x-\mu)/\sigma))} t^{\alpha-1}(1-t)^{\beta-1} dt, \quad x \in I\!\! R, 
$$ 
where $\alpha>0$ and $\beta>0$.
When $\alpha=1$ and $\beta=1$, the beta Gumbel distribution reduces to a Gumbel distribution, and it reduces to an exponentiated Gumbel distribution (EGu) when $\alpha=1$.

If $X \sim {\rm BG}(\mu,\sigma,\alpha,1)$, then
$$
F_{{\rm BG}}(x; \mu, \sigma,\alpha,1)
=\alpha \int_0^{\exp(-\exp(-(x-\mu)/\sigma))} t^{\alpha-1} dt
=F_{{\rm BG}}(x; \mu^*, \sigma,1,1)
=F_{{\rm EV}}(x; \mu^*, \sigma),
$$
where $\mu^*=\mu+\sigma \ln \alpha$. 
Therefore the beta Gumbel family of distributions BG$(\mu,\sigma,\alpha,\beta)$ 
with $\mu \in I\!\! R$, $\sigma>0$, $\alpha>0$ and $\beta>0$ is nonidentifiable. It coincides with the Gumbel family of distributions EV$(\mu^*,\sigma)$ with $\mu^* \in I\!\! R$ and $\sigma>0$, when $\beta=1$.
Therefore, this distribution will not be studied hereafter.

\paragraph{Kummer beta generalized Gumbel distribution.}  
\citet{PESCIM2012} defined a class of distributions known as the Kummer beta generalized family (KBG).
From an arbitrary cdf $G(x)$, the KGB family of distributions is defined by
$$
F(x)=K \int_0^{G(x)} t^{\alpha-1}(1-t)^{\beta-1}\exp(-\gamma t) dt,
$$
where $\alpha>0$, $\beta>0$ and $\gamma \in I\!\! R$ are shape parameters 
and 
$$
K^{-1}=\frac{\Gamma(a)\Gamma(b)}{\Gamma(a+b)} \; _1 F_1(a;a+b;-c),
$$
where 
$_1 F_1(a;a+b;-c)=\sum_{k=0}^\infty[(a)_k(-c)_k]/[(a+b)_k k \;!]$
is the confluent hypergeometric function, 
$(d)_k = d(d + 1) . . . (d + k - 1)$ denotes the ascending factorial, and $(d)_0=1$.
When $G(x)$ is a Gumbel cdf, it is known as the KGB-Gumbel distribution (KGBGu).

Let $X \sim {\rm KBGGu}(\mu,\sigma,\alpha,\beta,\gamma)$ be a KGB-Gumbel distributed random variable with cdf
$$
F_{{\rm KBGGu}}=K \int_0^{\exp( \exp(-(x-\mu)/\sigma))} t^{\alpha-1}(1-t)^{\beta-1}\exp(-\gamma t) dt,
$$
where 
%$\mu \in I\!\! R$ is the location parameter, $\sigma > 0$ is the dispersion parameter, 
$\alpha>0$, $\beta>0$ and $\gamma \in I\!\! R$.
When $\alpha=1$, $\beta=1$ and $\gamma=0$, the KGB-Gumbel distribution reduces to the Gumbel distribution and it reduces to a beta Gumbel distribution when $\gamma=0$.

If $X \sim {\rm KBGGu}(\mu,\sigma,\alpha,1,0)$, then
$$
F_{{\rm KBGGu}}(x; \mu, \sigma,\alpha,1,0)
=\alpha \int_0^{\exp(-\exp(-(x-\mu)/\sigma))} t^{\alpha-1} dt
=F_{{\rm KBGGu}}(x; \mu^*, \sigma,1,1,0)
=F_{{\rm EV}}(x; \mu^*, \sigma),
$$
where $ $ $\mu^*=\mu+\sigma \ln \alpha$. $ $ Hence, the $ $ Kummer beta $ $ Gumbel $ $ family of $ $ distributions KBGGu$(\mu,\sigma,\alpha,\beta,\gamma)$ 
with $\mu \in I\!\! R$, $\sigma>0$, $\alpha>0$, $\beta>0$ and $\gamma \in I\!\! R$ is nonidentifiable. It coincides with the Gumbel family distributions EV$(\mu^*,\sigma)$ with $\mu^* \in I\!\! R$ and $\sigma>0$, when $\beta=1$ and $\gamma=0$.
Therefore, this distribution will not be examined in the following discussion.

\paragraph{Two-component extreme value distribution (TCEV).} 
In studying annual flood series, \citet{ROSSI}  considered an approach to account for both the presence of outliers and high skewness. That approach results from assuming that flood peaks do not all arise from one and the same distribution but, instead, from a two-component extreme value mixture (TCEV).
One of the components generates ordinary (more frequent and less severe in the mean) floods. The other exhibits much greater variability and tends to generate more rare but more severe floods.

Let $X \sim {\rm TCEV}(\mu,\sigma,\mu_1,\sigma_1,\alpha)$ be a two-component extreme value distributed random variable. Its pdf and cdf are, respectively, 
\begin{eqnarray*}
f_{{\rm TCEV}}(x; \mu,\sigma,\mu_1,\sigma_1,\alpha)&=& 
\frac{1-\alpha}{\sigma}\exp\left(-\frac{x-\mu}{\sigma}\right)\exp\left(-\exp\left(-\frac{x-\mu}{\sigma}\right) \right) \\
&&+\frac{\alpha}{\sigma_1} \exp\left(-\frac{x-\mu_1}{\sigma_1}\right) \exp\left(- \exp\left(-\frac{x-\mu_1}{\sigma_1}\right) \right), 
\;\; x \in I\!\! R,
\end{eqnarray*}
and
\begin{eqnarray*}
F_{{\rm TCEV}}(x; \mu,\sigma,\mu_1,\sigma_1,\alpha)&=&
(1-\alpha)\exp\left(-\exp\left(-\frac{x-\mu}{\sigma}\right) \right) \\
&&+\alpha\exp\left(- \exp\left(-\frac{x-\mu_1}{\sigma_1}\right) \right), \;\; x \in I\!\! R,
\end{eqnarray*}
where $\mu \in I\!\! R$ and $\mu_1 \in I\!\! R$ are location parameters, $\sigma > 0$ and $\sigma_1 > 0$ are dispersion parameters and $0 < \alpha < 1$. Greater values of $\alpha$ increase the weight of the second component.

If $0<\alpha<1$, $F(x;\mu,\sigma,\mu_1,\sigma_1,\alpha)=F(x;\mu_1,\sigma_1,\mu,\sigma,(1-\alpha))$, then consequently, the mixture is nonidentifiable.
The lack of identifiability due to the label-switching effect is overcome by imposing identifiability constraints on the parameters.
It is sufficient to consider $0<\alpha<0.5$ to achieve identifiability\footnote{For purposes of parameter estimation, we follow \cite{AITKIN}, who suggest theoretical parameter constraints but no parameters constraints for estimation.}.
When $\alpha \rightarrow 0$, the two-component extreme value distribution reduces to a Gumbel distribution. 

Data from $X \sim {\rm TCEV}(\mu,\sigma,\mu_1,\sigma_1,\alpha)$ may be generated from the conditional distributions  $X | Z=0 \sim EV(\mu,\sigma)$
and $X | Z=1 \sim EV(\mu_1,\sigma_1)$, where  $Z \sim  {\rm Bernoulli}(\alpha)$.  

\begin{figure}[!ht]
\centering
\subfigure{\includegraphics[height=38mm,width=38mm]{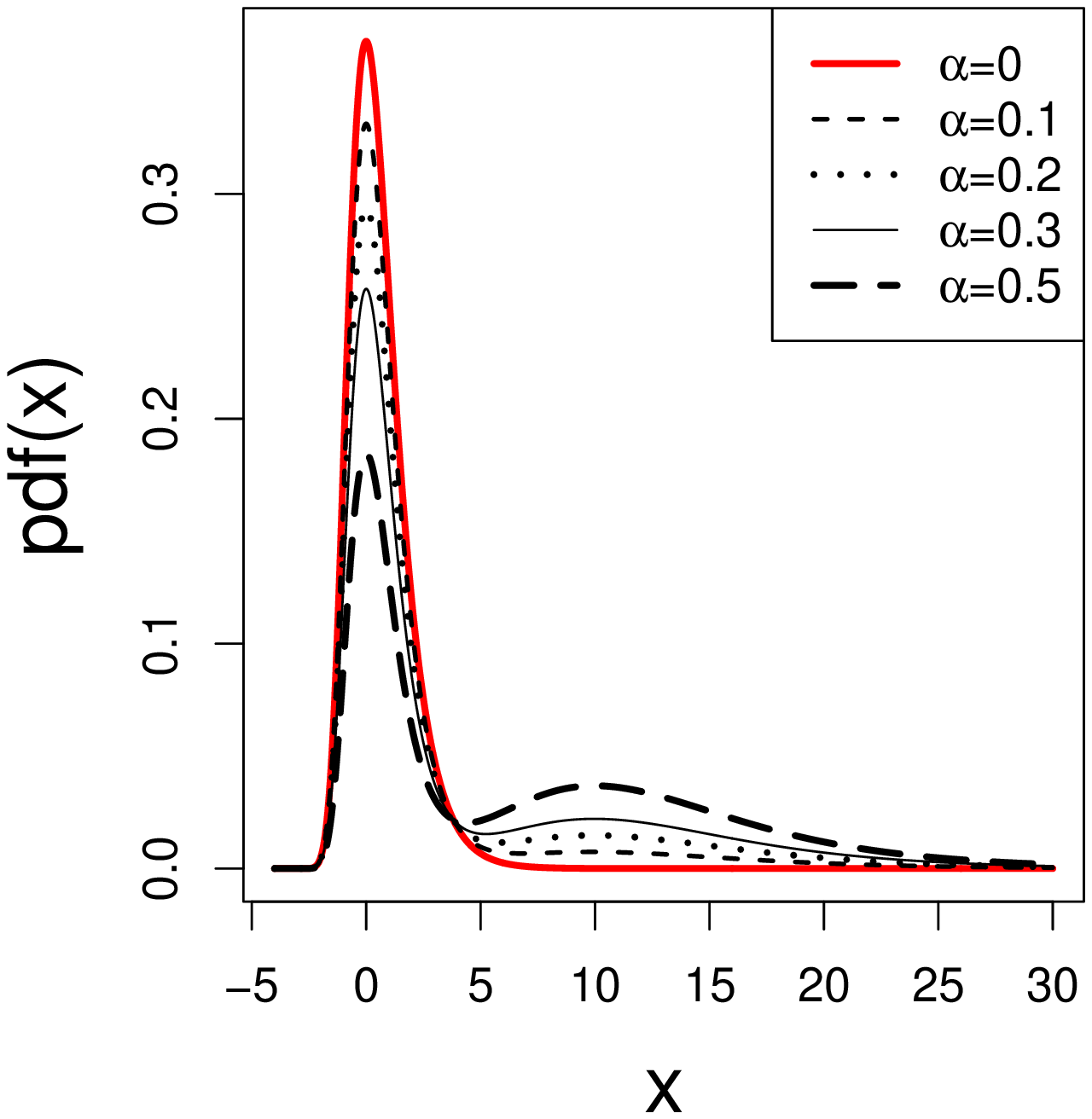}} \quad
\subfigure{\includegraphics[height=38mm,width=38mm]{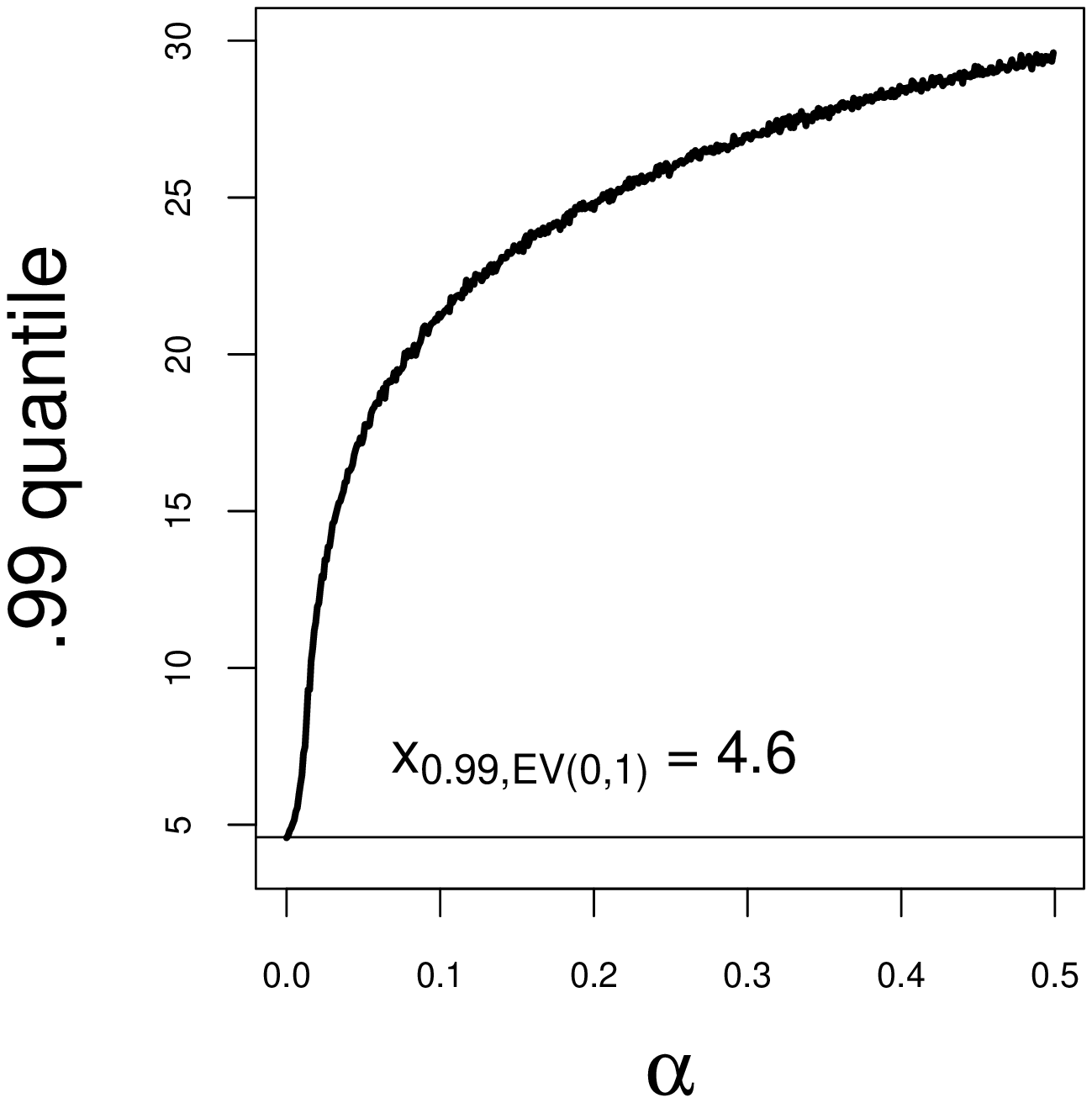}} \quad
\subfigure{\includegraphics[height=38mm,width=38mm]{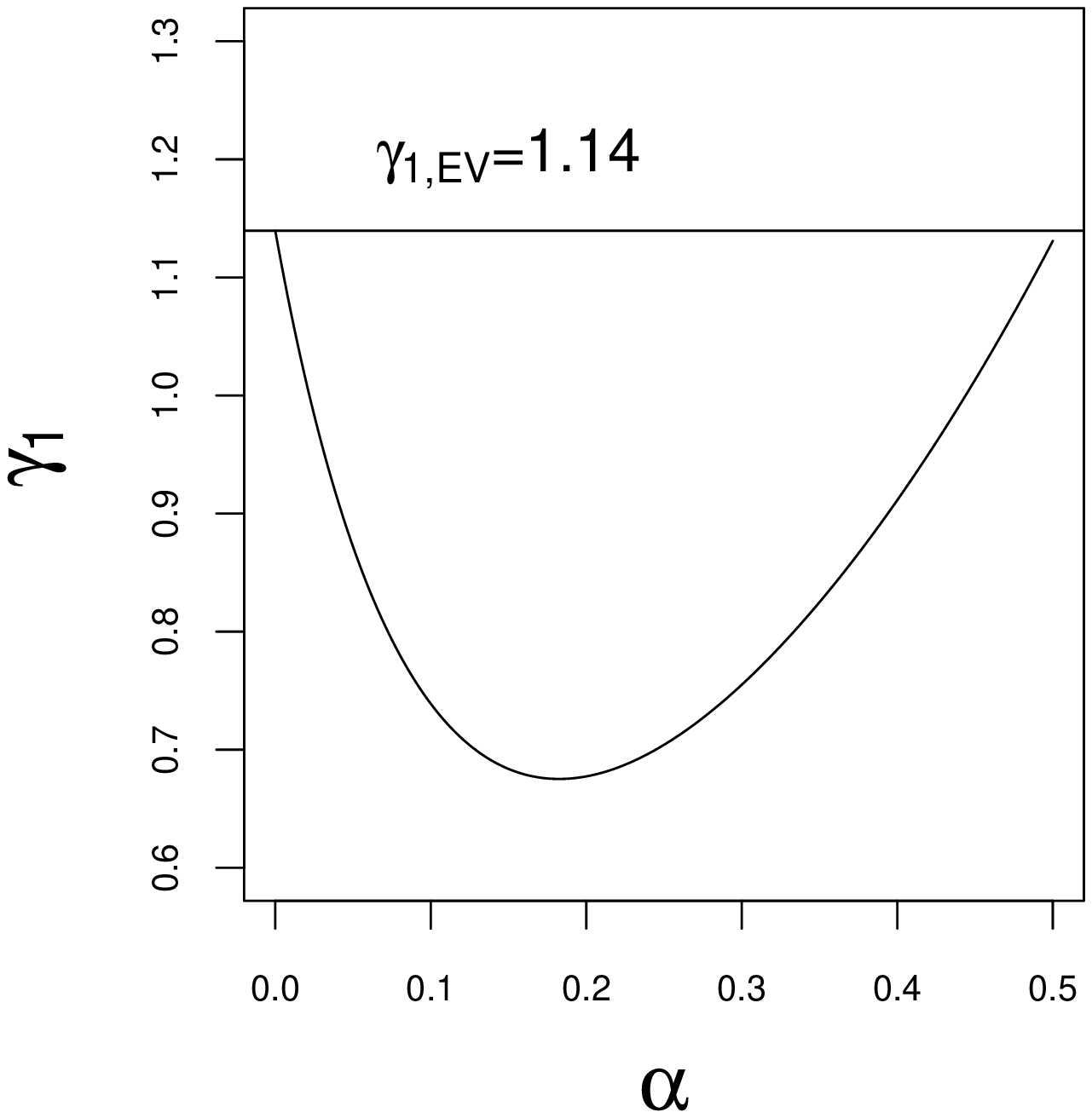}} \quad
\subfigure{\includegraphics[height=38mm,width=38mm]{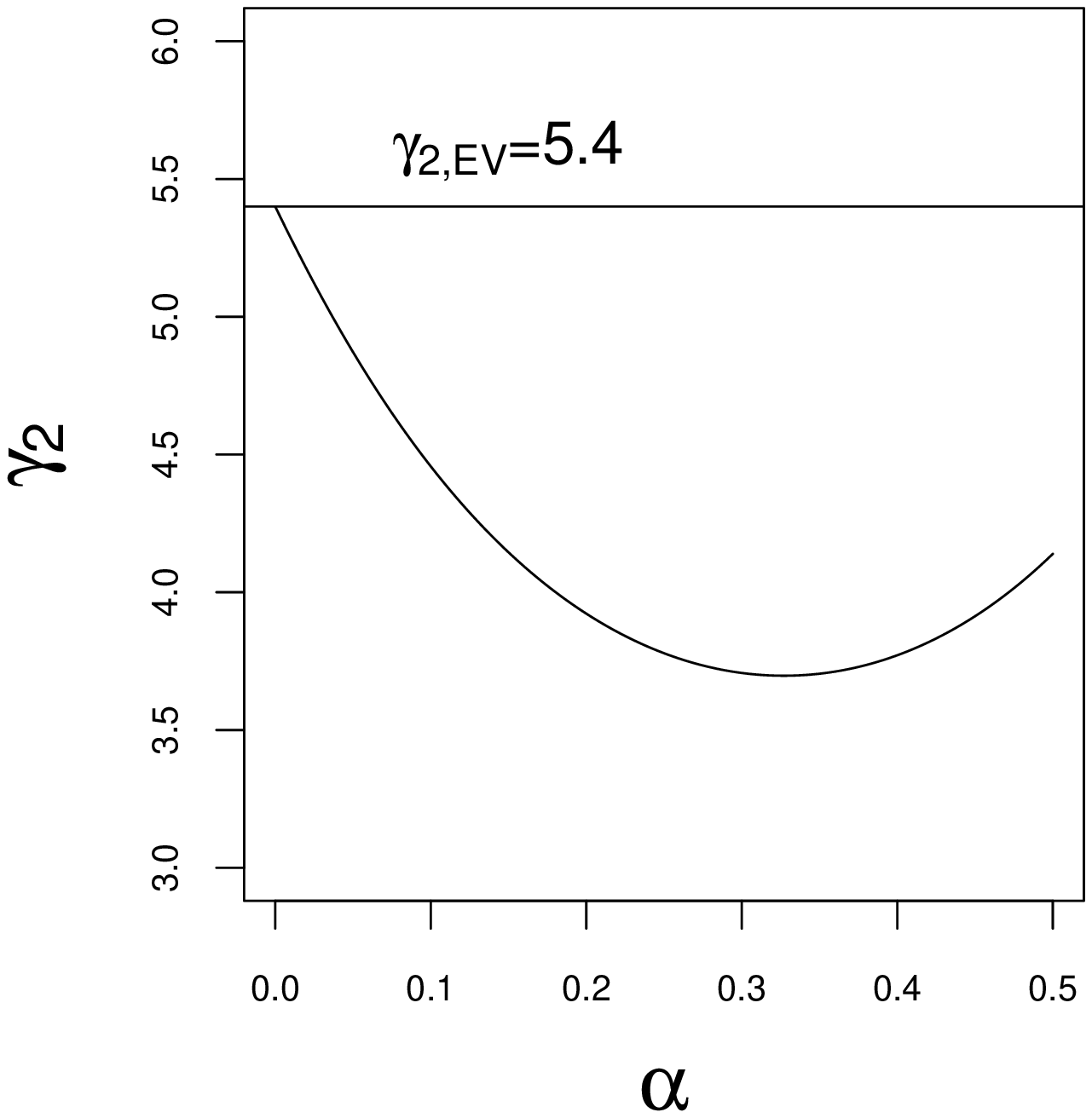}}
%\vspace{-0.6cm}
\caption{Density function, .99 quantile, skewness and kurtosis - TCEV$(0,1,10,5,\alpha)$}
\label{fig:TCEV}
\end{figure}

Figure~\ref{fig:TCEV} shows the plots of the pdf for selected parameters, and the .99 quantile, skewness and kurtosis of TCEV$(0,1,10,5,\alpha)$.
Note that the probability of  high values of the random variable and the $.99$ quantile grow with $\alpha$. 
The skewness and kurtosis coefficients are smaller than the corresponding Gumbel values for all $\alpha$.

\vspace{0.6cm}
As a summary, Table~\ref{pinheiro:distributions} presents the generalizations of the Gumbel presented above; 
the nonidentifiable distributions are marked with an asterisk. 
In the following sections, we will consider all the identifiable family of distributions, namely EV, GEV, EGu, TEV, GTIEV3, EGa, GGu, GLIV, and TCEV. 
\begin{table}[!ht]
\centering
\scriptsize
\caption{Generalizations of the Gumbel distribution.}
\label{pinheiro:distributions}
\medskip
\begin{tabular}{l r }
\hline
Distribution & Proposed by \\
\hline
Generalized extreme value GEV($\mu,\sigma,\alpha$) & \cite{JENKINSON} \\
Type IV generalized logistic GLIV($\mu,\sigma,\alpha$) & \cite{PRENTICE1975} \\
Two-component extreme value TCEV($\mu,\sigma,\mu_1,\sigma_1,\alpha$) & \cite{ROSSI} \\
Three parameter exponential-gamma EGa($\mu,\sigma,\alpha$) & \cite{OJO} \\
Exponentiated Gumbel EGu($\mu,\sigma,\alpha$) & \cite{NADARAJAH2006} \\
Transmuted extreme value TEV($\mu,\sigma,\alpha$) & \cite{ARYAL} \\
Generalized Gumbel GGu($\mu,\sigma,\alpha$) & \cite{COORAY} \\
Generalized three-parameter Gumbel GTIEV3($\mu,\sigma,\alpha$) &  \cite{JEONG} \\
Generalized type I extreme value GTIEV($\mu,\sigma,\alpha,\beta$) $^*$ & \cite{DUBEY} \\
Exponential-gamma ExpGamma($\mu,\sigma,\alpha,\beta$) $^*$ & \cite{ADEYEMI}\\
Beta Gumbel BG($\mu,\sigma,\alpha,\beta$) $^*$ & \cite{NADARAJAH2004}\\
Kummer beta generalized Gumbel KBGGu($\mu,\sigma,\alpha,\beta,\gamma$) $^*$ & \cite{PESCIM2012} \\
Kumaraswamy Gumbel KumGum($\mu,\sigma,\alpha,\beta$) $^*$ & \cite{CORDEIRO2012} \\
Exponentiated generalized Gumbel EGGu($\mu,\sigma,\alpha,\beta$) $^*$ & \cite{CORDEIRO2013} \\
\hline
\end{tabular}
\end{table}

\section{Right-tail heaviness}\label{tail}

Heavy right-tailed distributions have been used to model phenomena in economics, ecology, bibliometrics, and biometry, among others; see, for instance, \citet{MARKOVICH} and \citet{RESNICK}.
We next describe two criteria used to evaluate the right-tail heaviness of a distribution.

Informally, a regular variation function is asymptotically equivalent to a power function.
Formally, a Lebesgue measurable function $U: I\!\! R^+ \rightarrow I\!\! R^+$ is regularly varying at infinity with index $\rho$ ($U \in RV_\rho$ ), if $\lim\limits_{t \to \infty}U(tx) / U(t)=x^{\rho}$, for $x>0$.
If $\rho=0$, $U$ is referred to as slowly varying. 
The function $U$ varies rapidly at infinity (or is rapidly varying at infinity with index $\infty$ ($-\infty$), or  $U \in RV_\infty$ ($U \in RV_{-\infty}$) (\citealp[p. 4]{HAAN}), if $\forall x$
$$
\begin{matrix}
\lim\limits_{t\rightarrow \infty}\frac{U(tx)}{U(t)}:=x^{\infty}=
\left\{
\begin{matrix}
0 & {\rm if} & 0<x<1 \\
1 & {\rm if} & x=1 \\
\infty & {\rm if} & x>1.
\end{matrix}
\right.
&
\left(
\lim\limits_{t\rightarrow \infty}\frac{U(tx)}{U(t)}:=x^{-\infty}=
\left\{
\begin{matrix}
\infty & {\rm if} & 0<x<1 \\
1 & {\rm if} & x=1 \\
0 & {\rm if} & x>1.
\end{matrix}
\right.
\right).
\end{matrix}
$$

A distribution with cdf $F$ is said to have a heavy right tail whenever the survival function, $\overline{F}:=1-F$, is a regularly varying at infinity function with a negative index of regular variation $\rho=-1/\xi, \; \xi>0$, i.e., $\lim\limits_{t \rightarrow \infty} \overline{F}(tx) / \overline{F}(t)=x^{-1/\xi}$. 
The parameter $\xi$ is known as the tail index, and  is one of the primary parameters of rare events. 
The distribution is said to have light (non-heavy) right tail if the limit equals $x^{-\infty}$, and $\xi=0$.
When the limit equals 1, i.e. $\overline F$ is a slowly varying function, we will say that the distribution has a heavy right tail with
tail index $\xi=\infty$.
It follows from \citet[Corollary 1.2.1 - 2 and 3]{HAAN} that the index of regular variation is invariant under the location-scale transformation. It is thus sufficient to derive it from the standard form of the distribution.

The generalized extreme value distribution GEV($\mu,\sigma,\alpha$) has a heavy right tail with tail index $\alpha$ when $\alpha>0$ 
(Fréchet family).
It has a non-heavy right tail  when $\alpha=0$ (Gumbel family). 
Other heavy right-tailed distributions are, e.g., Student-t($\nu$) ($\xi=1/\nu$), Cauchy ($\xi=1$) and F($\alpha,\beta$) ($\xi=2/\beta$).
%whose tail indices are respectively $1/\nu$, 1 and $2/\beta$.
The other distributions addressed in this paper\footnote{Recall that we restrict our attention to the identifiable family of distributions only.}, viz., Gumbel (EV), exponentiated Gumbel (EGu), transmuted extreme value (TEV), generalized three-parameter Gumbel (GTIEV3), three-parameter exponential-gamma (EGa), type IV generalized logistic (GLIV), generalized Gumbel distribution (GGu), and two-component extreme value distribution (TCEV), are all non-heavy right-tailed distributions (see Supplement).
Hence, among the identifiable distributions addressed in this work, the GEV distribution is the only one with a potentially heavier right tail than that of the Gumbel distribution under the tail index approach.

\cite{RIGBY} ordered the heaviness of the tails of a continuous distribution based on the logarithm of the pdf. If random variables $X_1$ and $X_2$ have continuous pdf $f_{X_1}(x)$ and $f_{X_2}(x)$ and $\lim\limits_{x \to \infty}f_{X_1}(x)=\lim\limits_{x \to \infty}f_{X_2}(x)=0$, then $X_2$ has a heavier right tail than $X_1$ if and only if $\lim\limits_{x \to \infty}[\ln f_{X_2}(x) - \ln f_{X_1}(x)]=\infty$. There are three main forms for $\ln f_{X}(x)$ when $x \to \infty$ (right tail) or $x \to -\infty$ (left tail):  $\ln f_{X}(x) \approx -k_2(\ln|x|)^{k_1}$ (type I), $\ln f_{X}(x) \approx -k_4|x|^{k_3}$ (type II) or $\ln f_{X}(x) \approx-k_6 \exp(k_5|x|)$ (type III). 
The three types are in decreasing order of tail heaviness. 
For type I, decreasing $k_1$ results in a heavier tail while decreasing $k_2$ for fixed $k_1$ results in a heavier tail. Similarly, for
the two types.
If two distributions have the same values of $k_1$ and $k_2$ (analogously for $k_3$ and $k_4$, or $k_5$ and $k_6$), their right tails are not necessarily equally heavy. 
In this case, it is necessary to compare the second-order terms of the logarithm of the pdf to distinguish the distributions.

Table~\ref{tab:tail_heaviness} summarizes the right tail asymptotic form of the logarithm of the pdf for the distributions mentioned above.
The GEV distribution with $\alpha>0$ is of type I with $k_1=1$ and $k_2=1+1/\alpha$.
As expected, the GEV distribution is the only one that has a `Paretian type' right tail (type I with $k_1 = 1$ and $k_2 > 1$). 
Note that the Cauchy distribution has $k_1=1$ and $k_2=2$, 
and hence if $\alpha > 1$, the GEV distribution has a heavier right tail than that of the Cauchy distribution. 
The Student-t distribution with $\nu$ degrees of freedom has $k_1=1$ and $k_2=1+\nu$. If $\alpha > 1/2$, the right tail heaviness of the GEV distribution is greater than that of the Student-t distribution with two degrees of freedom, which is uncommon in real data.  

The Gumbel distribution is of type II with $k_3=1$ and $k_4=1/\sigma$. 
According to \cite{RIGBY}, distributions with $k_3=1$ are non-heavy tailed.
All of the other distributions are also of type II and non-heavy tailed distributions.
The EGu, EGa, GGu, and GLIV distributions have $k_4=\alpha / \sigma$, and hence they have heavier right tail than the Gumbel distribution when $\alpha<1$. 
The TEV, GTIEV3, and TCEV distributions have the same $k_4=1/\sigma$ as the Gumbel distribution.
To distinguish among the TEV, GTIEV3 and TCEV distributions it is necessary to compare the second-order terms of the logarithm of their pdf. 
Comparing the second-order terms, the TEV distribution with $\alpha<0$ and the TCEV distribution have heavier right tail than the Gumbel distribution. 
The right tail of the GTIEV3 distribution is lighter than that of the Gumbel distribution (see Supplement). 
These findings agree with the pdf plots shown in Figures~\ref{fig:TEV}, \ref{fig:GTIEV3}, and \ref{fig:TCEV}, respectively.
\begin{table}[!ht]
\begin{center}
\caption{Right tail asymptotic form of the logarithm of the pdf for the Gumbel distribution and its generalizations} 
\label{tab:tail_heaviness}
\begin{threeparttable}[c]
\small
\begin{tabular}
{c| c c c c c c c c c}
\hline \hline
& GEV & EGu &EGa & GLIV & GGu & EV & GTIEV3 & TEV & TCEV \\
\hline
parameter& $\alpha >0$   &  $\alpha >0$  & $\alpha >0$    &  $\alpha >0$  & $\alpha>0$ &  & $\alpha >0$ &  $-1<\alpha \leq 1$ & $0<\alpha<0.5$  \\
$k_1$& 1   & $\alpha>0$               &      &     &    &  &  &   &   \\
$k_2$& $1 + 1/\alpha$  &      &     &    &  &  &   &  & \\
$k_3$&                   &  1    &  1   &  1  & 1 & 1 &  1 & 1  & 1 \\
$k_4$&                   &  $\alpha/\sigma$    & $\alpha/\sigma$    &  $\alpha/\sigma$ &  $\alpha/\sigma$  & $1/\sigma$  & $1/\sigma$ & $1/\sigma$ & $1/\sigma_2$ \tnote{*} \\
\hline \hline
\end{tabular}
\begin{tablenotes}[flushleft]
\item[*] if $\sigma_1 < \sigma_2$. 
\end{tablenotes}
\end{threeparttable}
\end{center}
\end{table}

\section{Monte Carlo simulation results}\label{simulation}

We next compare the ability of the Gumbel distribution and its generalizations to model data taken from different distributions. 
To this end, we present a Monte Carlo simulation study in which 10,000 samples of size 500 are generated from and modeled with each of the 
identifiable distributions considered in this paper. For generating the data, we set $\mu=0$ and $\sigma=1$ (for the TCEV 
distribution $\mu_1=10$ and $\sigma_1=5$). The remaining parameters were chosen in such a way that the 
$.999$ quantile is close to 10 whenever possible.
Table~\ref{tab:samples} presents the distributions from which the samples were drawn and their $.999$ quantile and kurtosis.
\begin{table}[!ht]
\begin{center}
\caption{Distributions, $.999$ quantile, and kurtosis} \label{tab:samples}
\small
\begin{tabular}
{c| c c c c c c c c c c c c}
\hline \hline
        & EV & GEV & EGu & TEV & EGa & GGu & GLIV & TCEV \\
\hline
Parameters & $-$ & $\alpha=0.1$ & $\alpha=0.7$ & $\alpha=-0.99$ & $\alpha=0.7$ & $\alpha=0.7$ & $\alpha=0.65$ & $\alpha=0.0016$ \\
 & &  &  &  &  & &$\beta=15$ & $\mu_1=10$, $\sigma_1=5$ \\
$x_{.999}$ & 6.91  &9.95  &9.87  &7.60  &9.99  & 9.87 &10.38  & 10.28  \\
Kurtosis  &  5.40 & 10.98 & 6.28 &   5.39 &  6.22 & 5.72 & 6.26 & 5.38 \\
\hline \hline
\end{tabular}
\end{center}
\end{table}

As measures of model adequacy, we use the Akaike information criterion (AIC) and two modified Anderson Darling statistics (ADR and AD2R). 
AIC is a measure of dissimilarity between two distributions over the support; smaller AIC suggests that the fit is closer to the true density. 
ADR and AD2R \citep[Table 2 and B.1]{LUCENO} are sensitive to the lack of fit in the right tail of the distribution. 
AD2R puts more weight in the right tail than ADR. Smaller values of ADR and AD2R are indicative of a better fit.

A characteristic that is often of interest in extreme data modeling is the return level.
The return level with return period $1/p$ is the quantile $x_{1-p}$, and it is interpreted as the value that we expect 
to be exceeded once every $1/p$ periods on average. 
To evaluate the quantile goodness of fit, we compute the $.999$ quantile discrepancy, which is defined as the difference between the $.999$ quantile of the fitted model and the $.999$ quantile of the distribution from which samples are generated divided by the latter.

The estimates of the parameters were obtained by numerically maximizing the log-likelihood function. 
%For the maximization procedure, we used the nonlinear quasi-Newton BFGS method with numerical derivatives (see, e.g., \citealp{PRESS}), 
%and a  method that implements a sequential quadratic programming technique to maximize a nonlinear function subject to nonlinear constraints, similar to Algorithm 18.7 in \cite{WRIGHT}. These methods are individually implemented in the functions MaxBFGS and MaxSQP in the matrix programming language {\tt Ox} \citep{DOORNIK}. 
For the maximization procedure, we used a  method that implements a sequential quadratic programming technique to maximize a nonlinear function subject to non-linear constraints, similar to Algorithm 18.7 in \cite{WRIGHT}. This method is implemented in the function MaxSQP in the matrix programming language {\tt Ox} \citep{DOORNIK} and allows to establish bounds for the individual parameters. 
For the TEV distribution, we used the profile log-likelihood function for the parameter $\alpha$.
We used the bounds $-0.6<\alpha<0.6$, for the GEV distribution, and $0<\alpha<1$ and $0<\beta<20$, for the GLIV distribution.
The GTIEV3 distribution is not included in the simulation study because it behaves like the EV distribution with respect to the right tail. 
 
\begin{figure}[!htb]
\centering
\includegraphics[height=99.5mm,width=99.5mm]{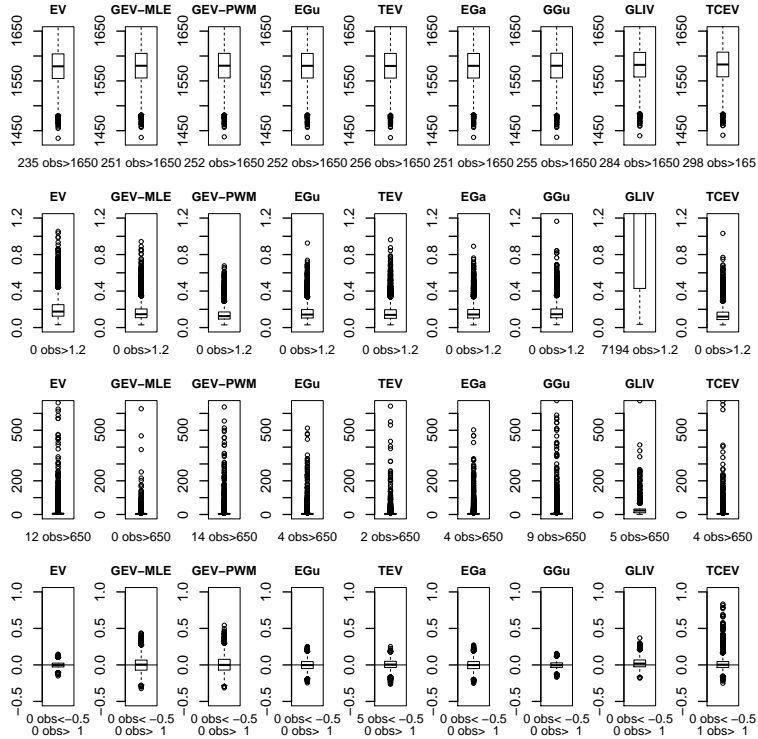}
%\vspace{-0.5cm}
\caption{Boxplots of AIC (first row), ADR (second row), AD2R (third row) and .999 quantile discrepancies (fourth row) - random samples generated from EV.}
\label{fig:discrepanciaEV}
\end{figure}

\begin{figure}[!h]
\centering
\includegraphics[height=99.5mm,width=99.5mm]{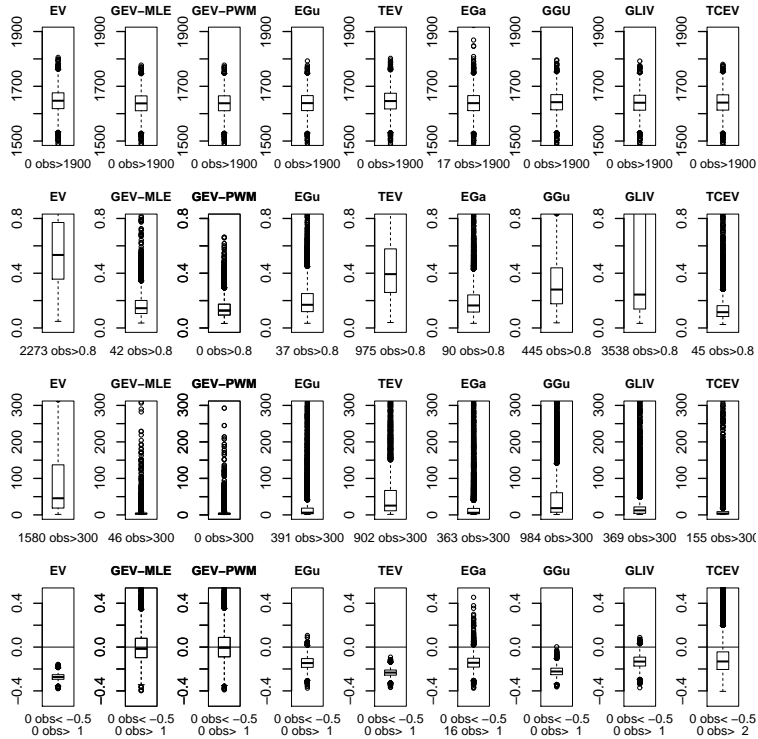}
%\vspace{-0.5cm}
\caption{Boxplots of AIC (first row), ADR (second row), AD2R (third row) and .999 quantile discrepancies (fourth row) - Random samples generated from GEV.}
\label{fig:discrepanciaGEV}
\end{figure}

Figures~\ref{fig:discrepanciaEV} and \ref{fig:discrepanciaGEV} present the boxplots of AIC, ADR, AD2R, and the .999 quantile discrepancies of the fitted models. 
%For these figures, the samples were generated from a standard Gumbel distribution, i.e., EV($0,1$), and a generalized extreme value GEV($0,1,0.1$) distribution, respectively.
%The figures corresponding to the cases where the samples were generated from the exponentiated Gumbel EGu($0,1,0.7$), transmuted extreme value TEV($0,1,-0.99$), three-parameter exponential-gamma EGa($0,1,0.7$), the generalized Gumbel GGu($0,1,0.7$), type IV generalized logistic GLIV($0,1,0.65,15$) and two-component extreme value TCEV($0,1,10,5,0.0016$) distributions are presented in the Supplement.
For these figures, the samples were generated from the EV($0,1$) and the GEV($0,1,0.1$) distributions, respectively.
The figures, corresponding to the cases where the samples were generated from the EGu($0,1,0.7$), TEV($0,1,-0.99$),  EGa($0,1,0.7$), GGu($0,1,0.7$), GLIV($0,1,0.65,15$) and TCEV($0,1,10,5,0.0016$) distributions, are presented in the Supplement.
For the GEV distribution, we show results for two estimation methods: maximum likelihood estimation (GEV-MLE) and probability-weighted moments (GEV-PWM) methods \citep[Section 5.3]{CASTILLO}.

For the Gumbel distributed samples (Figure~\ref{fig:discrepanciaEV}), the boxplots of AIC are quite similar and suggest that all generalizations of the Gumbel distribution can suitably fit Gumbel distributed data.
The goodness of fit at the right tail, illustrated by the boxplots of ADR, are also quite similar except for the Gumbel and the GLIV distributions. For the Gumbel distribution, the median and the dispersion are bigger than for the other distributions.
The GLIV fit is poor in the right tail, which is consistent with the quantile plot in Figure~\ref{fig:GLIV}, that suggests that a small difference in the estimated parameter $\alpha$ produces a significant difference in the upper quantiles. 
This characteristic makes the right tail goodness of fit dependent on the precision adopted for the parameter estimation.
Boxplots of AD2R emphasize the right-tail lack of fit. The boxplots of AD2R in Figure~\ref{fig:discrepanciaEV} are similar except for the GLIV fit, whose interquartile range (IQR) is the largest. 
The GEV-MLE right tail fit seems to be slightly better than the others.
The boxplots of $0.999$ quantile discrepancy show that the median is close to zero for all the distributions. 
The EV and GGu fits exhibit the smallest amplitudes. 
The TEV fit presents some cases of marked underestimation due to numerical problems.   

%For the GEV distributed samples (Figure~\ref{fig:discrepanciaGEV}), the boxplots of AIC of the GEV-MLE and GEV-PWM fits present the smallest medians and the GEV-PWM fit exhibits the smallest amplitude.
For the GEV distributed samples (Figure~\ref{fig:discrepanciaGEV}), the boxplots of AIC of the EV and TEV fits present the biggest medians, and the GEV-MLE, GEV-PWM, and TCEV fits exhibit the smallest amplitudes.
We recall that, theoretically, the AIC of the GEV-MLE fit can not be bigger than that of the GEV-PWM fit. 
%For the simulations, we use $\alpha=0.1$ that defines a very heavy-tailed distribution which perhaps requires a greater level of precision to estimate.   
The boxplots of ADR and AD2R highlight the GEV-PWM fit as the best right-tail fit.
The boxplots of quantile discrepancy show that the GEV-PWM and GEV-MLE fits have the closest to zero medians, and all the others underestimate the $.999$ quantile, markedly the EV and TEV fits.

Hereafter we analyze the fits when the data were generated from the other distributions (see boxplots in the Supplement).
The boxplots of AIC are similar to those in Figure~\ref{fig:discrepanciaEV}.
%, except when the data were generated from the TCEV distribution. In this case, the TCEV distribution presents the smallest median value.
The boxplots of ADR for the TCEV fit exhibit the smallest medians and IQR followed by the GEV-PWM fit except for samples generated from the TCEV distribution when it is followed by the EGu, EGa, and GEV-MLE fits. 
The boxplots of AD2R reveal that the GEV-MLE fit is among the best fits regardless of from which distribution the data were generated.
When generating the data from the GLIV distribution, all the distributions underestimate the $.999$ quantile, and this is the only case where the GEV-MLE and GEV-PWM fits underestimate the $.999$ quantile. 
%We recall that, for the parameters chosen to generate the distributions, the TCEV .99 quantile is the biggest one, i.e, it is the heavier tailed distribution. The median of the GEV-PWM fit is the closest to zero.

Summing up, the GEV-PWM, TCEV, and GEV-MLE fits are among the best fits in all of the simulated settings.

\section{Application to a wind speed data set}\label{application}

\begin{figure}[!ht]
\centering
\subfigure
{
\includegraphics[height=38mm,width=38mm]{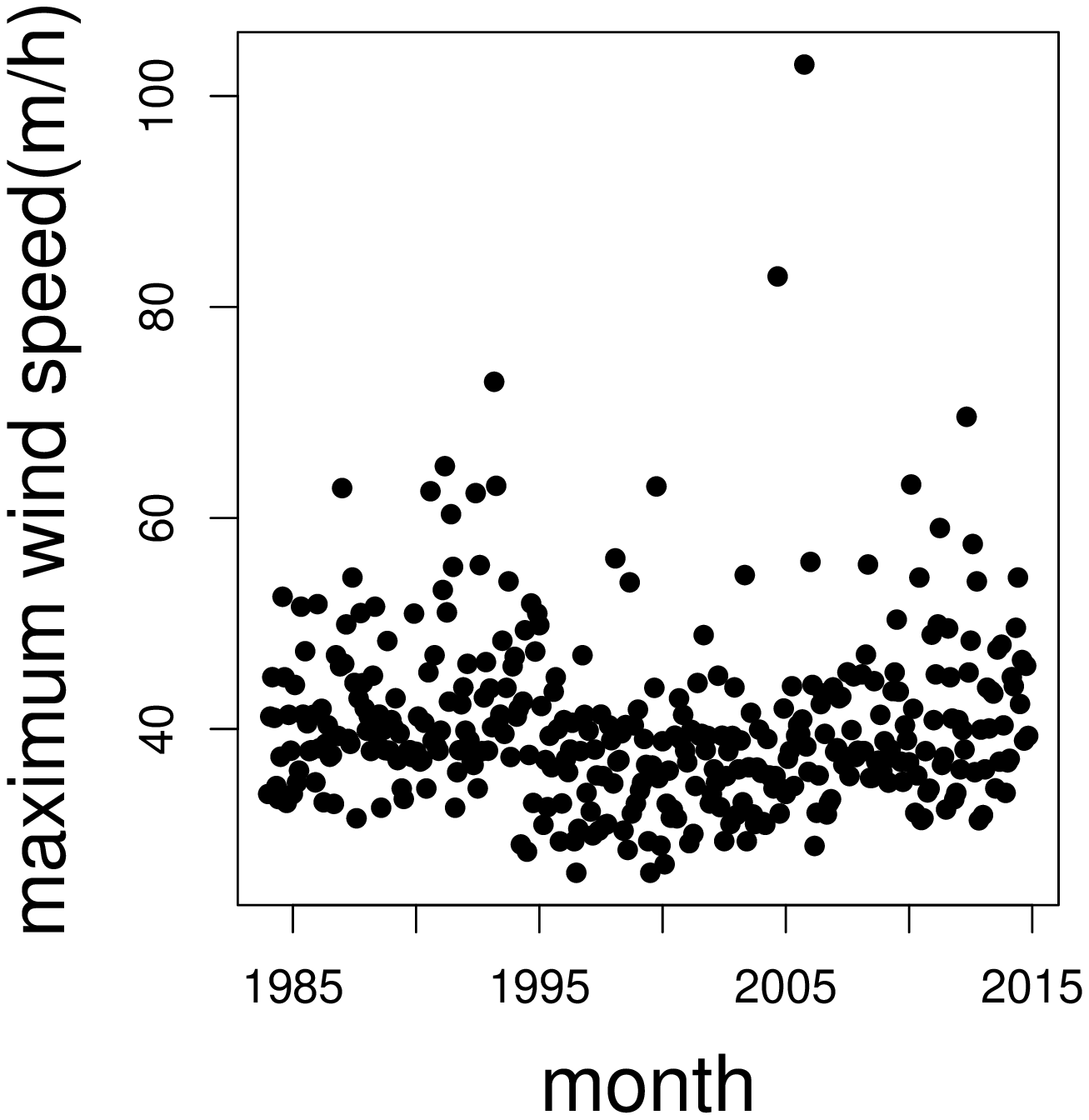}
}
\subfigure
{
\includegraphics[height=38mm,width=38mm]{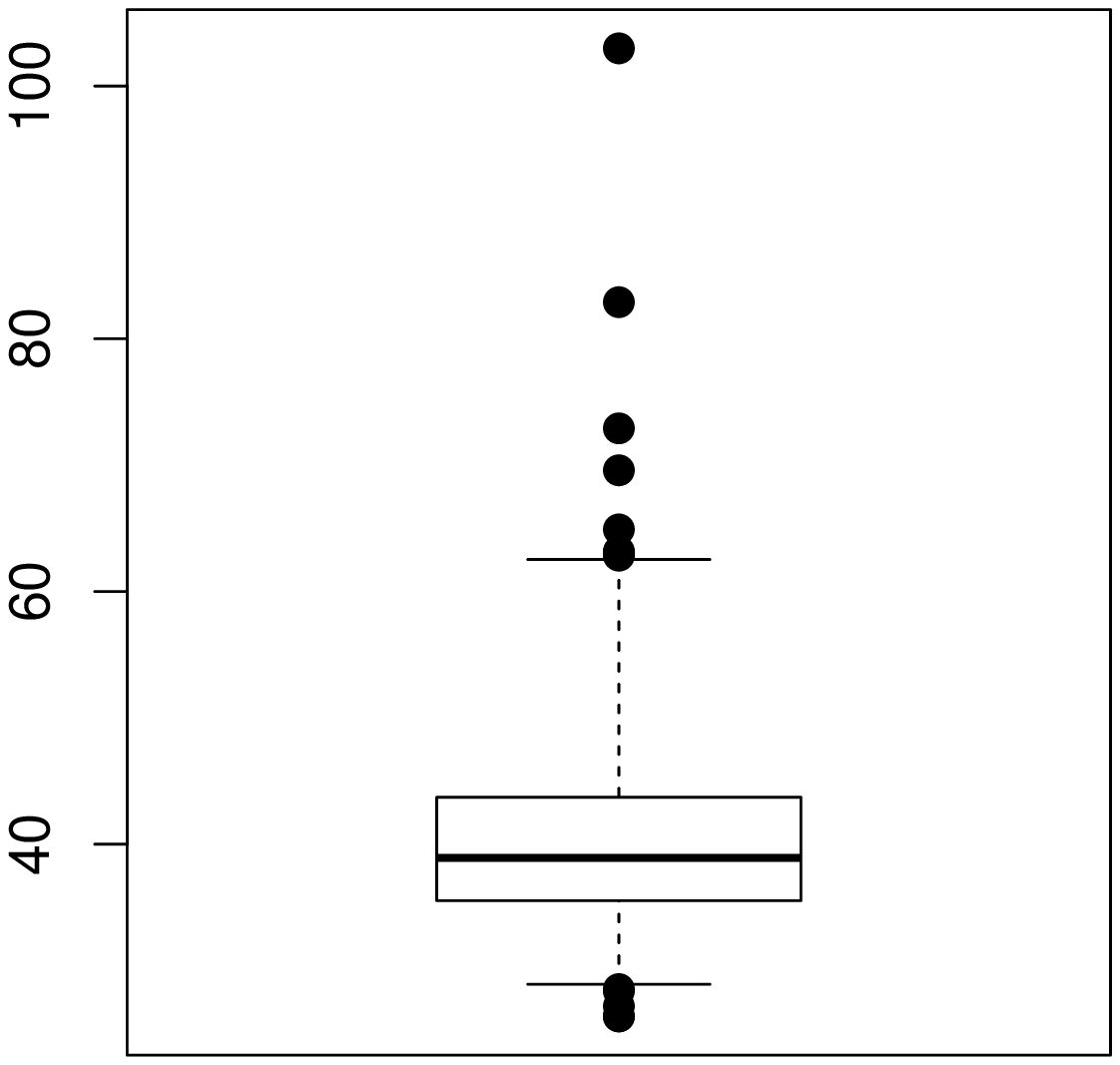}
}
\subfigure
{
\includegraphics[height=38mm,width=38mm]{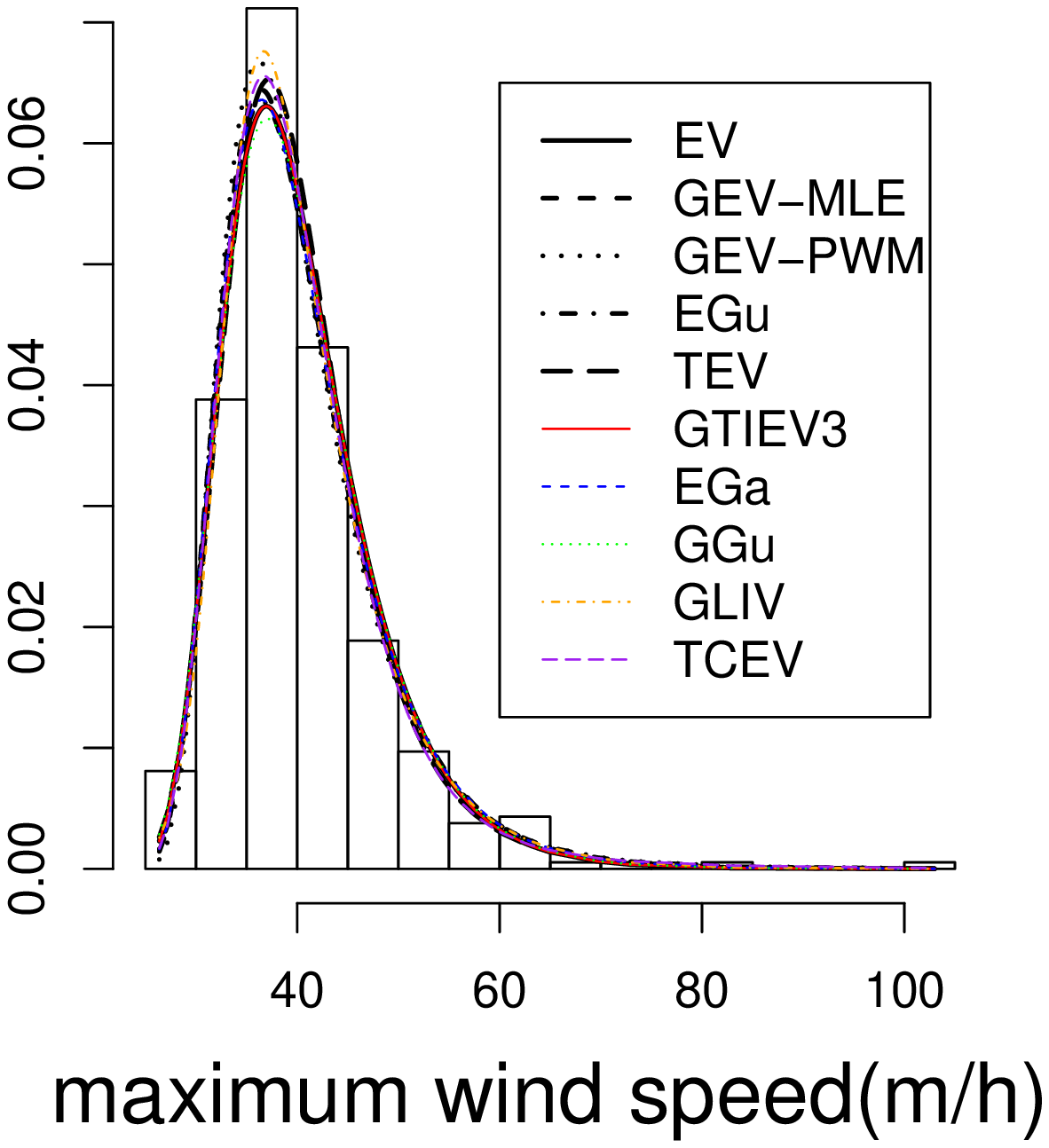}
}
%\vspace{-0.6cm}
\caption{Scatterplot, boxplot, adjusted boxplot, and histogram; wind speed data.}
\label{fig:pontos-WestPalmBeach}
\end{figure}

We analyze data on the maximum monthly peak gust wind speed (mph) in West Palm Beach, Florida (USA) for the months January, 1984 to November, 2014, with $n=371$ observations. 
The data are available online for download from the {\it National Climate Data Center (NCDC) -- National Oceanic and 
Atmospheric Administration (NOAA)} at {\small{\tt \url{http://www.ncdc.noaa.gov/}}}\footnote{
More specifically,
{\small{\tt \url{http://www.ncdc.noaa.gov/}}} $\rightarrow$  
{\it I want to search for data at a particular location. } $\rightarrow$ 
{\it Additional Data Access: Publications} $\rightarrow$ 
{\it Local Climatological Data}. (Last accessed on January 23, 2014)} and given in the Supplement.
%We fit the different models described in Section~\ref{gumbel_generalizations} to the seasonally adjusted wind speed data obtained with the function {\tt stl} of the R package {\tt stats}.
We fit the different models described in Section~\ref{gumbel_generalizations} to the seasonally adjusted wind speed data. 
The seasonally adjusted wind speed was calculated removing the seasonal component using a robust seasonal trend decomposition implemented in the functions {\tt stl} of the R package {\tt stats} and {\tt seasadj} of the package {\tt forecast} \citep{CLEVELAND}. 
The maximum likelihood estimates of the parameters were obtained similarly as in the Monte Carlo simulation (Section~\ref{simulation}).
\begin{sloppypar} 
Figure~\ref{fig:pontos-WestPalmBeach} shows the scatterplot, the adjusted boxplot for asymmetric distributions \citep{HUBERT} and the histogram of the data together with the fitted densities.
The outliers at the right tail of the adjusted boxplot are much more spread than those at the left one. 
The empirical skewness and kurtosis  coefficients are 2.26 and 13.42, respectively. Both are much higher than those expected from a 
Gumbel distribution (1.14 and 5.4, respectively), which suggests the fitting of the generalized distributions.
Recall that the only distribution for which the skewness can be higher than 2 and the kurtosis can be higher than 9 
is the GEV distribution.
\end{sloppypar}

The maximum likelihood estimates and the probability weighted moments estimates for the GEV distribution (standard errors in parentheses) and the estimated return levels of the seasonally adjusted series 
%and the original one 
of the return period of one thousand months from the selected distributions
%, and the return period ratio
 are summarized in Table~\ref{tab:estimativasventoWestPalmBeach}.
The additional parameter $\alpha$ estimate of the GTIEV3 distribution is notably high, i.e., the fitted GTIEV3 distribution nearly coincides with the fitted Gumbel distribution. Recall that the Gumbel distribution is a limiting case of the GTIEV3 distribution when $\alpha \to \infty$. 
Indeed, estimates of $\mu$ and $\sigma$ for these distributions are the same up to the third decimal places.
The estimated return levels of a return period of one thousand months (for the seasonally adjusted data) from the selected distributions differ by up to 20 mph. 
The Gumbel distribution produces the smallest estimated return level (77.23 mph), and the largest estimates are obtained from the GEV-PWM (89.52 mph) and TCEV (100.21 mph) fits. 
\begin{table}[htp]
\centering
\renewcommand{\arraystretch}{1.1}
\caption{Parameter estimates, standard errors (in parentheses) and return level for a thousand years - wind speed data.}
\label{tab:estimativasventoWestPalmBeach}
\footnotesize
\begin{tabular}
{l c c c c c c c}
\hline \hline
 & $\hat\mu$ & $\widehat\sigma$ & $\widehat\alpha$ & $\widehat\beta$ or $\widehat{\mu_1}$ & $\widehat{\sigma_1}$ & $\widehat{x_{.999}}$  \\
\hline            
EV      & 36.94(0.32) & 5.83(0.24) & --         & -- & -- & 77.23\\ 
GEV-MLE & 36.75(0.34) & 5.72(0.25) & 0.06(0.03) & -- & -- & 85.74\\ 
GEV-PWM & 36.67(0.34) & 5.53(0.25) & 0.09(0.03) & -- & -- & 89.52\\ 
EGu     & 35.78(0.95) & 5.07(0.63) & 0.79(0.16) & -- & -- & 80.34\\ 
TEV     & 39.19(0.35) & 6.95(0.29) & 0.61(0.16)    & -- & -- & 80.57\\ 
GTIEV3   & 36.94(0.29) & 5.83(0.22) & 16698.18(-)& -- & -- & 77.23 \\ 
EGa     & 35.10(1.24) & 4.89(0.68) & 0.75(0.16) & -- & -- & 80.83\\ 
GGU     & 36.89(0.38) & 6.07(0.43) & 1.03(0.11) & -- & -- & 77.69\\ 
GLIV    & 36.67(0.48) & 2.94(1.00) & 0.43(0.17) & 2.16(1.95) & -- & 81.98\\ 
TCEV    & 36.70(0.37) & 5.51(0.30) & 0.02(0.02) & 61.56(23.41) & 13.38(8.22) & 100.21\\ 
\hline
\end{tabular}
\end{table}

\begin{figure}[!ht]
\centering
\subfigure
{
\includegraphics[height=38mm,width=38mm]{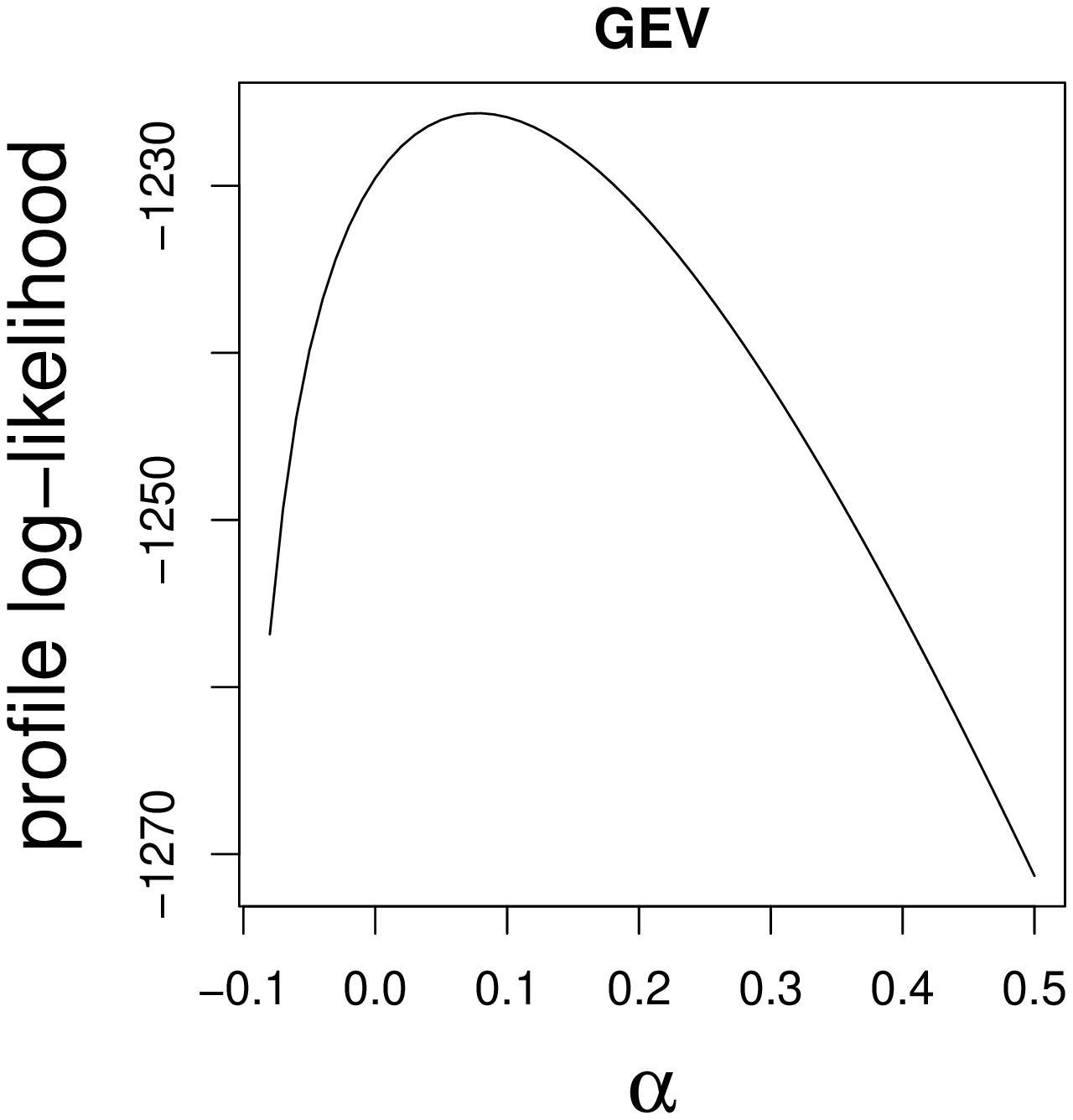}
}
\subfigure
{
\includegraphics[height=38mm,width=38mm]{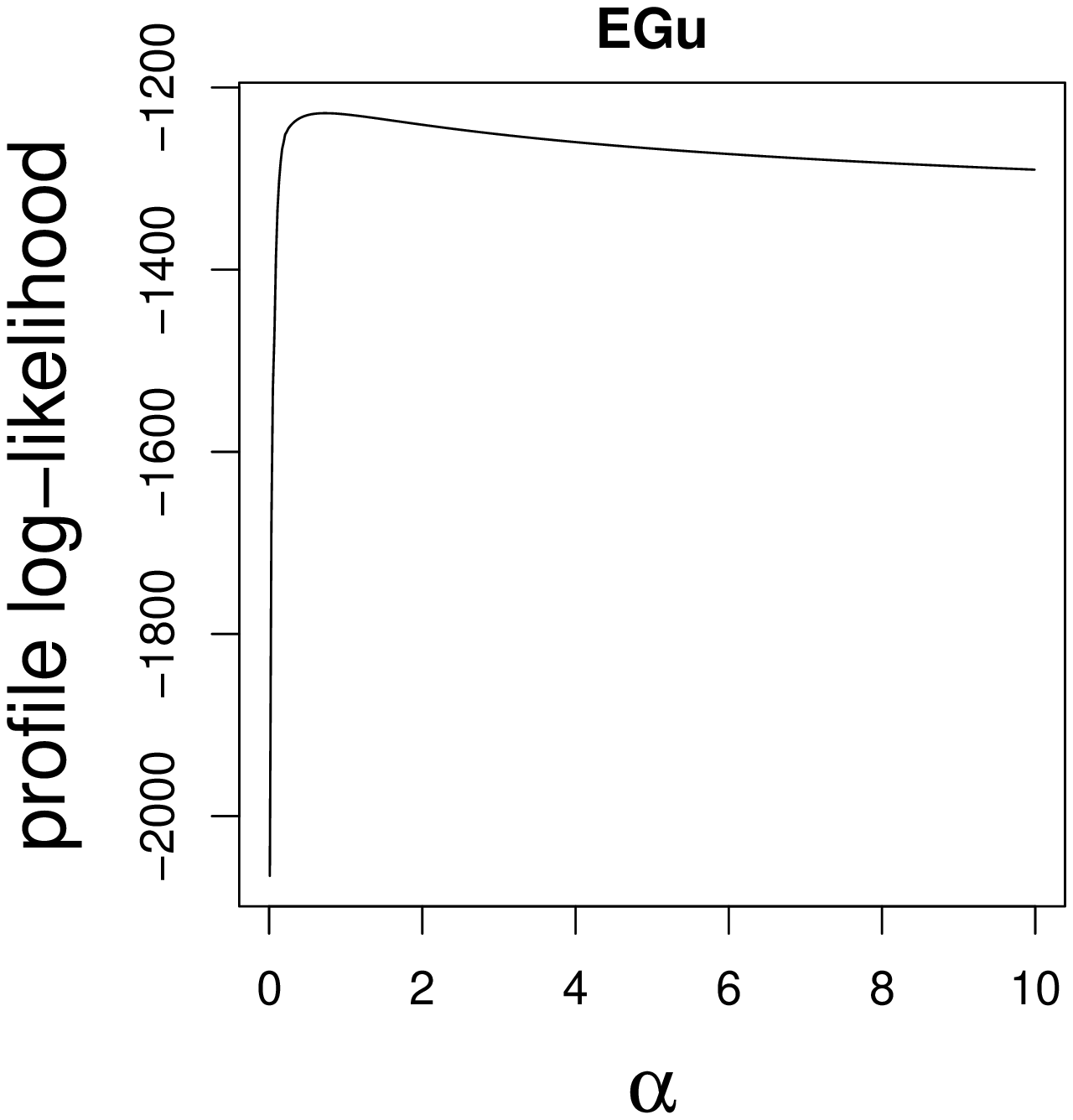}
}
\subfigure
{
\includegraphics[height=38mm,width=38mm]{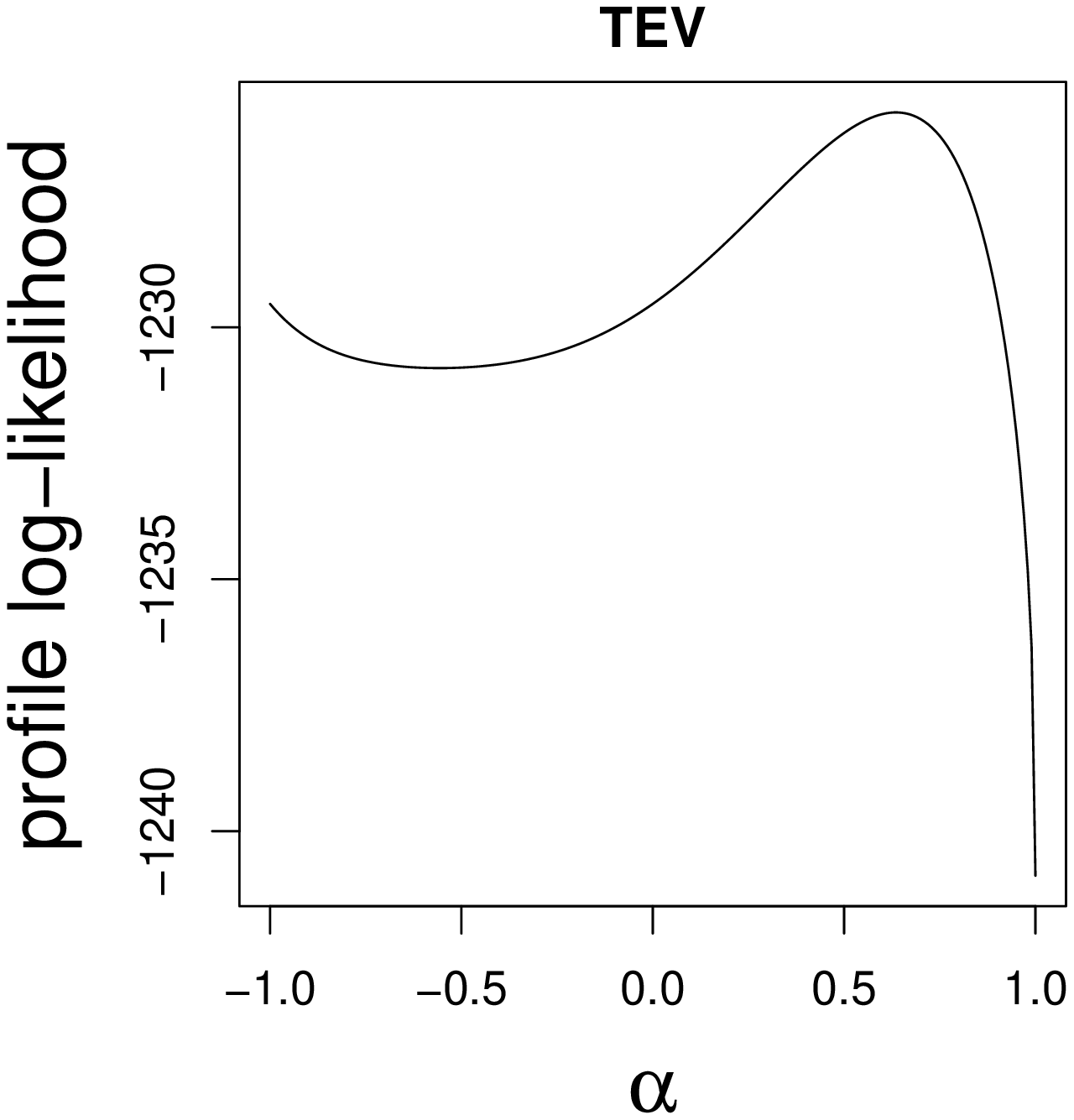}
}
\subfigure
{
\includegraphics[height=38mm,width=38mm]{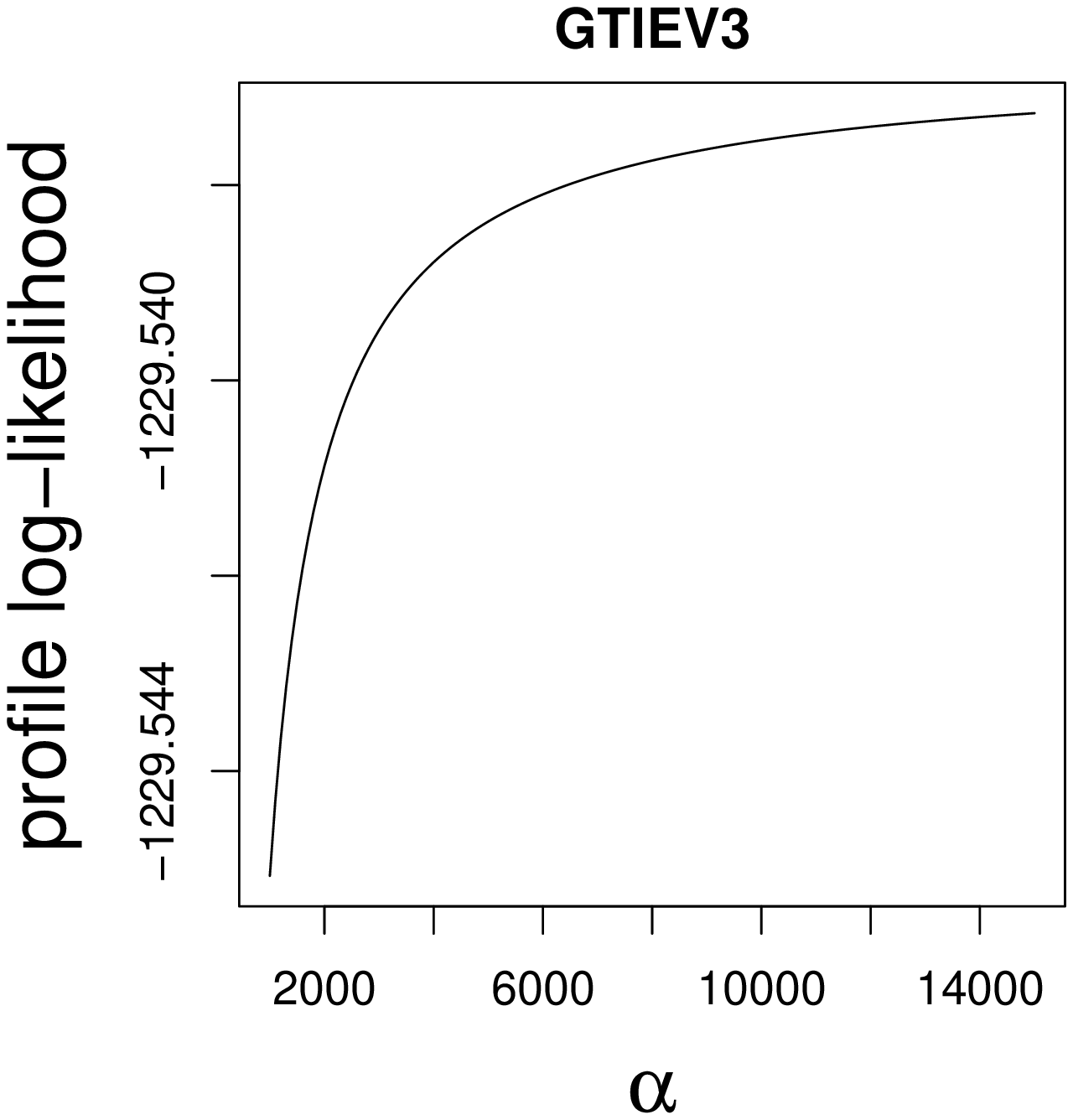}
}
\\
\centering
\subfigure
{
\includegraphics[height=38mm,width=38mm]{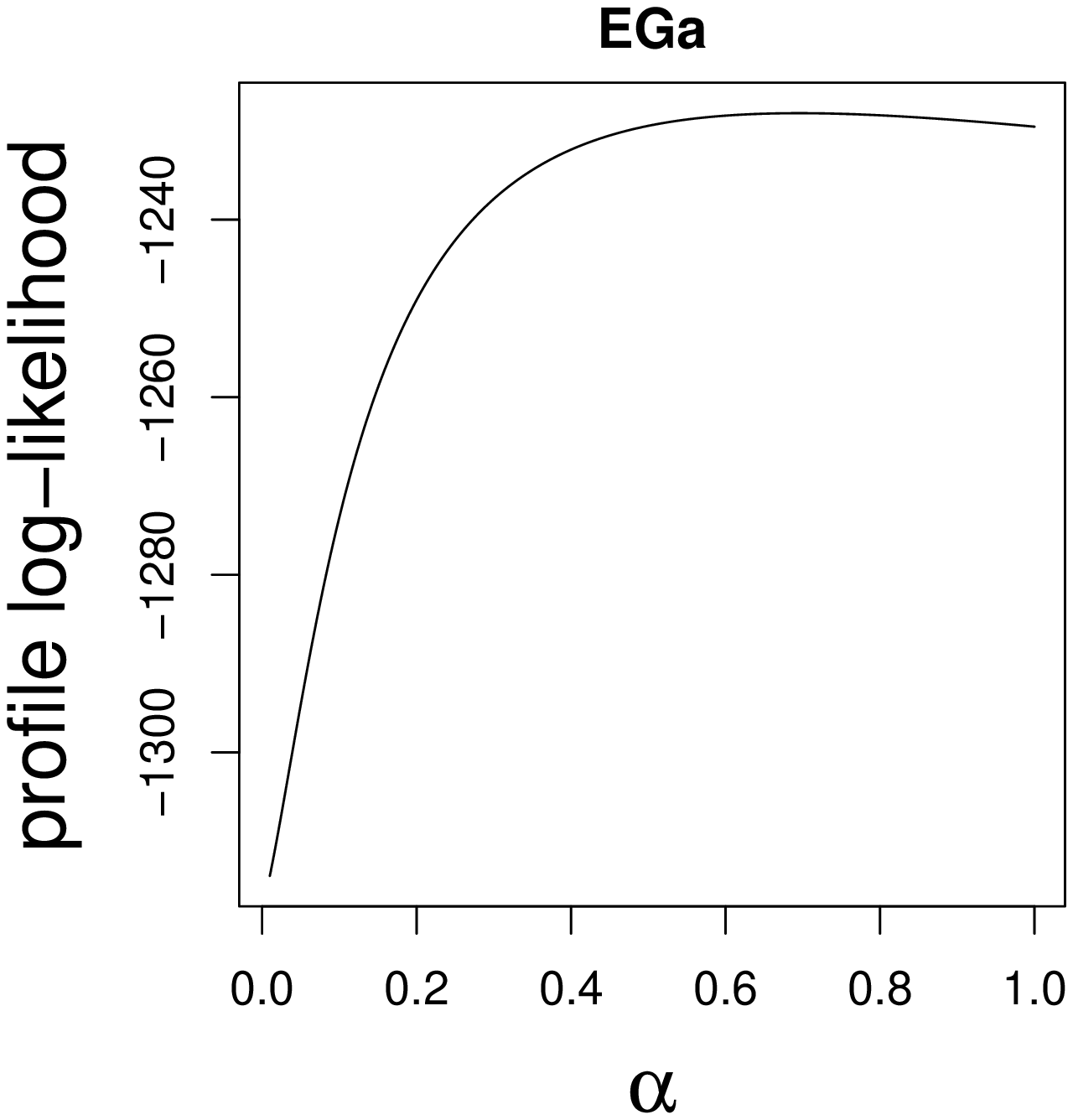}
}
\subfigure
{
\includegraphics[height=38mm,width=38mm]{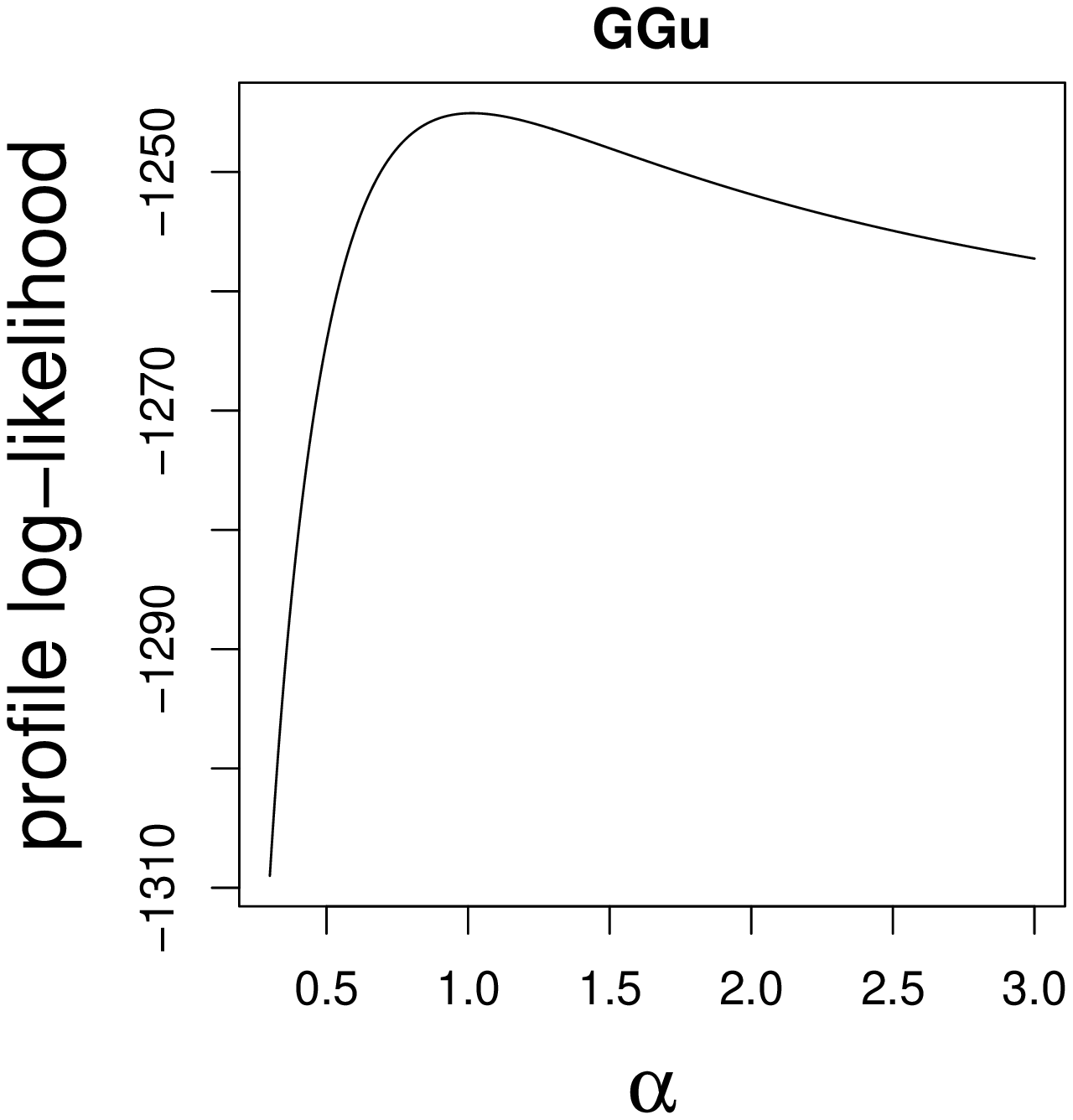}
}
\subfigure
{
\includegraphics[height=38mm,width=38mm]{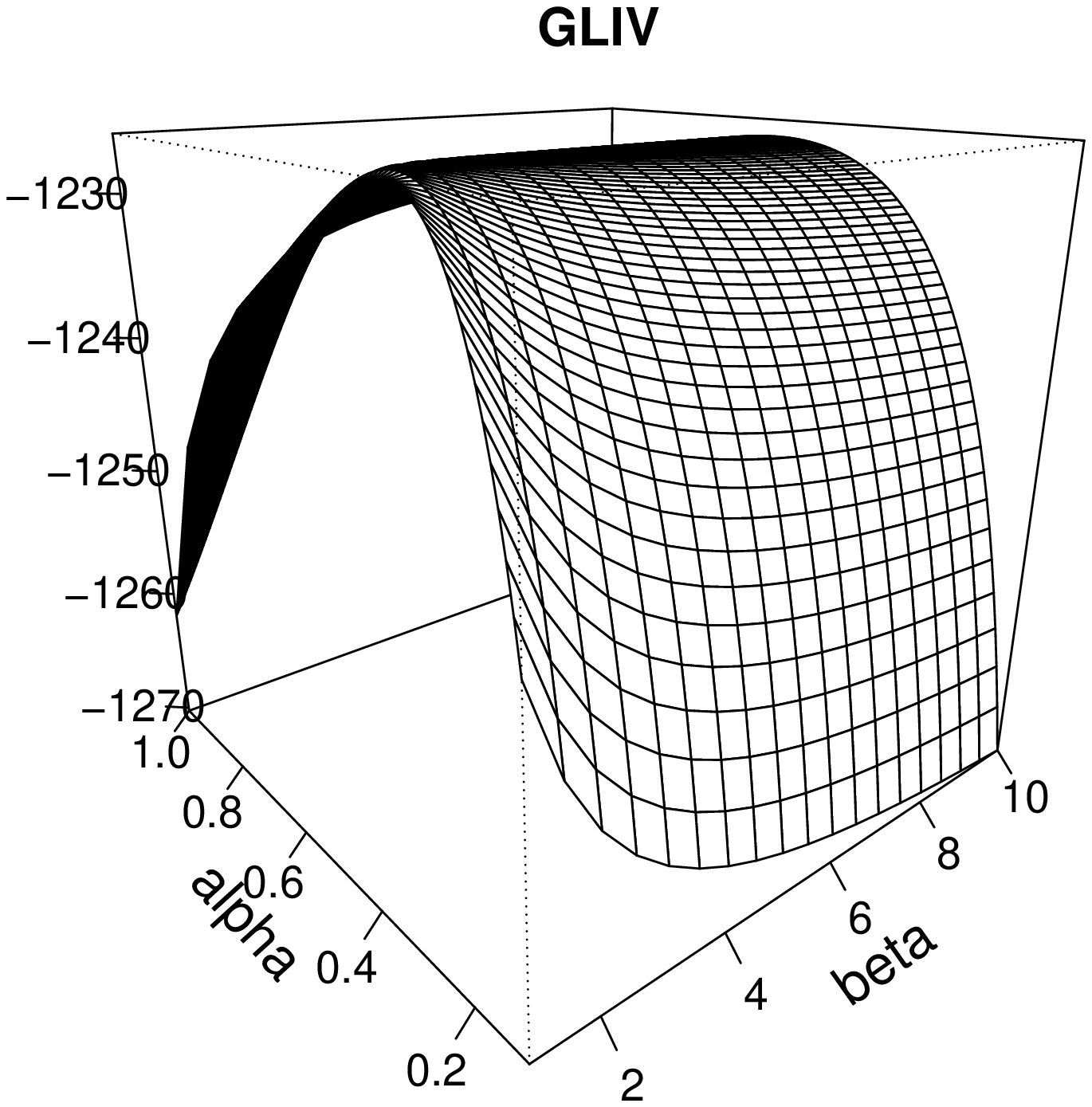}
}
\subfigure
{
\includegraphics[height=38mm,width=38mm]{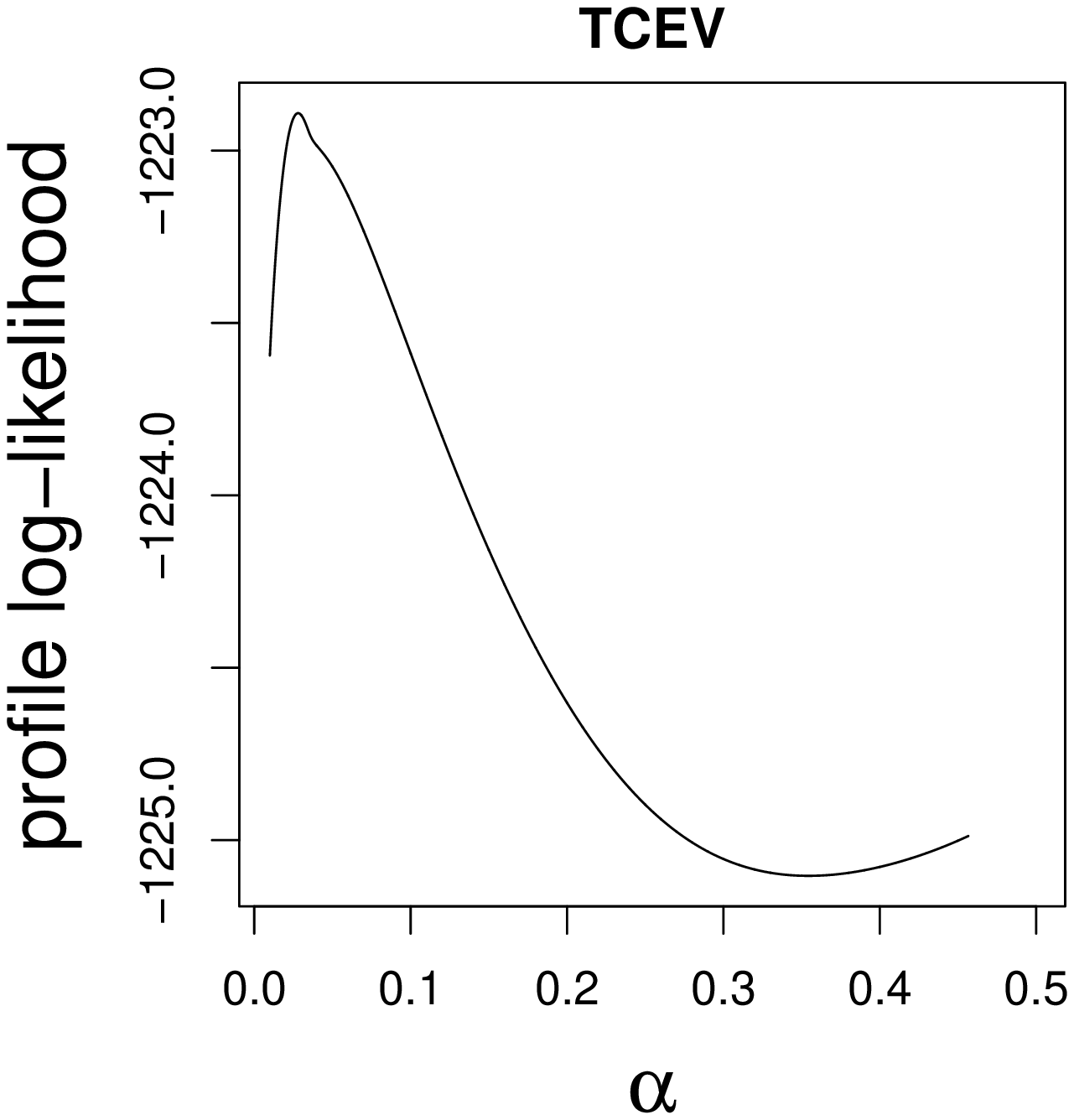}
}
%\vspace{-0.6cm}
\caption{Profile log-likelihood function - wind speed data.}
\label{fig:logvero-perfilada-NOAA-WestPalmBeach-vento-max-mensal-5segundos}
\end{figure}

Figure~\ref{fig:logvero-perfilada-NOAA-WestPalmBeach-vento-max-mensal-5segundos} displays the profile log-likelihood function for the additional parameter(s) of the fitted models. 
The profile log-likelihood function is well behaved if there is no inflection point, multimodality or lack of concavity. 
Note the slight concavity in the profile log-likelihood function for the EGu, EGa, and GGu models.
%, which casts doubt on the approximate standard error for the MLE of $\alpha$.
The profile log-likelihood function for the TEV model exhibits an inflection point, a local minimum, and two local maxima, and hence maximization can converge to a local maximum depending on the initial value. 
The profile log-likelihood function for the  GTIEV3 model is increasing and flat for large values of $\alpha$.
Hence, the estimate of $\alpha$ depends on the numerical precision specified for the maximization algorithm,  and the standard error estimate is very large. 
The profile log-likelihood function for the GLIV model also varies slowly in the $\beta$ parameter direction.
%, but its $\beta$ estimate is not as large as for the GTIEV3 model. Therefore, there is no difficulty with standard error estimation despite its large size due to lack of concavity of the curve.
The profile log-likelihood function for the TCEV model presents inflection points. This appears not to disturb the parameter estimation because of the highly concave profile log-likelihood function near its maximum.
The profile log-likelihood function for the GEV model is well behaved.
% produced by using estimação-perfilada-generalizações-dados.ox and gráficosParâmetrosLogveroGeneralizaçõesVE-NOAA-ventowest.R

%\textcolor{red}{incluir Allamano(2011)-return period ration (statistical measure of the error associated to the adoption of a homogeneous model) ???? }
%To quantify the error associated to the selection of the seasonally adjusted distribution, we consider the return level ratio between the return levels of the original data and the seasonally adjusted data. It provides an indication of the magnitude of the underestimation (overestimation) in design values related to neglecting seasonality. In particular, if the ratio is greater than 1, the adoption of the seasonally adjusted distribution corresponds to an underestimation of the real design value \cite{ALLAMANO}. 
%The .999 return levels for the original data for EV, GEV-MLE, GEV-PWM, EGu, TEV, GTIEV3, EGa, GLIV, TCEV fits are, respectively,
%??????. Hence, the return period ratio are, respectively, ???????

A summary of the goodness of fit measures is given in Table~\ref{tab:bondadeventoWestPalmBeach}.
%Unlike the other distributions, the GTIEV3 distribution does not exhibit better fit relative to the Gumbel distribution according 
%to the AIC criterion.
%The TEV fit produces the smallest AIC (2491.18), followed by the GEV-MLE (2491.96), while the GTIEV3 distribution exhibits the largest AIC.
%The lowest ADR value is achieved by the TCEV distribution (0.31), followed by the GLIV fit (0.35) and TEV fit (0.37). 
%The Gumbel and GTIEV3 distributions produce the greatest ADR (0.58).
The GEV-MLE, GEV-PWM, TEV, GLIV, and TCEV fits produce smaller AIC and ADR measures than the EV fit.
The AD2R measure highlights the EV and the GTIEV3 fits (235.75 and 235.82, respectively) as the worst fits. 
It points to the TCEV (2.33) and the GEV-PWM (7.13) as the best fits.
All the goodness of fit measures reveal that the Gumbel distribution is not the best choice for this data set.
Taking all of the goodness of fit criteria into account, we conclude that the GEV and TCEV distributions produce the best fits.

% produced by using estimação-generalizações-dados.ox and NOAA-WestPalmBeach-vento-max-mensal.R
%\begin{table}[!ht]
%\setlength{\tabcolsep}{4.5pt}
%\centering
%\caption{Goodness of fit measures - wind speed data.}\label{tab:bondadeventoWestPalmBeach}
%\footnotesize
%\begin{tabular}
%{l c c c c c c c c c}
%\hline \hline
	%&	EV	&	GEV	&	GEV & EGu	&	TEV	&	GTIEV3	&	EGa	&	GLIV	&	TCEV	\\
	%&		&	MLE	&	PWM & &		&	&		&		&		\\
%\hline
%$\ell(\widehat \theta)$ & $-$1261.02 & $-$1258.40 & $-$1258.51 & $-$1259.47 & $-$1259.43 & $-$1261.02 & $-$1259.40 & $-$1259.21 & $-$1257.65 \\ 
%AIC & 2526.04 & 2522.80 & 2523.03 & 2524.94 & 2524.86 & 2528.04 & 2524.79 & 2526.43 & 2525.30 \\ 
%ADR & 0.56 & 0.42 & 0.41 & 0.49 & 0.43 & 0.56 & 0.48 & 0.44 & 0.34 \\ 
%AD2R & 117.35 & 6.54 & 4.70 & 31.69 & 42.02 & 117.38 & 30.01 & 25.07 & 2.84 \\ 
%\hline \hline
%\end{tabular}
%\end{table}
\begin{table}[!ht]
\setlength{\tabcolsep}{4.5pt}
\centering
\caption{Goodness of fit measures - wind speed data.}\label{tab:bondadeventoWestPalmBeach}
\footnotesize
\begin{tabular}
{l c c c c c c c c c c}
\hline \hline
	&	EV	&	GEV	&	GEV & EGu	&	TEV	&	GTIEV3	&	EGa	&	GGu & GLIV	&	TCEV	\\
	&		&	MLE	&	PWM & &		&	&		&		&		\\
\hline
$-\ell(\widehat \theta)$     & 1245.08 & 1242.98 & 1243.69 & 1244.34 & 1242.59 & 1245.08 & 1244.23 & 1245.20 & 1243.04 & 1241.24 \\ 
  AIC & 2494.15 & 2491.96 & 2493.39 & 2494.67 & 2491.18 & 2496.15 & 2494.47 & 2496.40 & 2494.08 & 2492.47 \\ 
  ADR & 0.58 & 0.46 & 0.39 & 0.55 & 0.37 & 0.58 & 0.54 & 0.65 & 0.35 & 0.31 \\ 
  AD2R & 235.75 & 15.84 & 7.13 & 91.88 & 70.75 & 235.82 & 84.84 & 205.59 & 56.41 & 2.33 \\ 
\hline \hline
\end{tabular}
\end{table}

\begin{figure}[!ht]
\centering
\subfigure
{
\includegraphics[height=38mm,width=38mm]{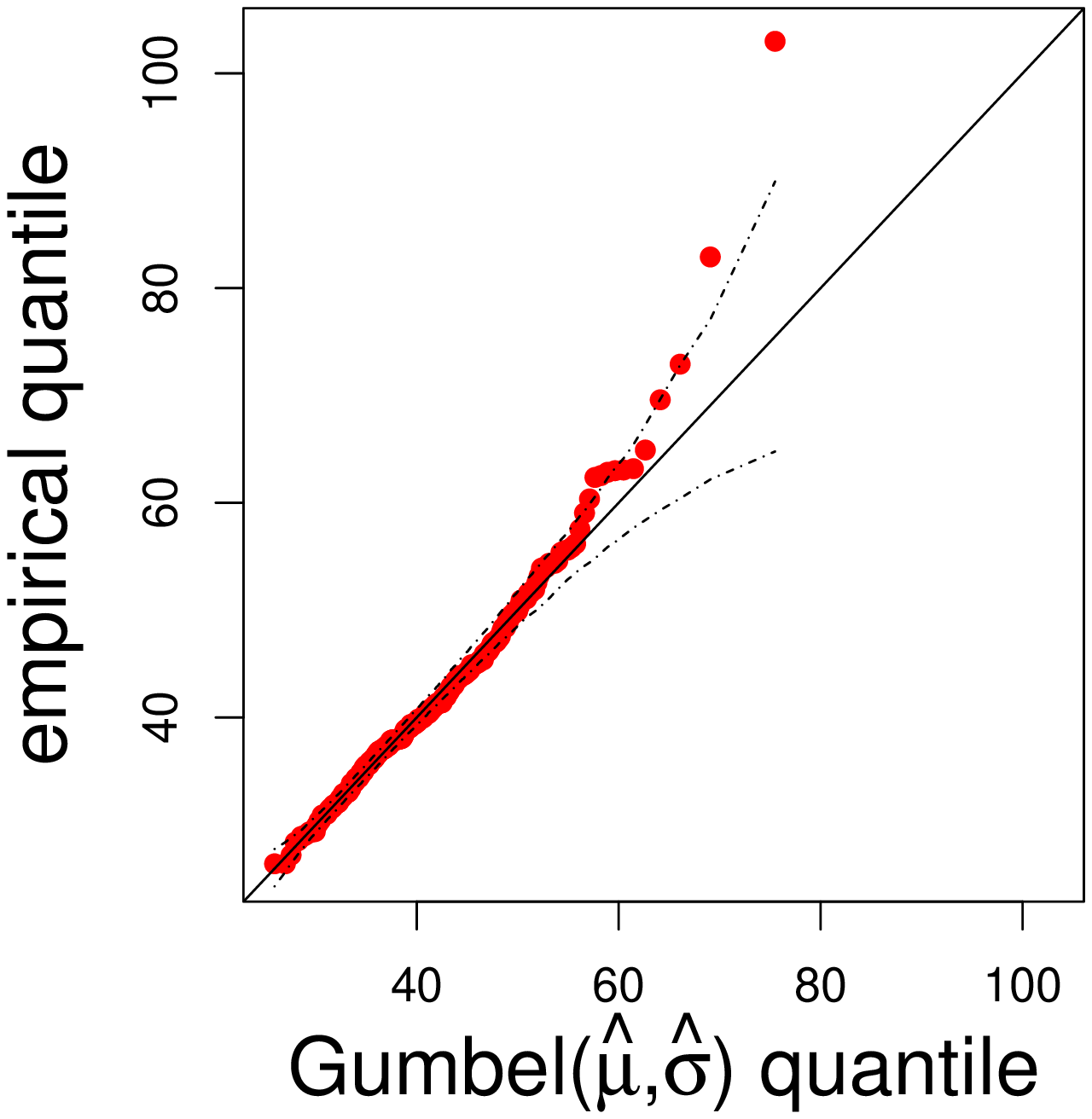}
}
\subfigure
{
\includegraphics[height=38mm,width=38mm]{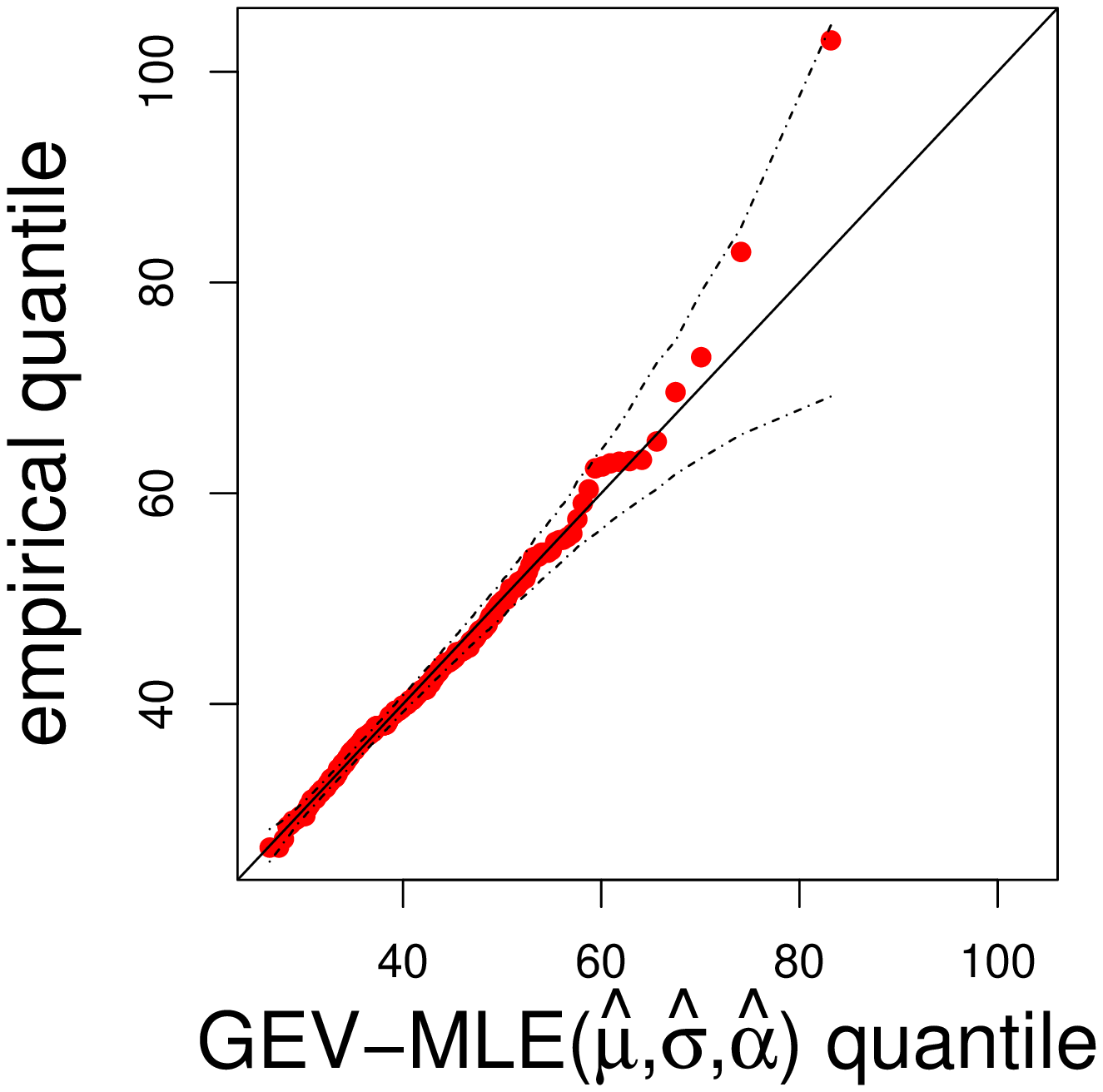}
}
\subfigure
{
\includegraphics[height=38mm,width=38mm]{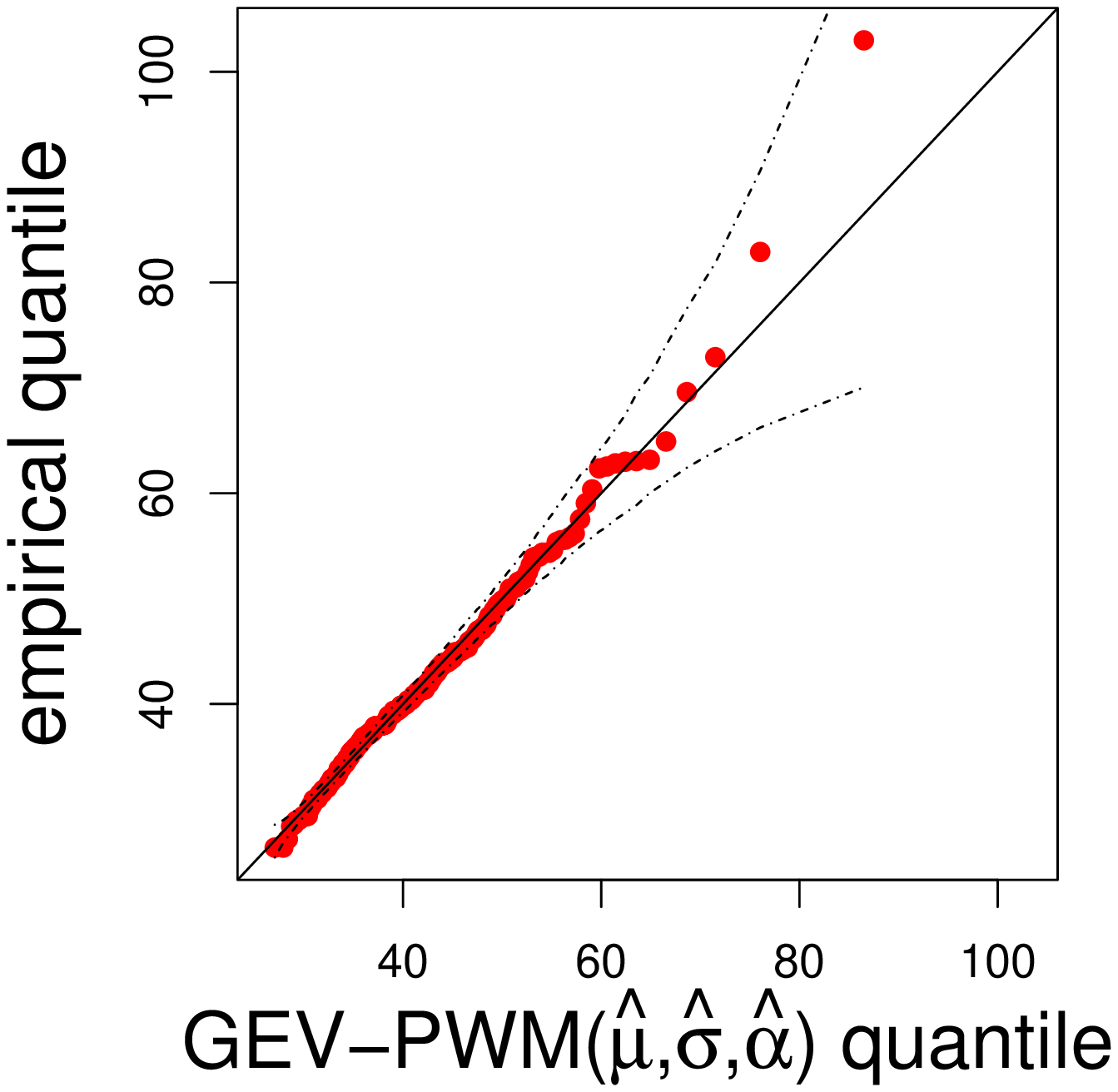}
}
\subfigure
{
\includegraphics[height=38mm,width=38mm]{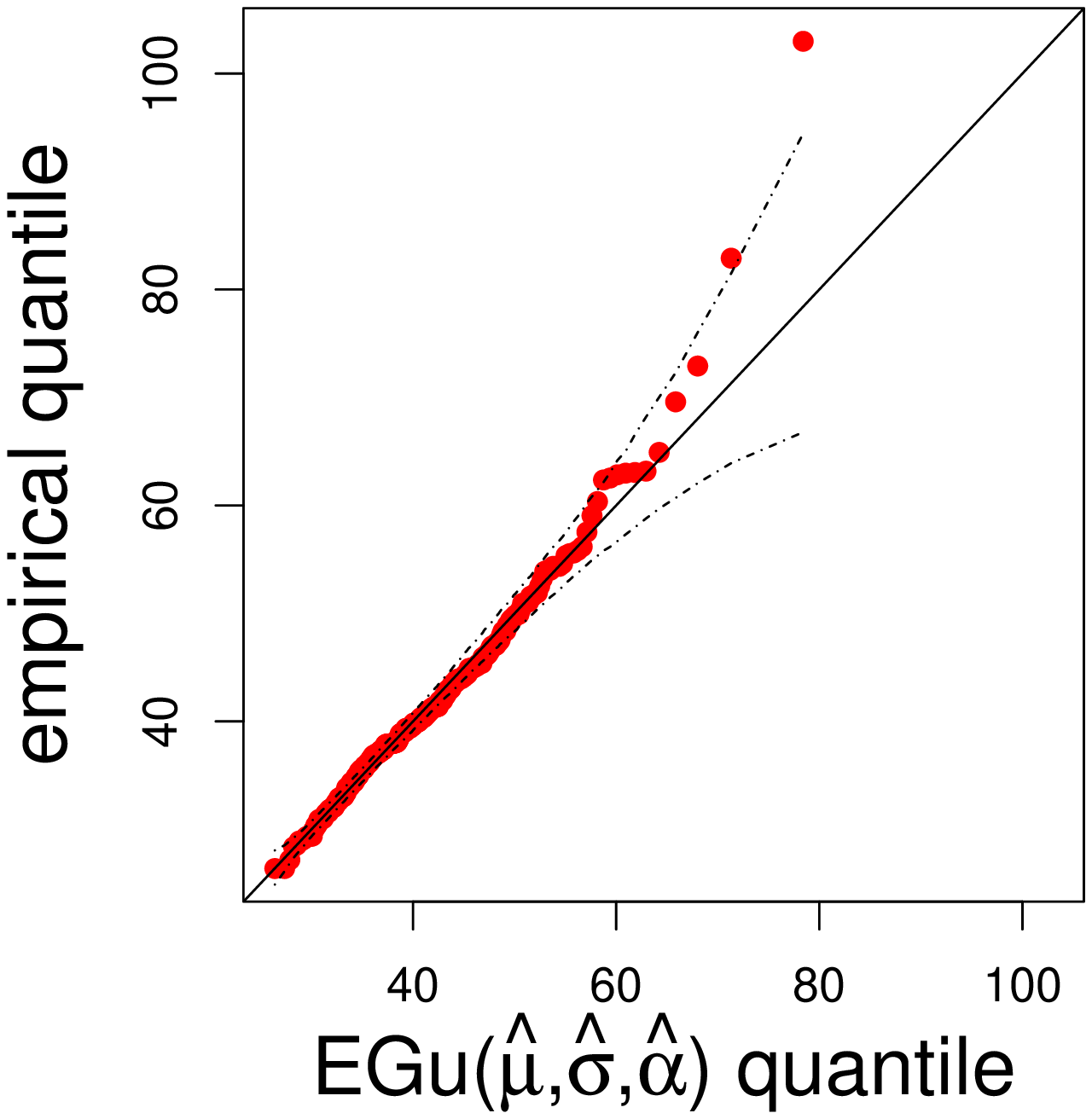}
}
\\
\centering
\subfigure
{
\includegraphics[height=38mm,width=38mm]{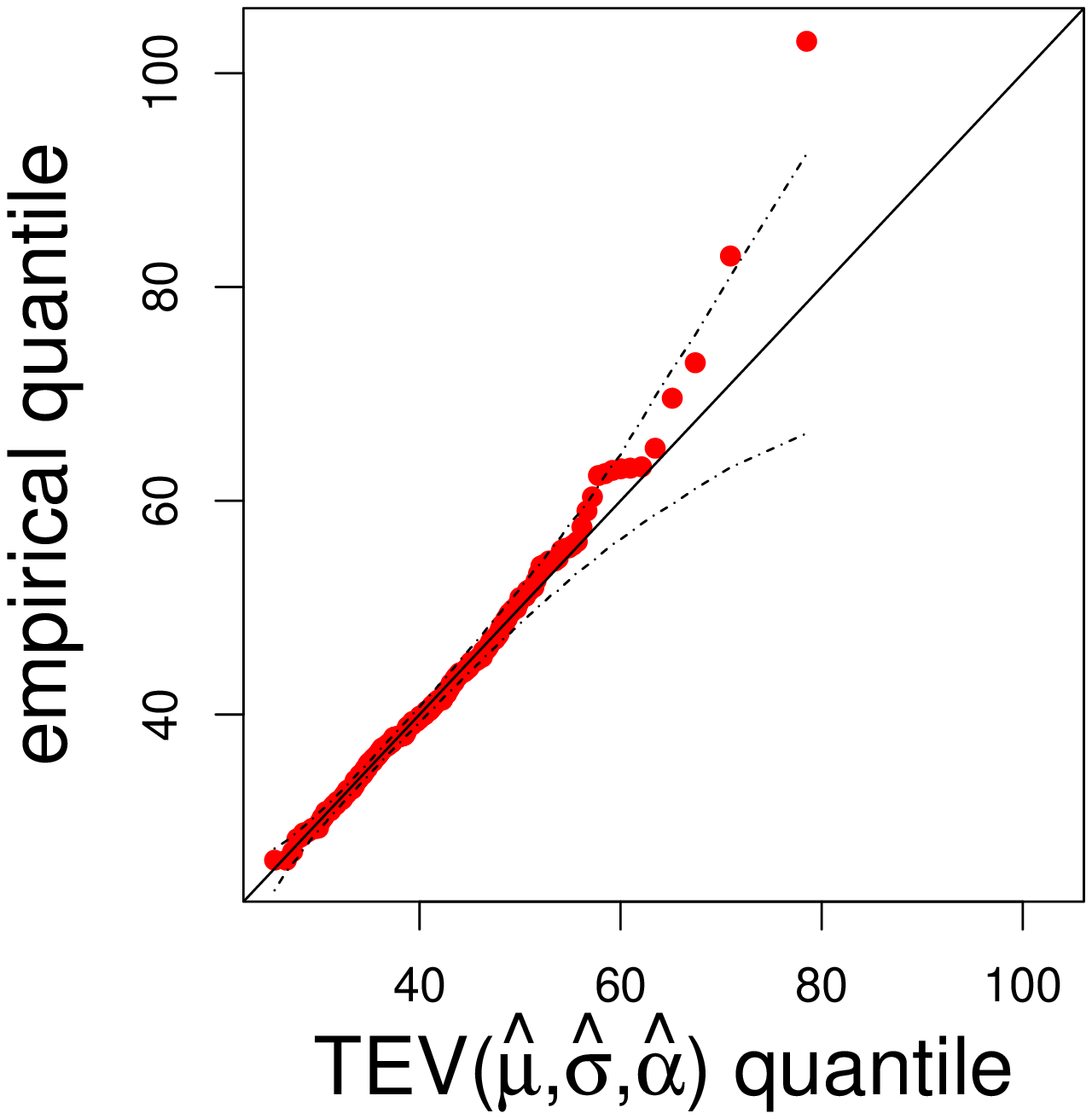}
}
\subfigure
{
\includegraphics[height=38mm,width=38mm]{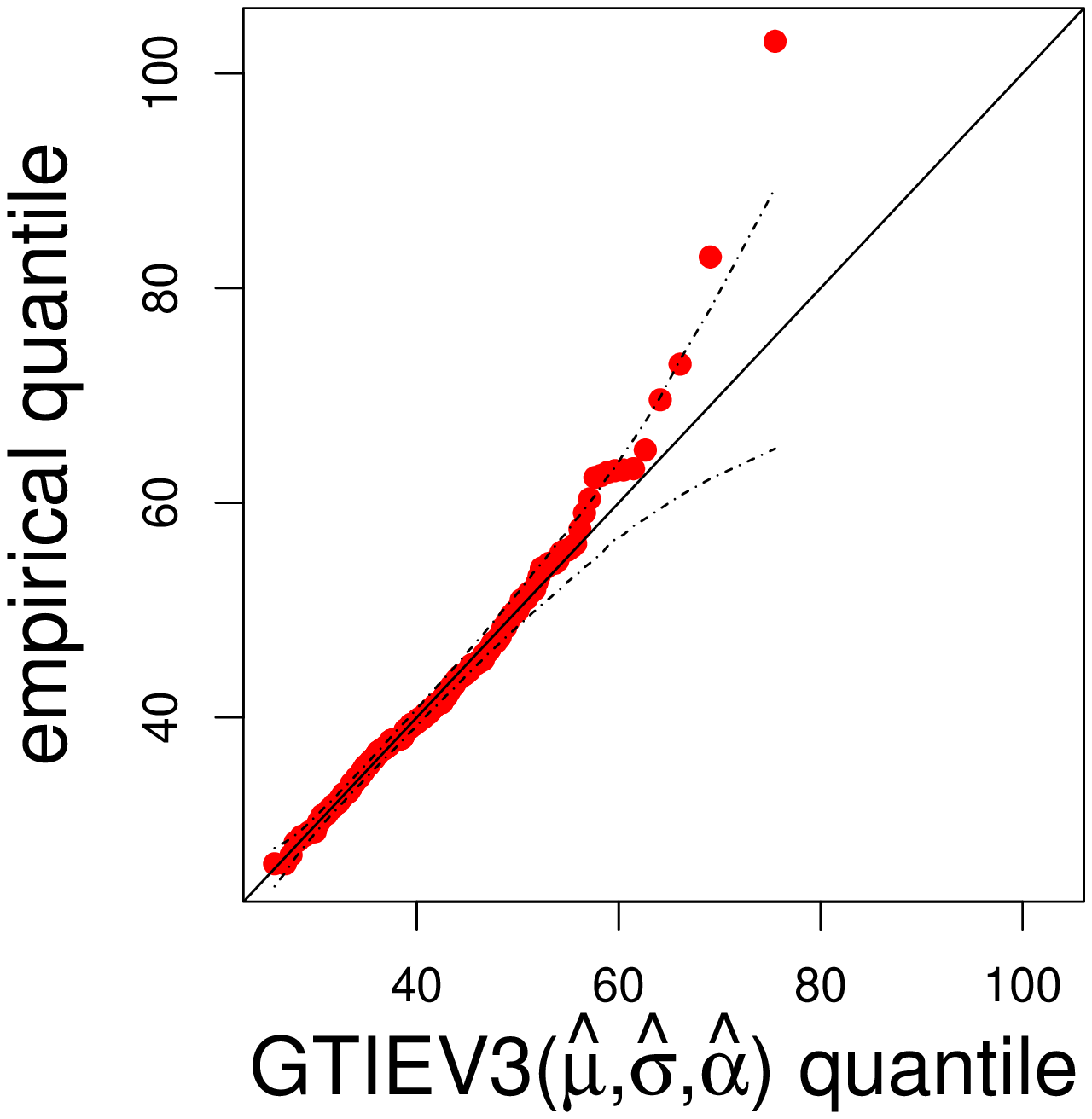}
}
\\
\centering
\subfigure
{
\includegraphics[height=38mm,width=38mm]{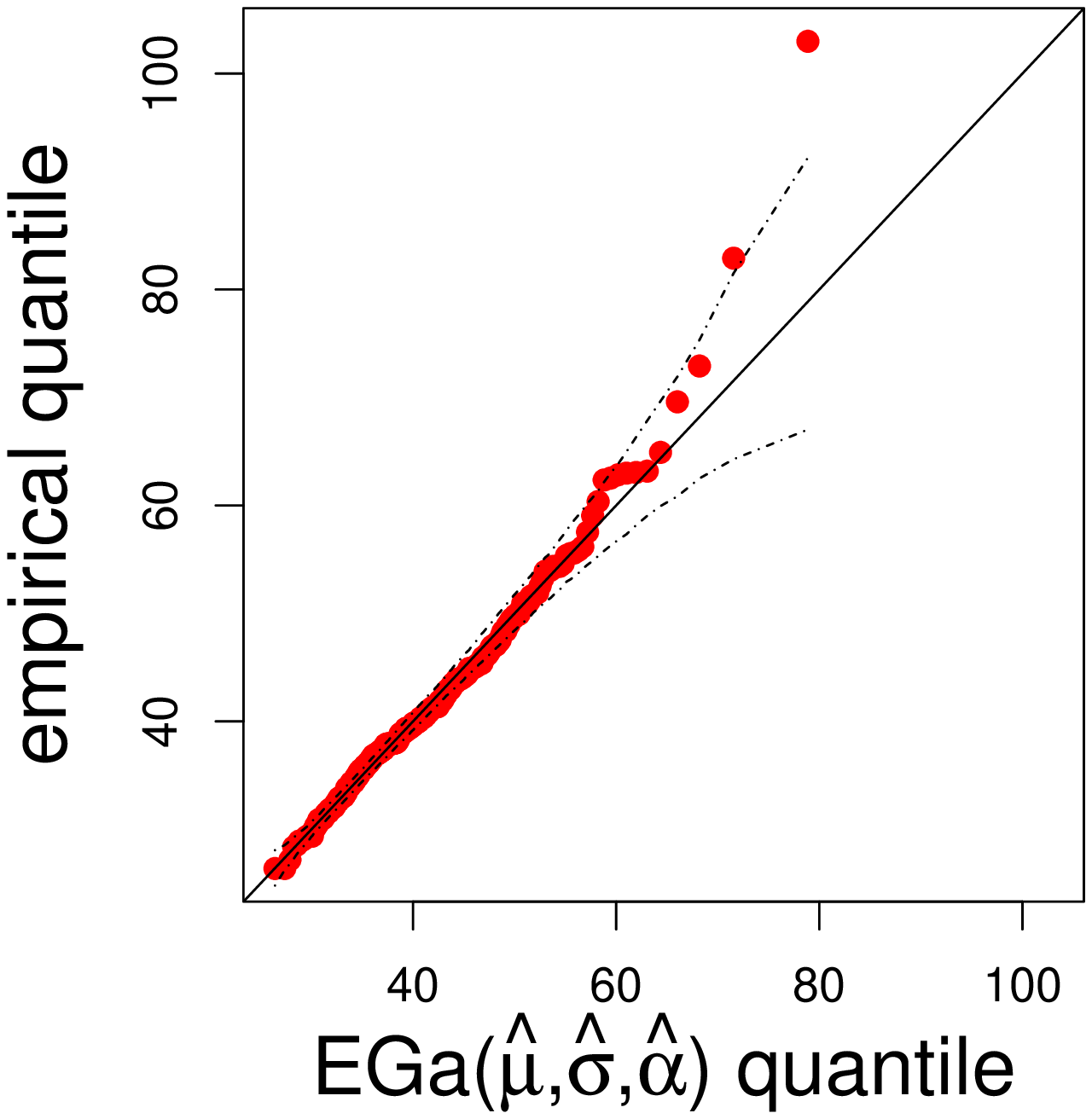}
}
\subfigure
{
\includegraphics[height=38mm,width=38mm]{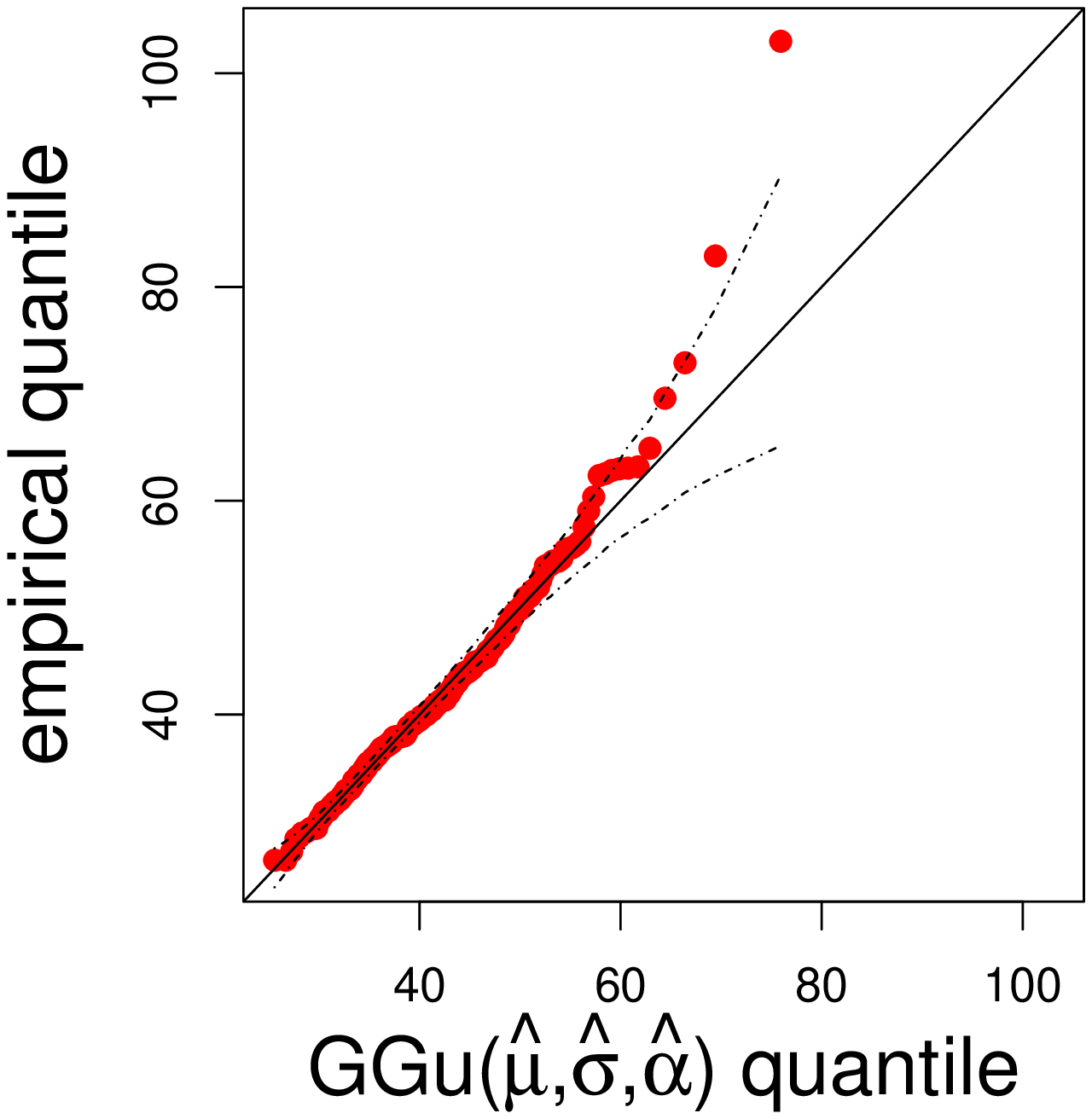}
}
\subfigure
{
\includegraphics[height=38mm,width=38mm]{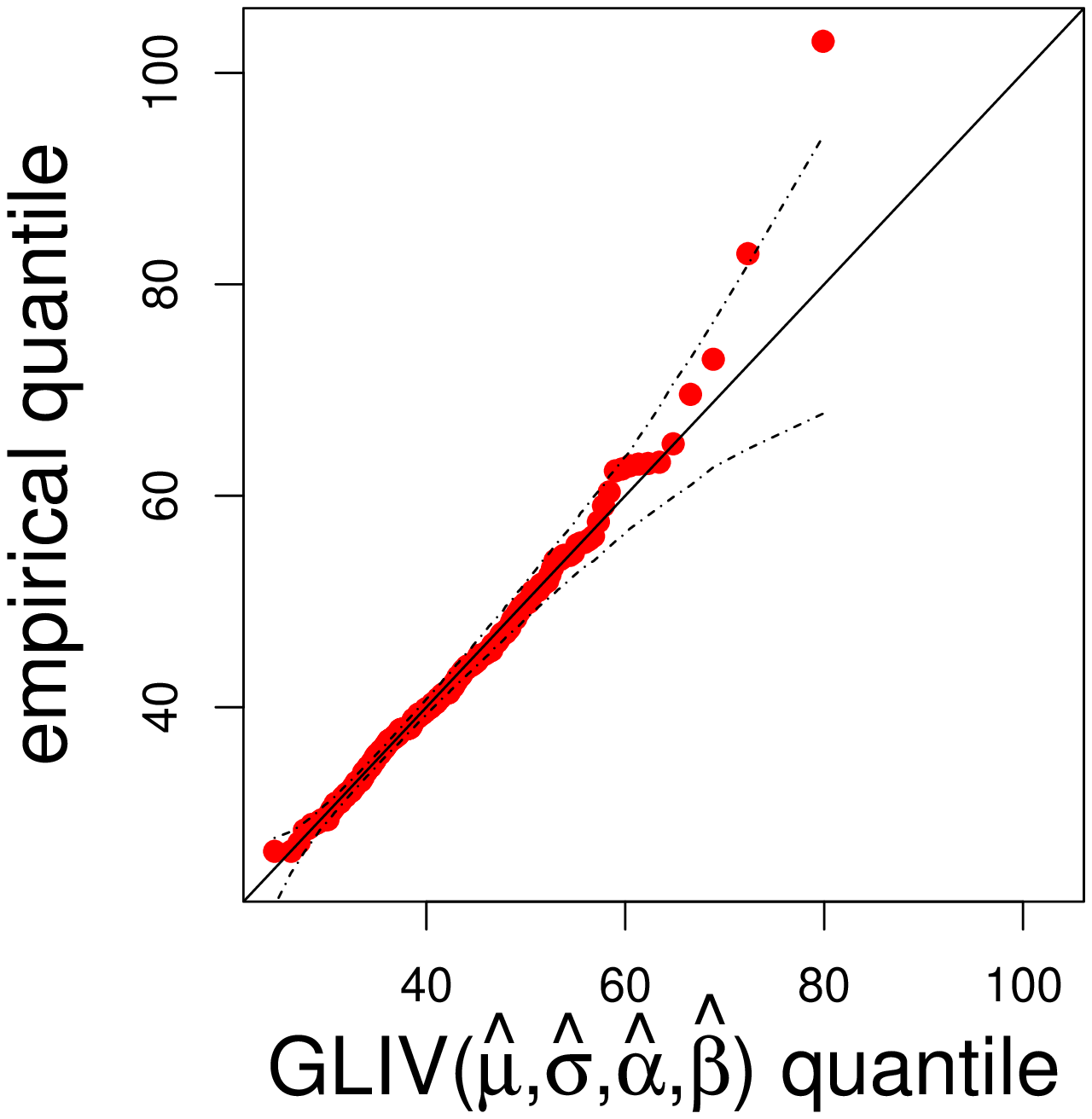}
}
\subfigure
{
\includegraphics[height=38mm,width=38mm]{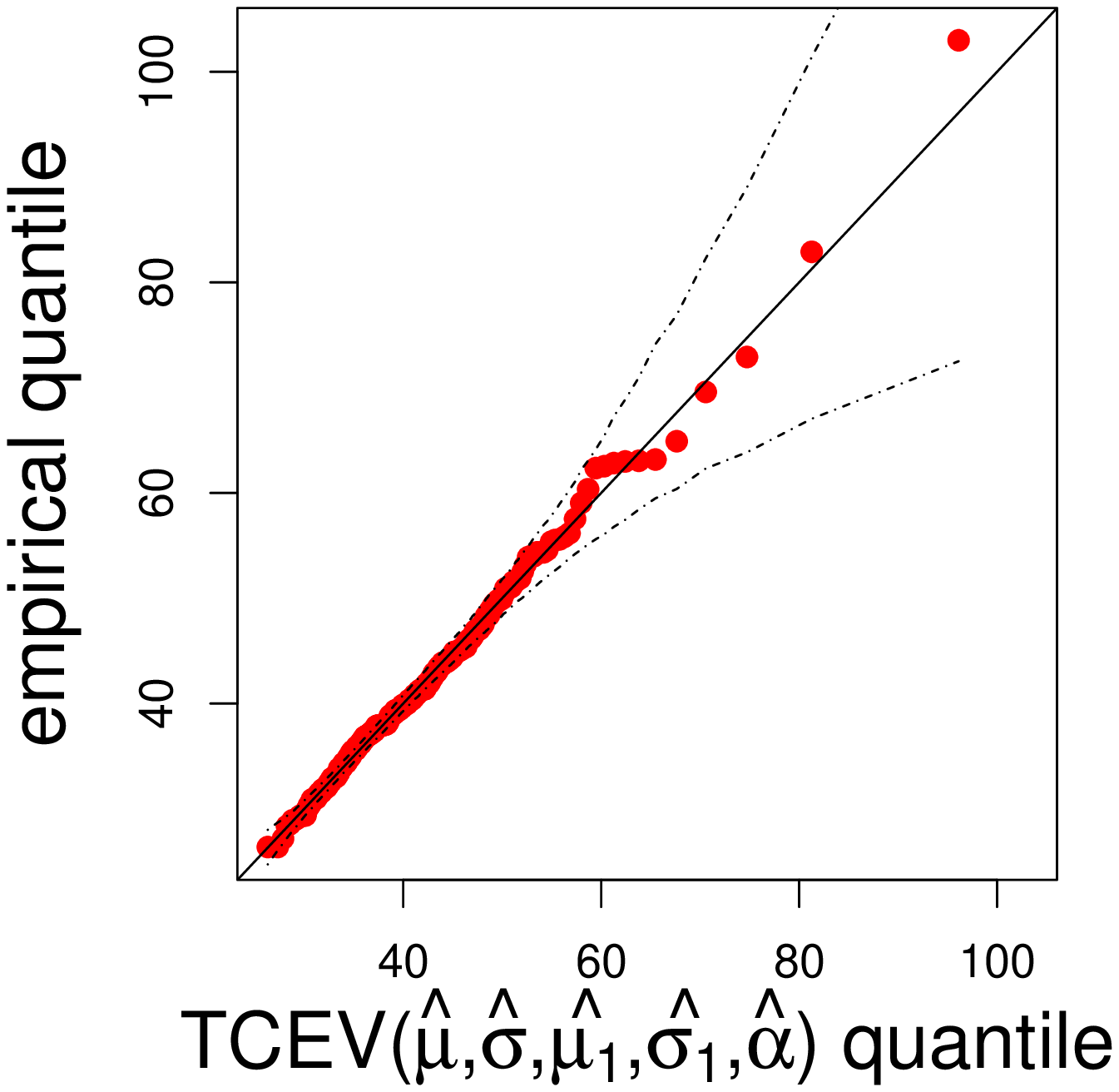}
}
%\vspace{-0.6cm}
\caption{QQplots - wind speed data.}
\label{fig:QQplot-NOAA-WestPalmBeach-vento-max-mensal-5segundos}
\end{figure}

Figure~\ref{fig:QQplot-NOAA-WestPalmBeach-vento-max-mensal-5segundos} shows the qqplots of the fitted models.
As an aid to interpretation, envelopes were generated by simulation.
The envelopes correspond to pointwise two-sided 90\% confidence intervals with the bootstrap replicates of each curve generated from the fitted model.
The qqplots for the Gumbel, EGu, TEV, GTIEV3, EGa, GGu, and GLIV distributions
clearly suggest a lack of fit at the extreme of the right tail.
However, the qqplots for the GEV (both estimation methods, MLE and PWM) and TCEV distributions accommodate all of the observations of the right tail inside the envelope.
Therefore, qqplots corroborate the previous conclusions that the GEV and TCEV models provide the best fits for this data set. 
% produced by using estimação-generalizações-dados.ox and NOAA-WestPalmBeach-vento-max-mensal.R
%\begin{figure}[!ht]
%\centering
%\subfigure
%{
%\includegraphics[height=38mm,width=38mm]{qqplotGu-NOAA-WestPalmBeach-vento-max-mensal-5segundos-1984-2014.pdf}
%}
%\subfigure
%{
%\includegraphics[height=38mm,width=38mm]{qqplotGEV_MLE-NOAA-WestPalmBeach-vento-max-mensal-5segundos-1984-2014.pdf}
%}
%\subfigure
%{
%\includegraphics[height=38mm,width=38mm]{qqplotGEV_PWM-NOAA-WestPalmBeach-vento-max-mensal-5segundos-1984-2014.pdf}
%}
%\\
%\centering
%\subfigure
%{
%\includegraphics[height=38mm,width=38mm]{qqplotEGu-NOAA-WestPalmBeach-vento-max-mensal-5segundos-1984-2014.pdf}
%}
%\subfigure
%{
%\includegraphics[height=38mm,width=38mm]{qqplotTEV-NOAA-WestPalmBeach-vento-max-mensal-5segundos-1984-2014.pdf}
%}
%\subfigure
%{
%\includegraphics[height=38mm,width=38mm]{qqplotGGu3-NOAA-WestPalmBeach-vento-max-mensal-5segundos-1984-2014.pdf}
%}
%\\
%\centering
%\subfigure
%{
%\includegraphics[height=38mm,width=38mm]{qqplotEGa-NOAA-WestPalmBeach-vento-max-mensal-5segundos-1984-2014.pdf}
%}
%\subfigure
%{
%\includegraphics[height=38mm,width=38mm]{qqplotGLIV-NOAA-WestPalmBeach-vento-max-mensal-5segundos-1984-2014.pdf}
%}
%\subfigure
%{
%\includegraphics[height=38mm,width=38mm]{qqplotTCEV-NOAA-WestPalmBeach-vento-max-mensal-5segundos-1984-2014.pdf}
%}
%\vspace{-0.6cm}
%\caption{QQplots - Monthly maximum wind speed data.}
%\label{fig:QQplot-NOAA-WestPalmBeach-vento-max-mensal-5segundos}
%\end{figure}

\section{Conclusion}
\label{conclusion}

Motivated by real problems with a probability of extreme events that is larger than usual, we investigated distributions that generalize the Gumbel extreme value distribution, frequently used to model extreme value phenomena.
We showed that some generalized Gumbel distributions proposed in the literature are nonidentifiable, which limits their usefulness in applications. 
We gathered the moments, quantiles, generating data methods, skewness and kurtosis coefficients, and classified their right-tail heaviness according to two criteria.
We provided a simulation study to evaluate the capacity of the selected distributions to fit data with kurtosis larger than that of the Gumbel distribution.
Our simulation results revealed that the generalized extreme value (GEV) distribution is more flexible in fitting this type of data and that the 
two-component extreme value (TCEV) distribution can also be a good choice. 
An application to an extreme wind speed data set in Florida confirmed the simulation study, with the GEV and TCEV models providing better fits than the other distributions.

As indicated by our simulations, practitioners should consider the GEV and TCEV distributions to model extreme value data with a heavy right tail.

\section*{Acknowledgments}
We gratefully acknowledge financial support from the Brazilian agencies CNPq, CAPES, and FAPESP.

\newpage

\Hide
\part{Supplement}{}{}{}{}{}
\titleclass{\part}{straight} % make part like a section (no newpage)
%\pagenumbering{gobble}% Remove page numbers (and reset to 1)

%\center{\Large{``A comparative review of generalizations of}} \\ \center{\Large{the Gumbel extreme value distribution}} \\ \center{\Large{with an application to wind speed data''}}

\setcounter{page}{1}
\setcounter{section}{0}
\setcounter{figure}{0}
\setcounter{table}{0}

\begin{center}
\LARGE\mdseries{A comparative review of generalizations of \\ the Gumbel extreme value distribution \\ with an application to wind speed data}
\end{center}

\section{Moments and quantiles}
\label{momentsquantiles}

The skewness ($\gamma_1$) and kurtosis ($\gamma_2$) coefficients of the distributions are obtained from the central moments (E($(X -{\rm E}(X))^n$)) or the moments (E($X^n$)), and the equations
$$
\gamma_1=\frac{{\rm E}((X-{\rm E}(X))^3)}{[{\rm E}((X-{\rm E}(X))^2)]^{3/2}}=\frac{{\rm E}(X^3)-3{\rm E}(X){\rm E}(X^2)+2{\rm E}(X)^3}{({\rm E}(X^2)-{\rm E}(X)^2)^{3/2}},
$$
$$
\gamma_2=\frac{{\rm E}((X-{\rm E}(X))^4)}{[{\rm E}((X-{\rm E}(X))^2)]^2}=\frac{{\rm E}(X^4)-4{\rm E}(X){\rm E}(X^3)+6{\rm E}(X)^2{\rm E}(X^2)+3{\rm E}(X)^4}{({\rm E}(X^2)-{\rm E}(X)^2)^2}.
$$
For the GEV distribution, we used that
$$
%\label{gradshteyn33814}
\int_{0}^{\infty} x^{\nu -1}\exp(-\mu x) dx = 1/\mu^\nu \Gamma(\nu),
$$ 
if Re $\mu>0$ and Re $\nu>0$ \cite[equation 3.381.4]{GRADSHTEYN}.

For the standard EGu distribution, the moment of order $n$ is 
$$
{\rm E}(X^n)=\int_0^\infty \alpha (-\ln(-\ln y))^n \; _2 F_1(1-\alpha,1,1,y)dy.
$$
For each value of $\alpha$, the $n$-th moment can be obtained by numerical integration with the computer algebra software Mathematica \citep{WOLFRAM} as follows:
\small\begin{verbatim}
Clear[a];Clear[EX1];Clear[EX2];Clear[EX3];Clear[EX4];
Clear[skewness];Clear[kurtosis];
EX1:=-a*NIntegrate[(Log[-Log[y]])^1*Hypergeometric2F1[(1-a),1,1,y],{y,0,1}]
EX2:=a*NIntegrate[(Log[-Log[y]])^2*Hypergeometric2F1[(1-a),1,1,y],{y,0,1}]
EX3:=-a*NIntegrate[(Log[-Log[y]])^3*Hypergeometric2F1[(1-a),1,1,y],{y,0,1}]
EX4:=a*NIntegrate[(Log[-Log[y]])^4*Hypergeometric2F1[(1-a),1,1,y],{y,0,1}]
a:=Table[i/100,{i,200}]
skewness=(EX3-3*EX2*EX1+2*EX1^3)/(EX2-EX1^2)^(3/2)//N
Clear[a];a := Table[i/100, {i, 300}]
kurtosis = (EX4 - 4*EX3*EX1 + 6*EX2*EX1^2 - 3*EX1^4)/(EX2 - EX1^2)^2//N
\end{verbatim}
\normalsize
For the TEV distribution, we used the moments in \citet[p. 1404]{ARYAL}.
For the EGa distribution, we used that
\begin{sloppypar}
$$
%\label{gradshteyn43585}
\int_{0}^{\infty} x^{\nu-1}\exp(-\mu x)(\ln x)^n dx = \partial ^n( \mu^{-\nu} \Gamma(\nu)) / \partial \nu^n,
$$
for $n=0,1,2,3,\ldots$ \cite[equation 4.358.5]{GRADSHTEYN}, where
$\Gamma(x)=\int_0^{\infty} \exp(-t) t^{x-1}dt$ and ${\rm Re} \; x>0$ is the Gamma function.
For the GLIV distribution, we used its moment generating function.
For the GGu distribution, the central moment of order $n$ is 
$$
{\rm E}((X - {\rm E}(X))^n)=\int_0^\infty \left(-\sigma \left(\ln\ln(1+y^{-1/\alpha}) + {\mathcal E}(1,\alpha) \right)\right)^n  (1+y)^{-2} dy,
$$
where ${\mathcal E}(n,\alpha)=-\int_0^\infty (\ln\ln(1+y^{-1/\alpha}))^n (1+y)^{-2} dy$ and ${\mathcal E}(1,1)={\mathcal E}$ is the Euler's constant.
For each value of $\alpha$, the skewness and kurtosis can be obtained by numerical integration with the computer algebra software Mathematica \citep{WOLFRAM} as follows:
\small\begin{verbatim}
Clear[alpha]; Clear[E1]; Clear[E2]; Clear[E3]; Clear[E4]
alpha = Table[i/10, {i, 7, 100}]
E1 = -NIntegrate[Log[Log[1 + x^(-1/alpha)]]/(1 + x)^2, {x, 0, Infinity}]
E2 = -NIntegrate[Log[Log[1 + x^(-1/alpha)]]^2/(1 + x)^2, {x, 0, Infinity}]
E3 = -NIntegrate[Log[Log[1 + x^(-1/alpha)]]^3/(1 + x)^2, {x, 0, Infinity}]
E4 = -NIntegrate[Log[Log[1 + x^(-1/alpha)]]^4/(1 + x)^2, {x, 0, Infinity}]
skewness = -(-E3 - 3 E2*E1 - 2 E1^3)/(-E2 - E1^2)^(3/2)
kurtosis = (-E4 - 4 E3*E1 - 6 E2*E1^2 - 3 E1^4)/(-E2 - E1^2)^2
\end{verbatim}
\normalsize
\end{sloppypar}

Table~\ref{tab:moments} presents moments, skewness and kurtosis coefficients and quantile functions for the Gumbel distribution (EV) and its generalizations.

\begin{sidewaystable}
\label{tab:moments}
\centering
\caption{Moments, skewness and kurtosis coefficients and quantile functions}
\begin{threeparttable}[c]
\footnotesize
\begin{tabular}
{l| c c c c}
\hline \hline
                         & E($X$)        & var($X$)  \\
\hline
EV$(\mu,\sigma)$	&	 $\mu+\sigma {\mathcal E}$ 	&	 $\sigma ^2 \, \pi^2/6$  	\\
GEV$(\mu,\sigma,\alpha), \, \alpha \neq 0$  &	 $\mu+(\sigma/\alpha)\left(\Gamma(1-\alpha)-1\right), \, \alpha<1$	&	 $(\sigma / \alpha )^2  \left(\Gamma(1-2\alpha)-\Gamma^2(1-\alpha)\right), \, 2\alpha<1$\\
EGu$(\mu,\sigma,\alpha)$ 	&	 $-$  	&	 $-$ 	\\
TEV$(\mu,\sigma,\alpha)$  	&	 $(\mu+ {\mathcal E} \sigma)-\alpha \sigma \ln 2$ 	&	 $\sigma^2( \pi^2/6 -\alpha (1+\alpha) (\ln 2)^2)$ 	\\
GTIEV3$(\mu,\sigma,\alpha)$ & $\mu-\sigma(-\mathcal{E}+\ln(\alpha)-\psi(\alpha))$ \tnote{I} & $\sigma^2(\pi^2 /6 +\psi^\prime(\alpha))$ \tnote{II} \\
EGa$(\mu,\sigma,\alpha)$  	&	 $\mu-\sigma\psi(\alpha)$ 	&	 $\sigma^2 \psi^\prime(\alpha)$	\\
GGu$(\mu,\sigma,\alpha)$  & $\mu+\sigma {\mathcal E}(1,\alpha)$ & $\sigma^2(-{\mathcal E}(2,\alpha)-{\mathcal E}^2(1,\alpha))$ \\
GLIV$(\mu,\sigma,\alpha,\beta)$  	&	 $\mu + \sigma (\psi(\beta) - \psi(\alpha)-\ln(\beta/\alpha) )$ 	&	 $\sigma^2 (\psi^\prime(\beta) + \psi^\prime(\alpha))$ 	\\
TCEV$(\mu,\sigma,\mu_1,\sigma_1,\alpha)$	&	 $(1-\alpha)(\mu+\sigma {\mathcal E}) + \alpha(\mu_1+\sigma_1 {\mathcal E})$ 	&	 $(1-\alpha) \sigma^2 \pi^2/6 +\alpha\sigma_1 ^2 \pi^2/6 +$ 	\\
	&	 &	 $\alpha(1-\alpha)(\mu+\sigma {\mathcal E} - \mu_1-\sigma_1 {\mathcal E})^2$ 	\\
\hline
	&	Quantile  $x_p$ 	&	E($X^n$)	\\
\hline
%EV$(\mu,\sigma)$	&	 $\mu-\sigma \ln(- \ln (p))$ 	&		$\sum_{i=0}^n { n \choose i }\mu^{n-i} \sigma^i (-1)^i \Gamma^{(i)}(1)$  \tnote{III}\\
EV$(\mu,\sigma)$	&	 $\mu-\sigma \ln(- \ln (p))$ 	&		$\sum_{i=0}^n \genfrac{(}{)}{0pt}{}{n}{i} \mu^{n-i} \sigma^i (-1)^i \Gamma^{(i)}(1)$  \tnote{III}\\
GEV$(\mu,\sigma,\alpha), \, \alpha \neq 0$ &	 $\mu+(\sigma/\alpha)((-\ln(p))^{-\alpha} - 1 )$ 	&	$\left( \frac{1}{\alpha}\right)^n\sum_{i=0}^n \genfrac{(}{)}{0pt}{}{n}{i} \left( \mu\alpha - \sigma \right)^{n-i} \sigma^i \Gamma\bigl(1 - \alpha i\bigr)$	\\
EGu$(\mu,\sigma,\alpha)$ 	&	 $\mu-\sigma\ln(-\ln(1-(1-p)^{1/\alpha}))$ &	$\alpha  \sum_{i=0}^n \genfrac{(}{)}{0pt}{}{n}{i} \mu^{n-i} \sigma^i (-1)^i \int_{0}^{1}  \left(\ln(- \ln y) \right)^i \; _2 F_1(1-\alpha,1,1,y) dy$	\\
TEV$(\mu,\sigma,\alpha)$ &	 $\mu-\sigma\ln\left(-\ln\left((1+\alpha - \sqrt{(1+\alpha)^2-4\alpha p})/2\alpha\right)\right)$ & $\sum_{i=0}^n \genfrac{(}{)}{0pt}{}{n}{i} \mu^{n-i}  \sigma^i (-1)^i \Biggl( (1+\alpha)\Gamma^{(i)}(1) -2\alpha\frac{\partial^i}{\partial \nu ^i}\biggl[2^{-\nu}\Gamma(\nu)\biggr]\biggl|_{\nu=1}\Biggr)
$\\	
GTIEV3$(\mu,\sigma,\alpha)$ & $\mu-\sigma\ln( \alpha (p^{-1/\alpha} -1) )$ & $\sum_{i=0}^n \genfrac{(}{)}{0pt}{}{n}{i}\mu^{n-i} \sigma^i (-1)^i \int_{0}^{1}  (\ln y)^i \bigl( 1+ (1/\alpha)y\bigr)^{-\alpha-1} dy$ \\
EGa$(\mu,\sigma,\alpha)$  	&	 $-$	&	$\frac{1}{\Gamma (\alpha)}\sum_{i=0}^n \genfrac{(}{)}{0pt}{}{n}{i} \mu^{n-i} \sigma^i (-1)^i \Gamma^{(i)}(\alpha)$	\\
GGu$(\mu,\sigma,\alpha)$  &  $\mu-\sigma\ln\ln(1+ (p/(1-p))^{-1/\alpha})$ & $-\sum_{i=0}^n \genfrac{(}{)}{0pt}{}{n}{i}  \mu^{n-i} \sigma^i (-1)^i {\mathcal E}(i,\alpha) $\\
GLIV$(\mu,\sigma,\alpha,\beta)$  	&	 $-$	&	$\frac{\partial^n}{\partial t^n} \left[ \left( \frac{\beta}{\alpha}\right)^{-t} 
\frac{\Gamma(\alpha-t) \Gamma(\beta+t)}{\Gamma(\alpha)\Gamma(\beta)} \right] \biggl|_{t=0} $	\\
%TCEV$(\mu_1,\sigma_1,\mu_2,\sigma_2,\alpha)$	&	 $-$	&	$(1-\alpha) (\sum_{i=0}^n { n \choose i }\mu_1^{n-i} \sigma_1^i (-1)^i \Gamma^{(i)}(1)) + $	\\
%&	 	&	$\alpha (\sum_{i=0}^n{ n \choose i }\mu_2^{n-i} \sigma_2^i (-1)^i \Gamma^{(i)}(1))$	\\
TCEV$(\mu,\sigma,\mu_1,\sigma_1,\alpha)$	&	 $-$	&	$\sum_{i=0}^n \genfrac{(}{)}{0pt}{}{n}{i} (-1)^i \Gamma^{(i)}(1) \biggl( (1-\alpha) \mu^{n-i} \sigma^i + \alpha \mu_1^{n-i} \sigma_1^i  \biggr) 
$	\\
\hline
	&	Skewness  $\gamma_1$	&	Kurtosis $\gamma_2$	\\
\hline
EV$(\mu,\sigma)$	&	$12\sqrt{6}\zeta(3)/\pi^3 \cong 1.139547$	&	5.4	\\
GEV$(\mu,\sigma,\alpha), \, \alpha \neq 0$ &	$\pm\frac{  \Gamma(1-3\alpha)-3\Gamma(1-2\alpha)\Gamma(1-\alpha) +2\Gamma^3(1-\alpha) }{\left( \Gamma(1-2\alpha)-\Gamma^2(1-\alpha)\right)^{3/2}}$	&	$\frac{\Gamma(1-4\alpha)-4\Gamma(1-3\alpha)\Gamma(1-\alpha)+6 \Gamma(1-2\alpha)\Gamma^2(1-\alpha)-3\Gamma^4(1-\alpha)   }{\left(\Gamma(1-2\alpha)-\Gamma^2(1-\alpha)\right)^2}$	\\
EGu$(\mu,\sigma,\alpha)$ 	&	$-$	&	$-$	\\
TEV$(\mu,\sigma,\alpha)$  	&	$\gamma_{1,EV}\frac{1-((\ln 2)^3/2\zeta(3))\alpha(1+\alpha)(1+2\alpha)}{(1-6(\ln 2/\pi)^2\alpha(1+\alpha))^{3/2}}$	&	$(\gamma_{2,EV}-3)\frac{1 - 15 (\ln 2 /\pi)^4 \alpha(1+\alpha) (1 + 6\alpha(1+\alpha)) }{(1-6(\ln 2 / \pi)^2\alpha(1+\alpha))^2}$	\\
GTIEV3$(\mu,\sigma,\alpha)$  & $(\psi^{\prime\prime}(\alpha)-\psi^{\prime\prime}(1)(\,/\,(\psi^\prime(\alpha)+\psi^\prime(1))^{3/2}$ & $(\psi^{\prime\prime\prime}(\alpha)+\psi^{\prime\prime\prime}(1)(\,/\,(\psi^\prime(\alpha) +\psi^\prime(1))^2 \,+\, 3$ \\
EGa$(\mu,\sigma,\alpha)$  	&	$-\psi^{\prime\prime}(\alpha)\,/\, \psi^\prime(\alpha)^{3/2}$	&	$\psi^{\prime\prime\prime}(\alpha) \,/\, \psi^\prime(\alpha)^2 \,+\, 3$	\\
GGu$(\mu,\sigma,\alpha)$  & $\frac{-{\mathcal E}(3,\alpha)-3{\mathcal E}(2,\alpha){\mathcal E}(1,\alpha)-2{\mathcal E}^3(1,\alpha)}{-{\mathcal E}(2,\alpha)-{\mathcal E}^2(1,\alpha)}$ &  $\frac{-{\mathcal E}(4,\alpha)-4{\mathcal E}(3,\alpha){\mathcal E}(1,\alpha)-6{\mathcal E}(2,\alpha){\mathcal E}^2(1,\alpha)-3{\mathcal E}^4(1,\alpha)}{-{\mathcal E}(2,\alpha)-{\mathcal E}^2(1,\alpha)}$ \\
GLIV$(\mu,\sigma,\alpha,\beta)$  	&	$(\psi^{\prime\prime}(\beta) - \psi^{\prime\prime}(\alpha))\,/\,(\psi^\prime(\beta) + \psi^\prime(\alpha))^{3/2}$	&	$(\psi^{\prime\prime\prime}(\beta)+\psi^{\prime\prime\prime}(\alpha))\,/\,(\psi^\prime(\beta) + \psi^\prime(\alpha))^{2} \,+\, 3$	\\
 TCEV$(\mu,\sigma,\mu_1,\sigma_1,\alpha)$	&	 $-$	&	 $-$	\\
\hline \hline
\end{tabular}
\begin{tablenotes}
%[para,flushleft]
%\centering
\raggedright
\item[I] $\psi(x)= d(\ln \Gamma(x)) / dx=\Gamma^{(1)}(x) / \Gamma(x)$ is the digamma function.
\item[II] $\psi^{(n)}(x)$ is the $n$-th derivative of the digamma function (the $n$-th polygamma function); $\psi^{(1)}(x)=\psi^\prime(x)$.
\item[III] $\Gamma^{(n)}(x)=\int_0^{\infty} \exp(-t) t^{x-1} (\ln t)^n dt$ and ${\rm Re} \; x>0$ is the $n$-th derivative of the Gamma function.
\item[IV] $\zeta(s)=\sum_{k=1}^\infty k^{-s}$, ${\rm Re} \, s >1$ is the Riemann zeta function and $\zeta(3) \approx 1,2$.
\end{tablenotes}
\end{threeparttable}
%\end{table}
%\end{landscape}
\end{sidewaystable}
\normalsize

\section{Right tail heaviness}
\label{tailheaviness}

\paragraph{Regular variation theory criterion.}

To obtain the index of regular variation, we use that
\begin{equation}\label{expalpha}
\lim_{t \rightarrow \infty} 
\frac{\exp\left( -\left[ 1+\alpha \left( \frac{tx-\mu}{\sigma}  \right) \right]^{-1/\alpha} \right) }
{\exp \left( -\left[ 1+\alpha \left( \frac{t-\mu}{\sigma}  \right) \right]^{-1/\alpha} \right) }=1, \quad {\rm if} \quad \alpha>0,
\end{equation}
\begin{equation}\label{expexp}
\lim_{t \rightarrow \infty} 
\frac{\exp(-\exp(-(tx-\mu)/\sigma))}{ \exp(-\exp(-(t-\mu)/\sigma))}=1 ,
\end{equation}
and
\begin{equation}\label{exp}
\lim_{t \rightarrow \infty} \frac
{\exp(-(tx-\mu)/\sigma)}
{\exp(-(t-\mu)/\sigma)}=
\left\{
\begin{matrix}
\infty, & {\rm if} & 0<x<1 \\
1, & {\rm if} & x=1 \\
0, & {\rm if} & x>1.
\end{matrix}
\right. 
\end{equation}

Let $X \sim {\rm GEV}(\mu,\sigma,\alpha)$ with $\alpha>0$.
From (\ref{expalpha}), we have
\begin{eqnarray*}
\lim_{t \rightarrow \infty} \frac{\overline{F}_{{\rm GEV}}(tx) }{ \overline{F}_{{\rm GEV}}(t)}
&=& \lim_{t \rightarrow \infty} \frac{1-F(tx) }{ 1-F(t)}=
\lim_{t \rightarrow \infty} \frac{xf(tx) }{ f(t)} \\
%= x \lim_{t \to \infty} \frac{f(tx) }{ f(t)} \\
&=&
x\lim_{t \rightarrow \infty} 
\frac{\frac{1}{\sigma}\exp \left( -\left[ 1+\alpha \left( \frac{tx-\mu}{\sigma}  \right) \right]^{-1/\alpha} \right)  \left[ 1+\alpha \left( \frac{tx-\mu}{\sigma}  \right) \right]^{-1/\alpha-1} }
{\frac{1}{\sigma}\exp \left( -\left[ 1+\alpha \left( \frac{t-\mu}{\sigma}  \right) \right]^{-1/\alpha} \right)   \left[ 1+\alpha \left( \frac{t-\mu}{\sigma}  \right) \right]^{-1/\alpha-1} } \\
&=& x^{-1/\alpha},
\end{eqnarray*}
%\noindent
i.e., the generalized extreme value distribution with $\alpha>0$ is regularly varying at infinity with index $-1/\alpha$ and tail index $\alpha$.
If $\alpha = 0$, from (\ref{expexp}) and (\ref{exp}), we have
\begin{eqnarray*}
\lim_{t \rightarrow \infty} \frac{\overline{F}_{{\rm EV}}(tx) }{ \overline{F}_{{\rm EV}}(t)}
&=&
x \lim_{t \rightarrow \infty} 
\frac
{(1/\sigma)\exp(-\exp(-(tx-\mu)/\sigma)) \exp(-(tx-\mu)/\sigma)}
{(1/\sigma)\exp(-\exp(-(t-\mu)/\sigma)) \exp(-(t-\mu)/\sigma)}
\\
&=&
\left\{
\begin{matrix}
\infty, & {\rm if} & 0<x<1 \\
1, & {\rm if} & x=1 \\
0, & {\rm if} & x>1,
\end{matrix}
\right.
\end{eqnarray*}
i.e., the Gumbel distribution is rapidly varying at infinity with index $-\infty$.

The index of regular variation of the other distributions can be obtained analogously. 
To obtain the index of regular variation of the TCEV$(\mu,\sigma,\mu_1,\sigma_1,\alpha)$ 
distribution we used the software MATHEMATICA \citep{WOLFRAM} as follows:
\small
\begin{verbatim}
Clear[a];Clear[s1];Clear[s2];Clear[x]
Clear[a];Clear[s1];Clear[s2];Clear[x]
Limit[((((1 + a)/s1)*Exp[-(t*x - m1)/s1]*
      Exp[-Exp[-(t*x - m1)/s1]] - (a/s2)*Exp[-(t*x - m2)/s2]*
      Exp[-Exp[-(t*x - m2)/s2]])/(((1 + a)/s1)*Exp[-(t - m1)/s1]*
      Exp[-Exp[-(t - m1)/s1]] - (a/s2)*Exp[-(t - m2)/s2]*
      Exp[-Exp[-(t - m2)/s2]])), {t -> Infinity}, 
 Assumptions :> {0 < a < 1, 0 < x < 1, 0 < s1 < s2, m1 < m2}]
{\[Infinity]}
Clear[a];Clear[s1];Clear[s2];Clear[x]
Limit[((((1 + a)/s1)*Exp[-(t*x - m1)/s1]*
      Exp[-Exp[-(t*x - m1)/s1]] - (a/s2)*Exp[-(t*x - m2)/s2]*
      Exp[-Exp[-(t*x - m2)/s2]])/(((1 + a)/s1)*Exp[-(t - m1)/s1]*
      Exp[-Exp[-(t - m1)/s1]] - (a/s2)*Exp[-(t - m2)/s2]*
      Exp[-Exp[-(t - m2)/s2]])), {t -> Infinity}, 
 Assumptions :> {0 < a < 1, x > 1, 0 < s1 < s2}]
{0}
\end{verbatim}
\normalsize

\paragraph{Criterion of Rigby et al. (2014).}
Let $X \sim {\rm EV}(\mu,\sigma)$.
%, i.e., $X$ has probability density function (pdf) (1) in the main article.
%Let $X$ follow a generalized extreme value distribution in the real line, which is denoted by
%$X \sim {\rm GEV}(\mu,\sigma,\alpha)$
%Let $X$ be a random variable with $X \sim {\rm GEV}(\mu,\sigma,\alpha)$, i.e., $X$ has a generalized extreme value
%distribution with probability density function (pdf)
The logarithm of the pdf, given in the main article, is
$$
\ln(f_{{\rm EV}}(x; \mu, \sigma,\alpha))=
-\ln(\sigma)-\frac{x-\mu}{\sigma}
-\exp\left(-\frac{x-\mu}{\sigma}\right) \approx -\frac{1}{\sigma} x,
$$
as $x \to \infty$. We have $k_3=1$ and $k_4=1/\sigma$.
%, $k_5=\exp(\mu/\sigma)$ and $k_6=-1/\sigma$.

Let $X \sim {\rm GEV}(\mu,\sigma,\alpha)$.
%, i.e., $X$ has probability density function (pdf) (5) in the main article.
%\begin{eqnarray*}
%&&f_{{\rm GEV}}(x; \mu, \sigma,\alpha)=\frac{1}{\sigma}
%{\rm exp} \left( -\left[ 1+\alpha \left( \frac{x-\mu}{\sigma}  \right) \right]^{-1/\alpha} \right)
%\left[ 1+\alpha \left( \frac{x-\mu}{\sigma}  \right) \right]^{-1/\alpha-1},\\
%&&\{x:1+\alpha ( x-\mu) / \sigma >0 \}.
%\end{eqnarray*}
The logarithm of the pdf, given in the main article, is
\begin{eqnarray*}
\ln(f_{{\rm GEV}}(x; \mu, \sigma,\alpha))&=&-\ln(\sigma)-\left[1+\alpha\left(\frac{x-\mu}{\sigma}\right)\right]^{-\frac{1}{\alpha}}-\left(\frac{1}{\alpha}+1\right)\ln\left[1+\alpha\left(\frac{x-\mu}{\sigma}\right)\right]\\
&\approx&-\left(\frac{1}{\alpha}+1\right)\ln x,
\end{eqnarray*}
as $x \to \infty$, if $\alpha>0$. We have $k_1=1$ and $k_2=1+ 1/\alpha$.

Let $X \sim {\rm EGu}(\mu,\sigma,\alpha)$.
%, i.e., $X$ has probability density function (pdf) (?) in the main article.
The logarithm of the pdf, given in the main article, is
\begin{eqnarray*}
\ln(f_{{\rm EGu}}(x; \mu, \sigma,\alpha))&=&
-\ln\left(\frac{\alpha}{\sigma}\right)-\frac{x-\mu}{\sigma}
-\exp\left(-\frac{x-\mu}{\sigma}\right)  \\
&& +(\alpha-1)\ln\left[ 1- \exp\left( -\exp\left(-\frac{x-\mu}{\sigma}\right)\right)\right] \\
&\approx& -\frac{1}{\sigma} x -\frac{\alpha -1}{\sigma}x=-\frac{\alpha}{\sigma}x,
\end{eqnarray*}
as $x \to \infty$. For the approximation above we used that $\exp(-(x-\mu)/\sigma) \to 0$ as $x \to \infty$ and the Taylor series approximation $\exp(-y) \approx (1-y)$ as $y \to 0$ and hence $(1- \exp\left( -\exp\left(-(x-\mu)/\sigma\right)\right)) \approx \exp(-(x-\mu)/\sigma)$ as $x \to \infty$. Therefore, $k_3=1$ and $k_4=\alpha/\sigma$.

Let $X \sim {\rm TEV}(\mu,\sigma,\alpha)$.
%, i.e., $X$ has probability density function (pdf) (?) in the main article.
The logarithm of the pdf, given in the main article, is
\begin{eqnarray*}
\ln(f_{{\rm TEV}}(x; \mu, \sigma,\alpha)) &=&-\ln(\sigma)-\frac{x-\mu}{\sigma}
-\exp\left(-\frac{x-\mu}{\sigma}\right) \\
&& + \ln\left[ 1 + \alpha -2\alpha \exp\left( -\exp\left(-\frac{x-\mu}{\sigma}\right)\right)\right] 
\approx
%-\ln(\sigma)-\frac{x-\mu}{\sigma}
%-\exp\left(-\frac{x-\mu}{\sigma}\right)  
%+ \ln(1-\alpha) \\
%&+& 2\left(\frac{\alpha}{\alpha-1}\right) \exp\left( -\exp\left(-\frac{x-\mu}{\sigma}\right)\right) \\
%&\approx&
-\frac{1}{\sigma} x, \nonumber
\end{eqnarray*}
as $x \to \infty$. 
%Using that $\exp( -\exp(-(x-\mu)/(\sigma))) \to 1$ as $x \to \infty$ and Taylor series expansion $\ln(1+ \alpha-2\alpha y)=\ln(1-\alpha)+2(\alpha/(\alpha - 1))y$ as $y \to 1$. 
We have $k_3=1$, $k_4=1/\sigma$. 
%Note that, for moderate values of $x$, the second order term is relevant. When $-1<\alpha\leq 1$ decreases, $\ln(1-\alpha)$ increases to approximate $0.7$, $\alpha/(\alpha-1)$ increases to $0.5$ and also $\ln(f_{{\rm TEV}}(x; \mu, \sigma,\alpha))$, but is limited.
The fourth term of $\ln(f_{{\rm TEV}}(x; \mu, \sigma))$ tends to $\ln(1-\alpha)$ as $x \to \infty$.  Therefore, as $x \to \infty$, $\ln(f_{{\rm TEV}}(x; \mu, \sigma,\alpha))$ can be bigger or smaller than $\ln(f_{{\rm EV}}(x; \mu, \sigma))$, since $\ln(1-\alpha)>0$ for $\alpha<0$, and $\ln(1-\alpha)<0$ for $\alpha>0$, i.e., the TEV distribution can have heavier or lighter right tail than the EV distribution.

Let $X \sim {\rm GTIEV3}(\mu,\sigma,\alpha)$.
%, i.e., $X$ has probability density function (pdf) (?) in the main article.
The logarithm of the pdf, given in the main article, is
\begin{eqnarray*}
\ln(f_{{\rm GTIEV3}}(x; \mu, \sigma,\alpha))&=&
-\ln(\sigma)-\frac{x-\mu}{\sigma}
-(\alpha+1) \ln\left[1+ \frac{1}{\alpha}\exp\left(-\frac{x-\mu}{\sigma}\right)\right] \\
&\approx&
-\ln(\sigma)-\frac{x-\mu}{\sigma}
-\frac{\alpha+1}{\alpha} \exp\left(-\frac{x-\mu}{\sigma}\right) 
\approx 
-\frac{1}{\sigma} x,
\end{eqnarray*}
as $x \to \infty$. 
For the first approximation we used that $\exp(-(x-\mu)/\sigma) \to 0$ as $x \to \infty$ and the Taylor series approximation $\ln(1+y/\alpha) \approx y/\alpha$ as $y \to 0$.
We have $k_3=1$ and $k_4=1/\sigma$.
For $\alpha>0$, $(\alpha +1)/\alpha >1$. Thus, as $x \to \infty$, $\ln(f_{{\rm GTIEV3}}(x; \mu, \sigma,\alpha))$ is smaller than $\ln(f_{{\rm EV}}(x; \mu, \sigma))$, i.e., the GTIEV3 distribution have lighter right tail than the EV distribution.

Let $X \sim {\rm EGa}(\mu,\sigma,\alpha)$.
The logarithm of the pdf, given in the main article, is
\begin{eqnarray*}
\ln(f_{{\rm EGa}}(x; \mu, \sigma,\alpha))&=&
-\ln(\Gamma(\alpha)\sigma)-\alpha \frac{x-\mu}{\sigma}-\exp\left(-\frac{x-\mu}{\sigma}\right) 
\approx -\frac{\alpha}{\sigma} x,
\end{eqnarray*}
as $x \to \infty$. We have $k_3=1$ and $k_4=\alpha/\sigma$.

Let $X \sim {\rm GGu}(\mu,\sigma,\alpha)$.
%, i.e., $X$ has probability density function (pdf) (?) in the main article.
The logarithm of the pdf, given in the main article, is
\begin{eqnarray*}
\ln(f_{{\rm GGu}}(x; \mu, \sigma,\alpha))&=&
\ln\left(\frac{\alpha}{\sigma}\right)-\frac{x-\mu}{\sigma}
+\exp\left(-\frac{x-\mu}{\sigma}\right) \\
&&-(\alpha+1)\ln\left[ \exp\left(\exp\left(-\frac{x-\mu}{\sigma}\right)\right) - 1 \right] \\
&& -2\ln\left[ 1 + \left(\exp\left(\exp\left(-\frac{x-\mu}{\sigma}\right)\right) - 1\right)^{-\alpha} \right] \\
&\approx& -\frac{1}{\sigma} x +\frac{\alpha +1}{\sigma}x -2 \frac{\alpha}{\sigma} x =-\frac{\alpha}{\sigma}x,
\end{eqnarray*}
as $x \to \infty$. For the approximation above we used that $\exp(-(x-\mu)/\sigma) \to 0$ as $x \to \infty$ and the Taylor series approximation $\exp(y) \approx (1+y)$ as $y \to 0$ and hence $(\exp\left(\exp\left(-(x-\mu)/\sigma\right)\right) - 1) \approx \exp(-(x-\mu)/\sigma)$ as $x \to \infty$. Therefore, $k_3=1$ and $k_4=\alpha/\sigma$.

Let $X \sim {\rm GLIV}(\mu,\sigma,\alpha)$.
The logarithm of the p.d.f., given in the main article, is
\begin{eqnarray*}
\ln(f_{{\rm GLIV}}(x; \mu, \sigma,\alpha))&=&
-\ln\left[ \left(\frac{\alpha}{\beta}\right)^{\alpha}\frac{1}{\sigma B(\alpha,\beta)}\right]-\alpha \frac{x-\mu}{\sigma}-(\alpha+\beta)\ln\left[1+\frac{\alpha}{\beta}\exp\left(-\frac{x-\mu}{\sigma}\right)\right] \\
&\approx& -\frac{\alpha}{\sigma} x,
\end{eqnarray*}
as $x \to \infty$. We have $k_3=1$ and $k_4=\alpha/\sigma$.

Let $X \sim {\rm TCEV}(\mu,\sigma,\alpha)$.
%, i.e., $X$ has probability density function (pdf) (?) in the main article.
The logarithm of the pdf, given in the main article, can be written as
\begin{eqnarray*}
&&\ln(f_{{\rm TCEV}}(x; \mu, \sigma,\alpha))=
-\ln(\sigma_1)-\frac{x-\mu}{\sigma_1}-\exp\left(-\frac{x-\mu_1}{\sigma_1}\right) +  \ln\Biggl[1-\alpha+ \alpha \frac{\sigma_1}{\sigma_2}\\
&& \exp\left(-\frac{x-\mu_2}{\sigma_2}\right)  \exp\left(\frac{x-\mu_1}{\sigma_1}\right)\exp\left( -\exp\left(-\frac{x-\mu_2}{\sigma_2}\right) + \exp\left(-\frac{x-\mu_1}{\sigma_1}\right) \right)      \Biggr] \\
&&=
-\ln(\sigma_1)-\frac{x-\mu}{\sigma_1}-\exp\left(-\frac{x-\mu_1}{\sigma_1}\right) 
+ \ln\Biggl[1-\alpha+ \alpha \frac{\sigma_1}{\sigma_2}\\
&& \exp\left(\frac{\mu_2}{\sigma_2} - \frac{\mu_1}{\sigma_1}\right)  \exp\left(\left(\frac{1}{\sigma_1} - \frac{1}{\sigma_2} \right) x\right)\exp\left( -\exp\left(-\frac{x-\mu_2}{\sigma_2}\right) + \exp\left(-\frac{x-\mu_1}{\sigma_1}\right) \right)   \Biggr] \\
%&& \approx -\frac{1}{\sigma_1} x + \ln\left(\frac{\alpha}{1-\alpha}\right)+\ln\left(\frac{\sigma_1}{\sigma_2}\right)+\left( \frac{\mu_2}{\sigma_2} - \frac{\mu_1}{\sigma_1} \right) + \left( \frac{1}{\sigma_1} - \frac{1}{\sigma_2} \right) x \\
%&&= -\frac{1}{\sigma_2} x + \ln\left(\frac{\alpha}{1-\alpha}\right)+\ln\left(\frac{\sigma_1}{\sigma_2}\right)+\left( \frac{\mu_2}{\sigma_2} - \frac{\mu_1}{\sigma_1} \right) \\
&&\approx -\frac{1}{\sigma_2} x ,
\end{eqnarray*}
as $x \to \infty$ if $\sigma_1 < \sigma_2$. 
%The first approximation considered 
%$\exp\left( -\exp\left(-(x-\mu_2)/\sigma_2\right) \right) \approx 1$, 
%$\exp\left(  \exp\left(-(x-\mu_1)/\sigma_1\right) \right) \approx 1$ 
%and
%$(\alpha/(1-\alpha))(\sigma_1/\sigma_2)\exp(\mu_2/\sigma_2 - \mu_1/\sigma_1)  \exp((1/\sigma_1 - 1/\sigma_2) x) >> 1$, for moderate values of $x$. 
We have $k_3=1$ and $k_4=1/\sigma_2$.
%Note that, for moderate values of $x$, the second order term is relevant. 
%When $0<\alpha<0.5$ increases, $0<\alpha / (1-\alpha)<1$ increases and so $\ln(f_{{\rm TCEV}}(x; \mu, \sigma,\alpha))$.
When $x \to \infty$, the fourth term of $\ln(f_{{\rm TCEV}}(x; \mu, \sigma,\alpha))$ tends to infinity. Thus, $\ln(f_{{\rm TCEV}}(x; \mu, \sigma))$ is bigger than $\ln(f_{{\rm EV}}(x; \mu, \sigma))$, i.e., the TCEV distribution have heavier right tail than the EV distribution.

\section{Additional Monte Carlo simulation results}\label{simulationsupplement}

\begin{figure}[!htb]
\centering
  \includegraphics[height=99.5mm,width=99.5mm]{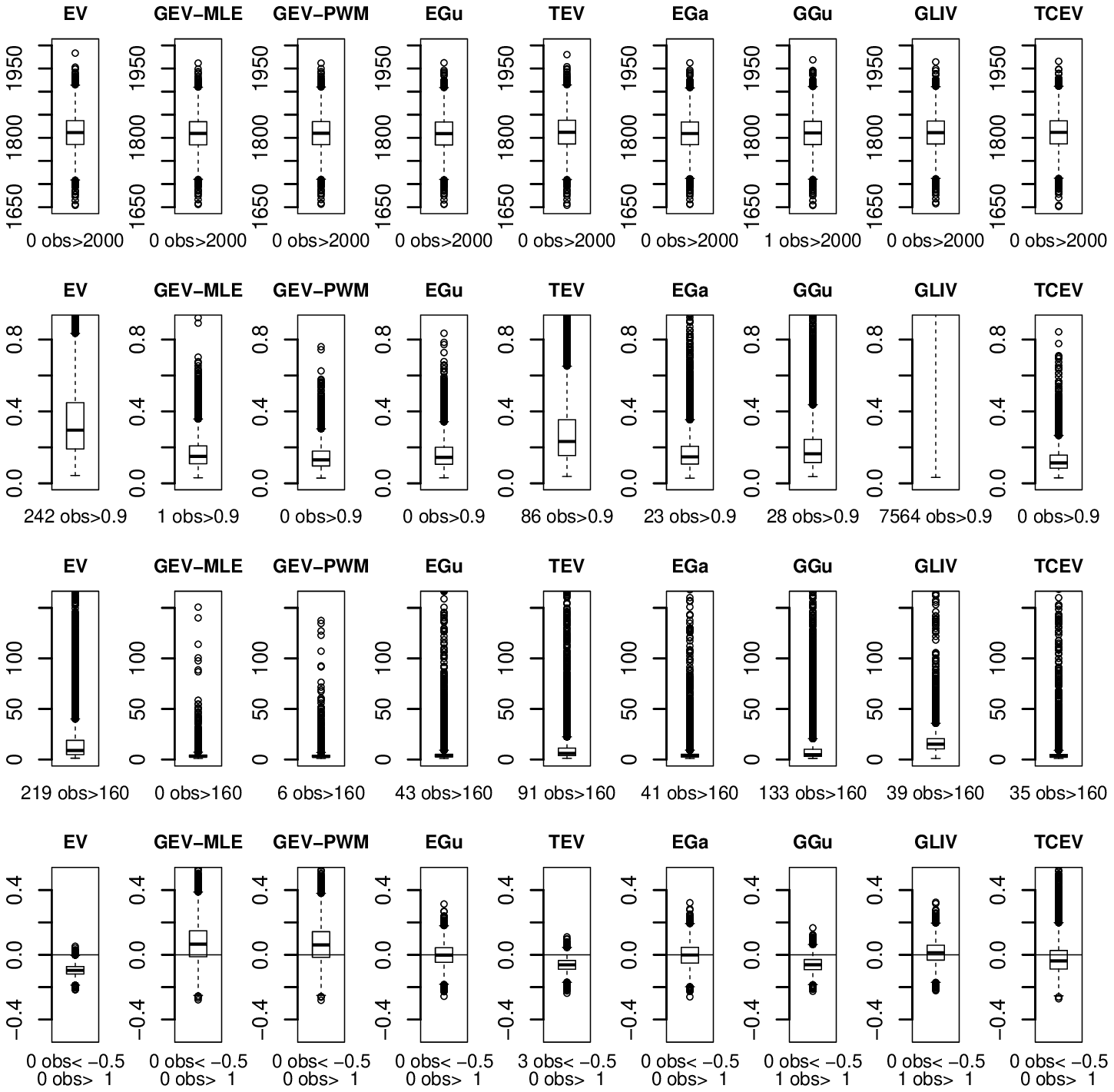}
  \caption{Boxplots of AIC (first row), ADR (second row), AD2R (third row) and quantile discrepancies (fourth row) - random samples generated from EGu.}
  \label{fig:discrepanciaEGu}
\end{figure}

\begin{figure}[!htb]
  \centering
  \includegraphics[height=95mm,width=99.5mm]{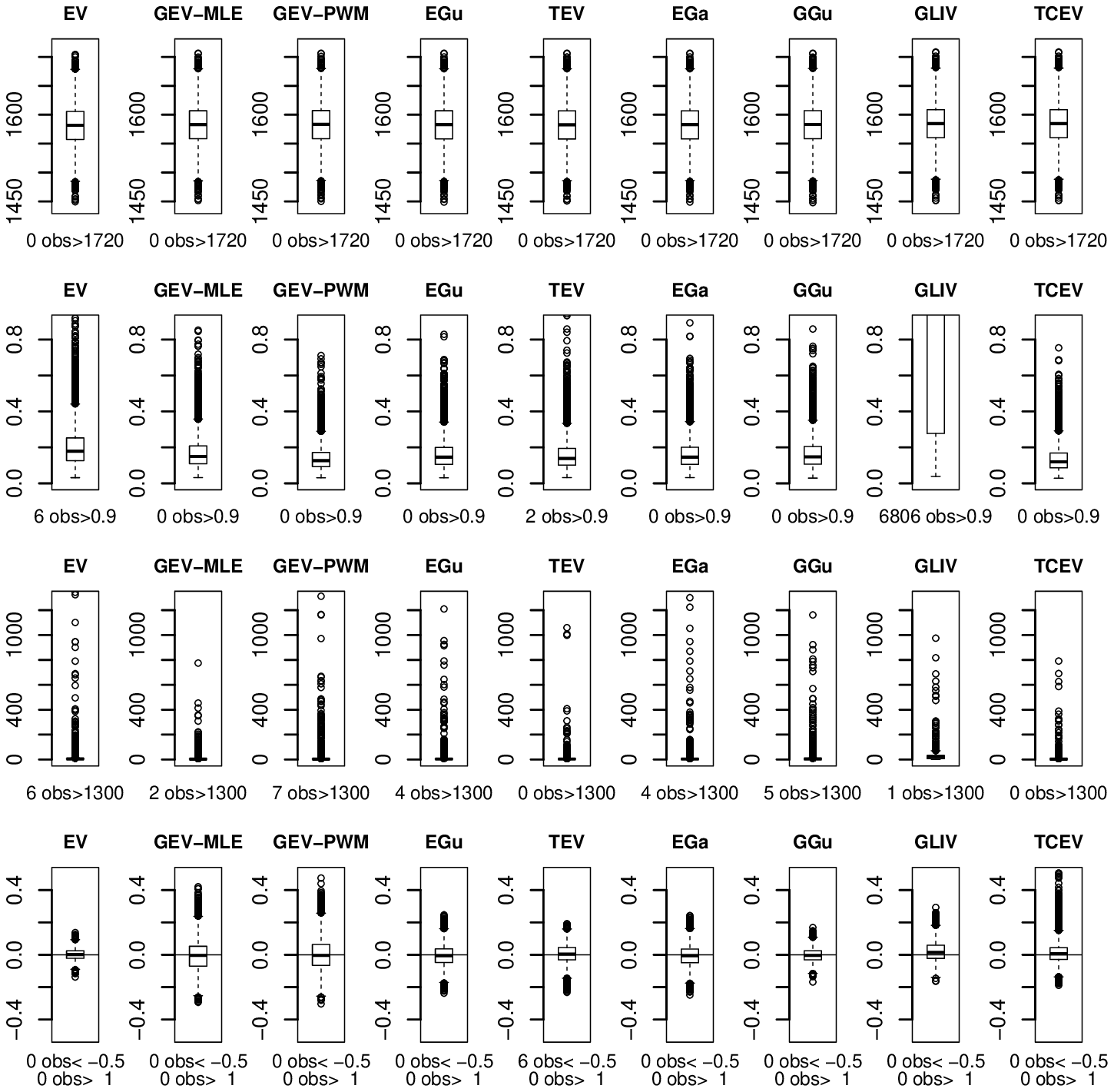}
  \caption{Boxplots of AIC (first row), ADR (second row), AD2R (third row) and quantile discrepancies (fourth row) - random samples generated from TEV.}
  \label{fig:discrepanciaTEV}
\end{figure}

\begin{figure}[!htb]
  \centering
  \includegraphics[height=95mm,width=99.5mm]{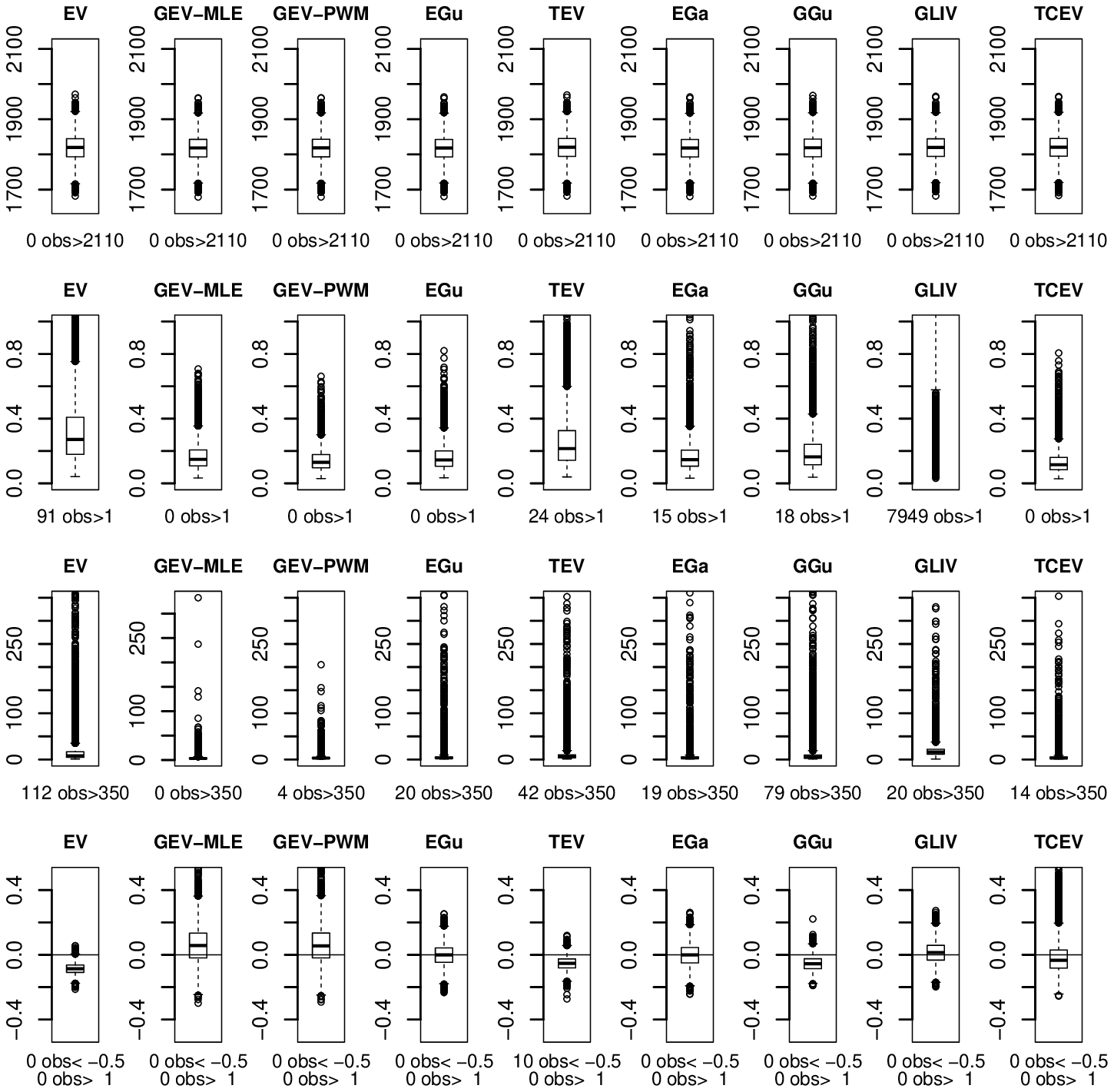}
  \caption{Boxplots of AIC (first row), ADR (second row), AD2R (third row) and quantile discrepancies (fourth row) - random samples generated from  EGa.}
  \label{fig:discrepanciaEGa}
\end{figure}

\begin{figure}[!htb]
  \centering
  \includegraphics[height=95mm,width=99.5mm]{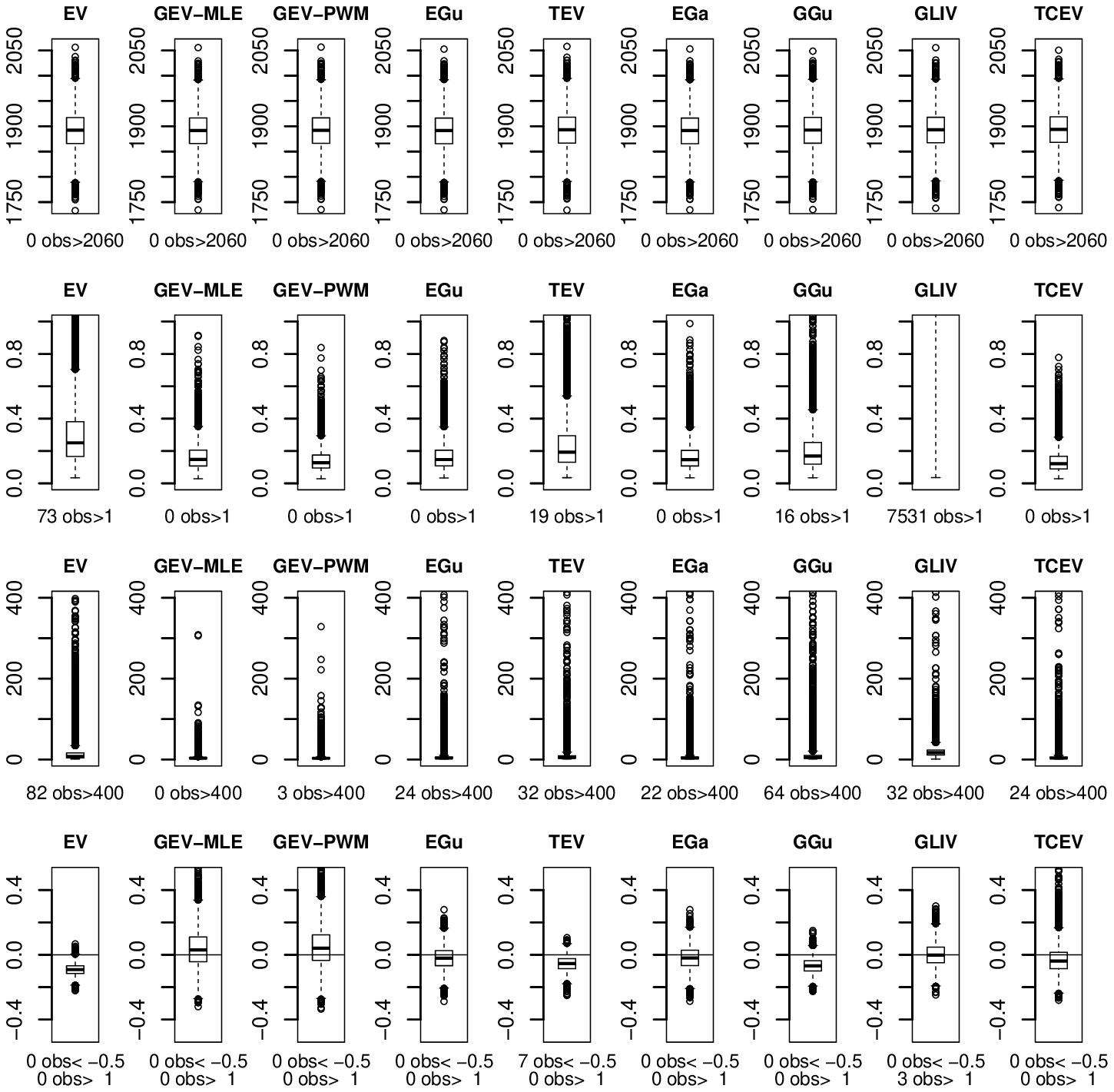}
  \caption{Boxplots of AIC (first row), ADR (second row), AD2R (third row) and quantile discrepancies (fourth row) - random samples generated from  GGu.}
  \label{fig:discrepanciaGGu}
\end{figure}

\begin{figure}[!htb]
  \centering
  \includegraphics[height=95mm,width=99.5mm]{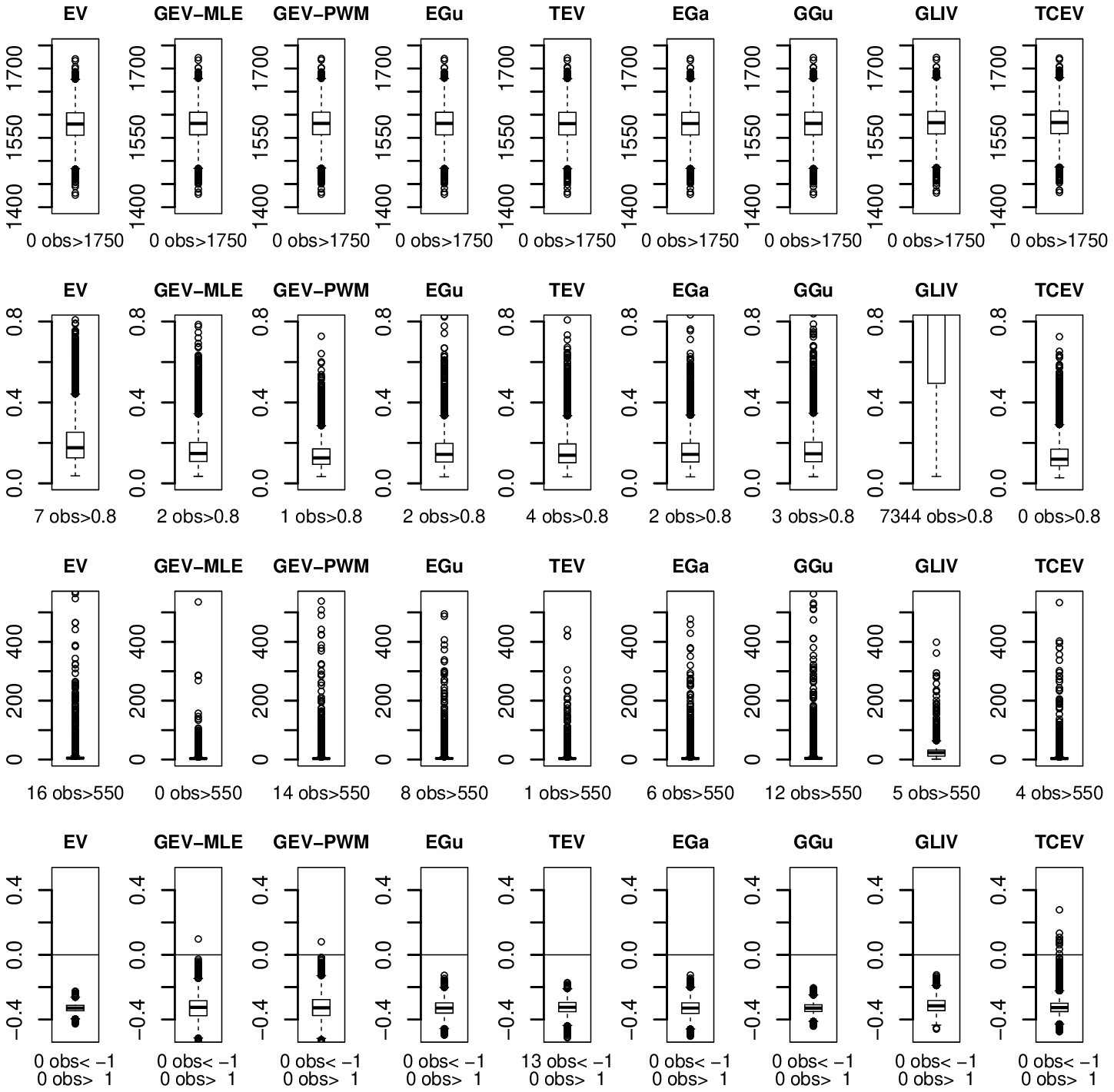}
  \caption{Boxplots of AIC (first row), ADR (second row), AD2R (third row) and quantile discrepancies (fourth row) - random samples generated from  GLIV.}
  \label{fig:discrepanciaGLIV}
\end{figure}

\begin{figure}[!htb]
  \centering
  \includegraphics[height=95mm,width=99.5mm]{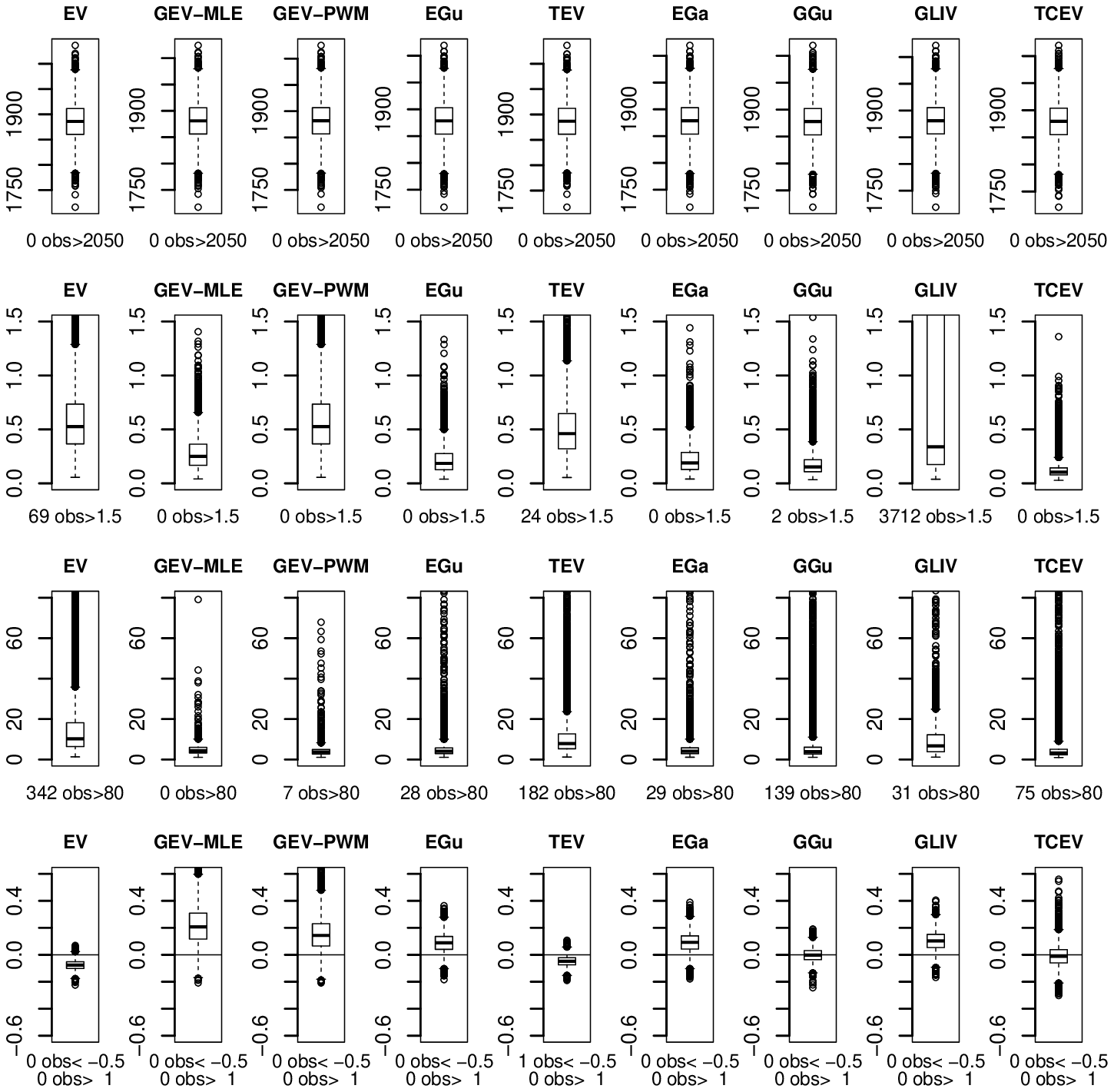}
  \caption{Boxplots of AIC (first row), ADR (second row), AD2R (third row) and quantile discrepancies (fourth row) - random samples generated from  TCEV.}
  \label{fig:discrepanciaTCEV}
\end{figure}

Figures~\ref{fig:discrepanciaEGu}-\ref{fig:discrepanciaTCEV} present the boxplots of AIC, ADR, AD2R, and the quantile discrepancies
of the fitted models, when the samples were generated from the exponentiated Gumbel EGu(0,1,0.6),
transmuted extreme value TEV(0,1,-0.99), three parameter
exponential-gamma EGa(0,1,0.6), generalized Gumbel GGu($0,1,0.7$), type IV generalized logistic GLIV(0,1,0.55,10) and two-component extreme value TCEV(0,1, 10,5,0.0125) distributions. Comments on these figures are given in Section 4 of the main article.


\section{Application}
\label{applicationdata}

Table~\ref{NOAA} presents the wind speed data.
\begin{table}[ht]
\centering
\caption{Wind speed data.}
\label{NOAA}
\begin{tabular}{rrrrrrrrrrrrrr}
\hline
 & year & Jan & Feb & Mar & Apr & May & Jun & Jul & Aug & Sep & Oct & Nov & Dec \\ 
\hline
1 & 1984 &  33 &  40 &  46 &  41 &  31 &  37 &  41 &  56 &  45 &  31 &  40 &  35 \\ 
  2 & 1985 &  33 &  43 &  36 &  36 &  48 &  45 &  51 &  44 &  38 &  36 &  40 &  32 \\ 
  3 & 1986 &  51 &  37 &  43 &  33 &  35 &  44 &  41 &  41 &  33 &  45 &  38 &  43 \\ 
  4 & 1987 &  62 &  45 &  51 &  39 &  35 &  58 &  48 &  35 &  43 &  49 &  43 &  39 \\ 
  5 & 1988 &  39 &  40 &  39 &  45 &  48 &  43 &  45 &  36 &  40 &  36 &  47 &  35 \\ 
  6 & 1989 &  40 &  39 &  44 &  37 &  36 &  38 &  37 &  41 &  38 &  36 &  36 &  48 \\ 
  7 & 1990 &  37 &  40 &  38 &  37 &  37 &  38 &  49 &  66 &  39 &  45 &  37 &  35 \\ 
  8 & 1991 &  39 &  52 &  66 &  51 &  39 &  64 &  59 &  36 &  36 &  36 &  41 &  41 \\ 
  9 & 1992 &  39 &  45 &  40 &  37 &  33 &  66 &  38 &  59 &  38 &  41 &  45 &  35 \\ 
  10 & 1993 &  43 &  39 &  74 &  63 &  37 &  45 &  52 &  43 &  44 &  52 &  36 &  43 \\ 
  11 & 1994 &  46 &  40 &  43 &  29 &  39 &  53 &  32 &  41 &  52 &  31 &  46 &  48 \\ 
  12 & 1995 &  49 &  41 &  32 &  37 &  29 &  43 &  40 &  47 &  45 &  38 &  28 &  30 \\ 
  13 & 1996 &  40 &  36 &  37 &  38 &  37 &  33 &  30 &  34 &  38 &  45 &  40 &  31 \\ 
  14 & 1997 &  39 &  31 &  31 &  38 &  32 &  34 &  45 &  39 &  31 &  29 &  39 &  36 \\ 
  15 & 1998 &  34 &  55 &  38 &  37 &  36 &  34 &  44 &  32 &  54 &  30 &  39 &  30 \\ 
  16 & 1999 &  41 &  33 &  36 &  39 &  33 &  33 &  30 &  40 &  44 &  61 &  34 &  26 \\ 
  17 & 2000 &  38 &  26 &  34 &  36 &  28 &  36 &  43 &  35 &  43 &  37 &  40 &  35 \\ 
  18 & 2001 &  36 &  28 &  41 &  30 &  31 &  48 &  43 &  43 &  49 &  36 &  38 &  30 \\ 
  19 & 2002 &  33 &  35 &  36 &  45 &  29 &  43 &  33 &  39 &  38 &  29 &  38 &  41 \\ 
  20 & 2003 &  31 &  35 &  40 &  33 &  51 &  33 &  40 &  45 &  32 &  29 &  35 &  37 \\ 
  21 & 2004 &  35 &  30 &  32 &  39 &  32 &  39 &  38 &  39 &  83 &  30 &  33 &  39 \\ 
  22 & 2005 &  33 &  36 &  39 &  44 &  31 &  43 &  44 &  43 &  41 & 101 &  37 &  33 \\ 
  23 & 2006 &  55 &  43 &  30 &  32 &  32 &  46 &  47 &  43 &  32 &  31 &  32 &  41 \\ 
  24 & 2007 &  37 &  37 &  44 &  43 &  33 &  41 &  49 &  39 &  40 &  43 &  36 &  35 \\ 
  25 & 2008 &  37 &  44 &  39 &  47 &  52 &  39 &  39 &  48 &  37 &  35 &  40 &  33 \\ 
  26 & 2009 &  38 &  36 &  36 &  38 &  40 &  49 &  54 &  47 &  37 &  33 &  39 &  36 \\ 
  27 & 2010 &  36 &  62 &  43 &  32 &  32 &  58 &  35 &  35 &  38 &  32 &  33 &  46 \\ 
  28 & 2011 &  40 &  44 &  51 &  59 &  33 &  41 &  36 &  53 &  45 &  39 &  32 &  31 \\ 
  29 & 2012 &  40 &  35 &  41 &  38 &  66 &  49 &  52 &  61 &  36 &  52 &  30 &  37 \\ 
  30 & 2013 &  31 &  35 &  45 &  40 &  40 &  47 &  38 &  51 &  37 &  46 &  39 &  31 \\ 
  31 & 2014 &  36 &  36 &  46 &  44 &  46 &  58 &  46 &  50 &  39 &  44 &  38 &  \\ 
\hline
\end{tabular}
\end{table}


\small
\begin{thebibliography}{9}

\bibitem[Adeyemi \& Ojo(2003)]{ADEYEMI}
Adeyemi, S. \& Ojo, M. O.~(2003). 
A generalization of the Gumbel Distribution. 
{\it Kragujevac Journal of Mathematics}, {\bf 25}, 19-29.

\bibitem[Aitkin \& Rubin(1985)]{AITKIN}
Aitkin, M. \& Rubin, D. B.~(1985).
Estimation and hypothesis testing in finite mixture models.
{\it Journal of the Royal Statistical Society B},{\bf 47}, 67-75.

\bibitem[Aryal \& Tsokos(2009)]{ARYAL}
Aryal, G. R. \& Tsokos, C. P.~(2009).
On transmuted extreme value distribution with application.
{\it Nonlinear Analysis: Theory, Methods \& Applications\/}, {\bf 71}, 1401-1407.

\bibitem[Balakrishnan \& Leung(1988)]{BALAKRISHNAN1988}
Balakrishnan, N. \& Leung, M. Y.~(1988).
Order statistics from the type I generalized logistic distribution
{\it Communications in Statistics - Simulation and Computation\/}, {\bf 17}, 25-50.

\bibitem[Bruxer et al.(2008)]{BRUXER}
Bruxer, J., Thompson, A., \& Eng, P. (2008). 
{\it St. Clair River Hydrodynamic Modelling Using RMA2 Phase 1 Report}.
IUGLS SCRTT, Canada. 

\bibitem[Castillo et al.(2005)]{CASTILLO}
Castillo, E., Hadi, A. S., Balakrishnan, N., \& Sarabia, J. M.~(2005).
{\it Extreme Value and Related Models with Applications in Engineering and Science.}
New Jersey: John Wiley \& Sons.

\bibitem[Catchpole \& Morgan(1997)]{CATCHPOLE}
Catchpole, E. A. \& Morgan, B. J. T.~(1997). 
Detecting parameter redundancy. 
{\it Biometrika}, {\bf 84}, 187-196.

\bibitem[Cleveland et al.(1990)]{CLEVELAND} 
Cleveland, R. B., Cleveland W. S., McRae J.E., \& Terpenning, I.~(1990).
STL: A Seasonal-Trend Decomposition Procedure Based on Loess. 
{\it Journal of Official Statistics}, {\bf 6}, 3-73.

\bibitem[Coles(2001)]{COLES}
Coles, S.~(2001).
{\it An Introduction to Statistical Modeling of Extremes.}
London: Springer-Verlag.

\bibitem[Cooley et al.(2006)]{COOLEY}
Cooley, D., Naveau P., Jomelli V., Rabatel A., \& Grancher D.~(2006).
A Bayesian hierarquical extreme value model for lichenometry.
{\it Environmetrics}, {\bf 17}, 555-574.

\bibitem[Cooray (2010)]{COORAY}
Cooray, K.~(2010). 
Generalized Gumbel distribution.
{\it Journal of Applied Statistics}, {\bf 37}, 171-179.

\bibitem[Cordeiro et al.(2012)]{CORDEIRO2012}
Cordeiro, G. M., Nadarajah, S., \& Ortega, E. M. M.~(2012). 
The Kumaraswamy Gumbel distribution. 
{\it Statistical Methods \& Applications}, {\bf 21}, 139-168.

\bibitem[Cordeiro et al.(2013)]{CORDEIRO2013}
Cordeiro, G. M., Ortega, E. M. M., \& da Cunha, D. C. C.~(2013).
The exponentiated generalized class of distributions.
{\it Journal of Data Science}, {\bf 11}, 1-27.

\bibitem[Danielsson et al.(2006)]{DANIELSSON}
Danielsson, J., Jorgensen, B. N., Sarma, M., \& de Vries, C. G.~(2006). 
Comparing downside risk measures for heavy tailed distributions. 
{\it Economics Letters}, {\bf 92}, 202-208. 

\bibitem[Doornik(2013)]{DOORNIK}
Doornik, J. A.~(2013).
{\it Object-Oriented Matrix Language Programming using Ox.} 
London: Timberlake Consultants Press.

\bibitem[Dubey(1969)]{DUBEY}
Dubey, S. D.~(1969). 
A new derivation of the logistic distribution. 
{\it Naval Research Logistics Quarterly}, {\bf 16}, 37-40.

\bibitem[Ferrari \& Pinheiro(2012)]{PINHEIRO2012}
Ferrari, S. L. P. \& Pinheiro, E. C.~(2012).
Small-sample likelihood inference in extreme-value regression models.
{\it Journal of Statistical Computation and Simulation}, {\bf 84}, 582-595.

\bibitem[Ferrari \& Pinheiro(2015)]{PINHEIRO2015}
Ferrari, S. L. P. \& Pinheiro, E. C.~(2015).
Small-sample one-sided testing in extreme value regression models.
{\it Advances in Statistical Analysis}.
{\small \tt{ \url{http://dx.doi.org/10.1007/s10182-015-0251-y}}}.

\bibitem[Gilleland(2005)]{GILLELAND}
Gilleland, E. \& Nychka D.~(2005).
Statistical models for monitoring and regulating ground-level ozone.
{\it Environmetrics}, {\bf 16}, 535-546. 


\bibitem[Gnedenko(1943)]{GNEDENKO}
Gnedenko, B.~(1943).
Sur la distribuition limite du terme maximum d'une série aléatoire.
{\it Ann. Math.}, {\bf   44}, 423-453. 
Translated and reprinted in {\it Breakthroughs in Statistics}, Vol.{\bf I}~(1992), eds. S. Kotz and N.L. Johnson, Springer-Verlag, pp. 195-225.

\bibitem[Gumbel(1935)]{GUMBEL1935}
Gumbel, E. J.~(1935). 
Les valeurs extrêmes des distributions statistiques. 
{\it Annales de l'Institut Henri Poincaré}, {\bf 5}, 115-158. 
Presses universitaires de France.

\bibitem[de Haan(1970)]{HAAN}
de Haan, L.~(1970).
{\it On Regular Variation and Its Application to the Weak Convergence of Sample Extremes}.
Mathematical Centre Tracts 32.
Amsterdam: Mathematics Centre.

\bibitem[Hald(1952)]{HALD}
Hald, A.~(1952). 
{\it Statistical Theory with Engineering Applications}.
7th ed. Canada: John Wiley \& Sons.

\bibitem[Hosking(1994)]{HOSKING}
Hosking, J. R. M.~(1994).
The four-parameter kappa distribution.
{\it IBM Journal of Research and Development}, {\bf 38}, 251-258.

\bibitem[Hosking \& Wallis(1997)]{HOSKING1997}
Hosking, J. R. M. \& Wallis, J. R.~(1997).
{\it Regional Frequency Analysis: An Approach Based on L-Moments.} 
Cambridge: Cambridge University Press.

\bibitem[Huang(2005)]{HUANG}
Huang, G.~(2005).
Model Identifiability.
{\it Encyclopedia of Statistics in Behavioral Science}, {\bf 3}, 1249-1251.
Chichester: John Wiley \& Sons.
{\small \tt{ \url{http://onlinelibrary.wiley.com/doi/10.1002/0470013192.bsa399/pdf}}}.

\bibitem[Hubert \& Vandervieren(2008)]{HUBERT}
Hubert, M. \& Vadervieren, E.~(2008).
An adjusted boxplot for skewed distributions.
{\it Computational Statistics \& Data Analysis\/}, {\bf   52}, 5186-5201.

%\bibitem[Hyndman \& Athanasopoulos(2013)]{HYNDMAN}
%Hyndman, R.J. \& Athanasopoulos, G.~(2013).
%Forecasting: principles and practice.
%{\small \tt{ \url{http://otexts.org/fpp/}}}.
%Accessed on April 15, 2015.

\bibitem[Jenkinson(1955)]{JENKINSON}
Jenkinson, A. F.~(1955). 
The frequency distribution of the annual maximum (or minimum) values of meteorological elements. 
{\it Quarterly Journal of the Royal Meteorological Society}, {\bf 81}, 348, 158-171.

\bibitem[Jeong et al.(2014)]{JEONG}
Jeong, B. Y., Murshed, M. S., Am Seo, Y., \& Park, J. S.~(2014). 
A three-parameter kappa distribution with hydrologic application: a generalized Gumbel distribution. 
{\it Stochastic Environmental Research and Risk Assessment}, {\bf 28}, 8, 2063-2074.

\bibitem[Kotz \& Nadajarah(2000)]{KOTZ2000}
Kotz, S. \& Nadajarah, S.~(2000).
{\it Extreme Value Distributions: Theory and Applications.\/}
London: Imperial College Press.

%\bibitem[Lawles(2003)]{LAWLESS}
%Lawless, J. F.~(2003).
%{\it Statistical Models and Methods for Lifetime Data.\/}
%2nd ed. New Jersey: John Wiley \& Sons.

\bibitem[Luce\~no(2005)]{LUCENO}
Luce\~no, A.~2005.
Fitting the generalized Pareto distribution to data using maximum goodness of fit estimators.
{\it Computational Statistics \& Data Analysis}, {\bf 51}, 904-917.

\bibitem[Markovich(2007)]{MARKOVICH}
Markovich, N.~(2007).
{\it Nonparametric Analysis of Univariate Heavy-Tailed Data.}
Wiley Series in Probability and Statistics, 311-318.
%John Wiley \& Sons.

\bibitem[Nadarajah \& Kotz(2004)]{NADARAJAH2004}
Nadarajah, S. \& Kotz, S.~(2004). 
The beta Gumbel distribution. 
{\it Mathematical Problems in Engineering}, {\bf 4}, 323-332.

\bibitem[Nadarajah(2006)]{NADARAJAH2006}
Nadarajah, S.~(2006).
The exponentiated Gumbel distribution with climate application.
{\it Environmetrics}, {\bf 17}, 13-23.

\bibitem[Ojo(2001)]{OJO}
Ojo, M. O.~(2001).
Some relationships between the generalized Gumbel and other distributions. 
{\it Kragujevac Journal of Mathematics}, {\bf 23}, 101-106.

\bibitem[Parida(1999)]{PARIDA}
Parida, B. P.~(1999).
Modeling of Indian summer monsoon rainfall using a four-parameter kappa distribution. 
{\it International journal of climatology}, {\bf 19}, 1389-1398.

\bibitem[Park \& Jung(2002)]{PARK}
Park, J. S. \& Jung, H. S.~(2002).
Modeling Korean extreme rainfall using a kappa distribution and maximum likelihood estimate. 
{\it Theoretical and Applied Climatology}, {\bf 72}, 55-64.

\bibitem[Pescim et al.(2012)]{PESCIM2012}
Pescim, R. R., Cordeiro, G. M., Demétrio, C. G. B., Ortega, E. M. M., \& Nadarajah, S.~(2012).
The new class of Kummer beta generalized distributions.
{\it SORT-Statistics and Operations Research Transactions}, {\bf 36}, 153-180.

\bibitem[Prentice(1975)]{PRENTICE1975}
Prentice, R. L.~(1975).
Discrimination among some parametric models.
{\it Biometrika}, {\bf 62}, 607-614.

\bibitem[Prentice(1976)]{PRENTICE1976}
Prentice, R. L.~(1976).
A generalization of the Probit and logit methods for dose response curves.
{\it Biometrics}, {\bf 32}, 761-768.
 
%\bibitem[Press et al.(1992)]{PRESS}
%Press, W. H., Teulosky, S. A., Vetterling, W. T., \& Flannery, B. P.~(1992).
%{\it Numerical Recipes in C: The Art of Scientific Computing.\/}
%2nd ed. London: Prentice Hall.

\bibitem[Reed \& Robson(1999)]{REED}
Reed, D. W. \& Robson, A. J.~(1999). 
Statistical procedures for flood frequency estimation.
In {\it  Flood Estimation Handbook}, volume 3.
Wallingford: Institute of Hydrology.

\bibitem[Resnick(2007)]{RESNICK}
Resnick, S.~(2007). 
{\it Heavy-Tail Phenomena: Probabilistic and Statistical Modeling}. 
New York: Springer.

\bibitem[Rigby et al.(2014)]{RIGBY}
Rigby, R. A., Stasinopoulous, D. M., Heller, G., \& Voudouris, V.~(2014).
{\it The Distribution Toolbox of GAMLSS}.
www.gamlss.org

\bibitem[Rossi et al.(1984)]{ROSSI}
Rossi, F., Fiorentino, M., \& Versace, P.~(1984). 
Two-component extreme value distribution for flood frequency analysis. 
{\it Water Resources Research}, {\bf 20}, 847-856.

\bibitem[Sanford(1997)]{SANFORD}
Sanford, L. P.~(1997). 
Turbulent mixing in experimental ecosystem studies. 
{\it Marine Ecology Progress Series}, 161, 265-293.

\bibitem[Shaw \& Buckley(2009)]{SHAW}
Shaw, W. T. \& Buckley, I. R.~(2009). 
The alchemy of probability distributions: beyond Gram-Charlier \& Cornish-Fisher expansions, and skew-normal or kurtotic-normal distributions. 
arXiv:0901.0434v1.

\bibitem[Sing \& Deng(2003)]{SINGH}
Singh, V. P. \& Deng, Z. Q.~(2003).
Entropy-based parameter estimation for kappa distribution. 
{\it Journal of Hydrologic Engineering}, {\bf 8}, 81-92.

\bibitem[Wright \& Nocedal(1999)]{WRIGHT}
Wright, S. J. \& Nocedal, J.~(1999).
{\it Numerical Optimization. Vol. 2}
New York: Springer.

\end{thebibliography}

\begin{thebibliography}{9}
\footnotesize

%\bibitem[Aryal \& Tsokos(2009)]{ARYAL}
%Aryal, G. R. \& Tsokos, C. P.~(2009).
%On transmuted extreme value distribution with application.
%{\it Nonlinear Analysis: Theory, Methods \& Applications\/}, {\bf 71}, 1401-1407.

\bibitem[Gradshteyn \& Ryzhik(2000)]{GRADSHTEYN}
Gradshteyn, I.S. \& Ryzhik, I.M.~(2000).
{\it Table os Integrals, Series and Products\/}.
Massachusetts: Academic Press.
%ver em http://en.wikipedia.org/wiki/Academic_Press

\bibitem[Wolfram Research(2012)]{WOLFRAM}
Wolfram Research, Inc.~(2012)
{\it Mathematica Edition: Version 9.0.\/}
Champaign, Illinois: Wolfram Research, Inc.
%Mathematica 9.0
%Author: Wolfram Research, Inc. Title: Mathematica Edition: Version 9.0 Publisher: Wolfram Research, Inc. Place of publication: Champaign, Illinois Date of publication: 2012

\end{thebibliography}
\end{document}